\documentclass[prd,showpacs,showkeys,floatfix,nofootinbib,eqsecnum,
                  preprint,12pt,tightenlines,fleqn]{revtex4}  
 
\usepackage{amsmath,amssymb,revsymb,graphicx,dcolumn}

\begin{document}
\title{The Superfluid Universe}
\author{
G.E.~Volovik 
}
\affiliation
{Low Temperature Laboratory, Aalto University, Finland\\
L.D. Landau Institute for Theoretical Physics, Moscow, Russia}

\date{\today}
\begin{abstract}
We discuss phenomenology of quantum vacuum. Phenomenology of macroscopic systems has three sources: thermodynamics,
topology and symmetry.  Thermodynamics of the self-sustained vacuum allows us to treat the problems
related to the vacuum energy: the cosmological constant problems. The 
natural value of the energy density of the equilibrium  self-sustained vacuum is zero.  Cosmology is discussed as the process of relaxation of vacuum towards the equilibrium state. The present value of the cosmological constant is very small compared to the Planck scale, because the present Universe is very old and thus is close to equilibrium.

Momentum space topology determines the universality classes of fermionic vacua. The Standard Model vacuum both in its massless and massive states is topological medium.
The  vacuum in its massless state shares the properties of superfluid  $^3$He-A, which is  topological superfluid. It belongs to the Fermi-point universality class, which has topologically protected fermionic quasiparticles. At low energy they behave as relativistic massless Weyl fermions. Gauge fields and gravity emerge together with Weyl fermions at low energy. This allows us to treat the hierarchy problem in Standard Model: the masses of elementary particles are very small compared to the Planck scale because the natural value of the quark and lepton masses is zero. The small nonzero masses appear in the infrared region, where the quantum vacuum acquires the properties of another topological superfluid, $^3$He-B, and 3+1 topological insulators. The other topological media in dimensions 2+1 and 3+1 are also discussed. In most cases,  topology is supported by discrete symmetry of the underlying microscopic system, which indicates the important role of discrete symmetry in Standard Model.

\end{abstract}

\newcommand{\beq}{\begin{equation}}
\newcommand{\eeq}{\end{equation}}
\newcommand{\beqa}{\begin{eqnarray}}
\newcommand{\eeqa}{\end{eqnarray}}
\newcommand{\bsubeqs}{\begin{subequations}}
\newcommand{\esubeqs}{\end{subequations}}
\newcommand{\dd}{\mathrm{d}}                    
\newcommand{\half}{{\textstyle \frac{1}{2}}}    



\maketitle
\tableofcontents


\section{Introduction. Phenomenology of quantum vacuum}
\subsection{Vacuum as macroscopic many-body system}

The aether of the 21-st century is the quantum vacuum. The quantum aether is a new form of matter. This substance has a very peculiar properties  strikingly different from the other forms of matter (solids, liquids, gases,  plasmas, Bose condensates, radiation, etc.) and from all the old aethers. The new aether has equation of state $p=-\epsilon$; it is Lorentz invariant;  and  as follows from the recent  cosmological observations its energy density is about $10^{-29}$g/cm$^3$ (i.e. the quantum aether by 29 orders magnitude lighter than water) and  it  is actually anti-gravitating. 

Quantum vacuum can be viewed as a macroscopic many-body system. 
 Characteristic energy scale in our vacuum
(analog of atomic scale in quantum liquids) is Planck energy $E_{\rm P}=(\hbar c^5/G)^{1/2} \sim 10^{19}$ GeV $\sim 10^{32}$K.
Our present Universe has extremely low energies and  temperatures compared to the Planck scale:
even the highest energy in the nowadays  accelerators  is extremely small compared to Planck energy: $E_{\rm max} \sim  10$ TeV $\sim 10^{17}$K$\sim 10^{-15}E_{\rm P}$. 
The temperature of cosmic background radiation is much smaller, $T_{CMBR} \sim  1$ K$\sim 10^{-32}E_{\rm P}$.

Cosmology belongs to ultra-low frequency physics. Expansion of Universe is extremely slow: the Hubble parameter compared to the characteristic Planck frequency $\omega_P=(c^5/G\hbar)^{1/2} $ is  $H  \sim 10^{-60}\omega_P$. This also means that at the moment our Universe is extremely close to equilibrium. This is natural for any many-body system: if there is no energy flux from environment the energy will be radiated away and the system will be approaching the equilibrium state with vanishing temperature and motion.

According to Landau, though the macroscopic many-body system can be very complicated,  at low energy and temperatures its description is highly simplified. Its behavior can be described in a fully phenomenological way, using the symmetry and thermodynamic consideration. Later it became clear that another factor also governs the low energy properties of a macroscopic system --  topology. 
The quantum vacuum is probably a very complicated system. However, using these three sources --  thermodynamics, symmetry and topology -- we may try to construct  the phenomenological theory of the quantum vacuum near its equilibrium state.

\subsection{3 sources of phenomenology: thermodynamics, symmetry and topology}

Following Landau, at low energy $E\ll E_{\rm P}$ the macroscopic quantum system  --  superfluid liquid  or our Universe --   contains two main components: vacuum (the ground state) and matter (fermionic and bosonic quasiparticles above the ground state). The physical laws which govern the matter component are more or less clear to us, because we are able to make experiments in the low-energy region and construct the theory. The quantum vacuum  occupies the Planckian and  trans-Planckian energy scales and it is governed by the microscopic (trans-Planckian) physics which is  still unknown.
However,  using our experience with a similar two-component quantum liquid we can expect  that the quantum vacuum component should also obey the  thermodynamic laws, which emerge in any macroscopically large system, relativistic or non-relativistic. This approach allows us to treat 
the   cosmological constant problems. Cosmological constant was introduced by Einstein \cite{Einstein1917}, and was interpreted as the energy density of the quantum vacuum \cite{Bronstein1933, Zeldovich1967}.  Astronomical observations (see, e.g., Refs.~\cite{Riess-etal1998,Perlmutter-etal1998,Komatsu2008}) confirmed the existence of cosmological constant which   value corresponds to the energy density of order $\Lambda_{\rm obs}\sim E_{\rm obs}^4$ with the characteristic  energy scale   $E_{\rm obs}\sim 10^{-3}\;\text{eV}$.
However, naive and intuitive theoretical estimation of the vacuum energy density as the zero-point energy of quantum fields  suggests that vacuum energy must have the Planck energy scale: $\sim E_{\rm P}^4 \sim 10^{120}E_{\rm obs}$. The huge disagreement between the naive expectations and observations is naturally resolved using the thermodynamics of quantum vacuum discussed in Sections II--V of this review. We shall see that the intuitive estimation for the vacuum energy density as $\sim E_{\rm P}^4$ is correct, but the relevant vacuum energy which enters Einstein equations as cosmological constant is somewhat different and its value in the fully equilibrium vacuum is zero.

The second element of the Landau phenomenological approach to macroscopic systems is symmetry.
It is in the basis of the modern theory of particle physics -- the Standard Model, and its extension to higher energy -- the Grand Unification  (GUT).  The vacuum of Standard Model and GUT  obeys the fundamental symmetries which become spontaneously broken at low energy, and are restored when the Planck energy scale is approached from below.  

This approach contains another huge disagreement between the naive expectations and observations. It concerns masses of elementary particles. The naive and intuitive estimation tells us that these masses should be on the order of  the characteristic energy scale in our vacuum, which is the Planck energy scale,  $M_{\rm theor} \sim E_{\rm P}$. However, the  masses of observed particles are many orders of magnitude smaller being below the electroweak energy scale $M_{\rm obs}<E_{\rm ew}\sim 1$  TeV $\sim 10^{-16}E_{\rm P}$.  This  is called the hierarchy problem.
There should be a general principle, which could resolve this paradoxes. This is the principle of emergent physics based on the topology in momentum space. This approach supports our  intuitive estimation 
of fermion masses of order $E_{\rm P}$, but this estimation is not valid for such vacua where the massless fermions are topologically protected. 

Let us consider fermionic quantum liquid $^3$He.  In laboratory we have four different states of this liquid. These are the normal liquid $^3$He, and three superfluid phases: $^3$He-A, $^3$He-B and $^3$He-A$_1$.  Only one of them, $^3$He-B, is fully gapped, and for this liquid the intuitive estimation of the gap (analog of mass) in terms of the characteristic energy scale is correct. However, the other liquids are gapless. This gaplessness is protected by the momentum space topology and thus is fundamental: it does not depend much on microscopic physics being robust to the perturbative modification of the interaction between the atoms of the liquid.

\subsection{Vacuum as topological medium}

Topology operates in particular with integer numbers -- topological charges -- which do not change under small deformation of the system. The conservation of these topological charges protects the Fermi surface and another object in momentum space -- the Fermi point -- from destruction.  They survive when the interaction between the fermions is introduced and modified. When the momentum of a particle approaches  the Fermi surface or  the Fermi point its energy necessarily vanishes. Thus the topology is the main reason why there are gapless   quasiparticles in quantum liquids  and (nearly) massless  elementary particles in our Universe. 

Topology provides the complementary anti-GUT approach in which the `fundamental' symmetry and `fundamental'  fields of GUT  gradually emerge together with `fundamental' physical laws when the Planck energy scale is approached from above \cite{Froggatt1991,Volovik2003}. The emergence of the `fundamental'  laws of physics is provided by the general property of topology -- robustness to details of the microscopic trans-Planckian physics. As a result, the physical laws which emerge at low energy  together with the matter itself are generic. They do not depend much on the details of the trans-Planckian subsystem, being  determined by the universality  class,  which the whole system  belongs to. 

In this scheme, fermions are primary objects. Approaching the Planck energy scale from above, they are transformed to the Standard Model chiral fermions  and  give rise to the secondary objects: gauge fields and gravity. Below the Planck scale,  the GUT scenario  intervenes giving rise to symmetry breaking at low energy. This is accompanied by formation of  composite objects, Higgs bosons, and tiny Dirac masses of quark and leptons.
 
In the GUT scheme, general relativity is assumed to be as fundamental as quantum mechanics, while in the second scheme  general relativity  is a secondary phenomenon. In the anti-GUT scheme, general relativity is the effective theory describing the dynamics of the effective metric experienced by the effective low-energy fields. It is a side product of quantum field theory or of the quantum mechanics in the vacuum with Fermi point.

Vacua with topologically protected gapless (massless) fermions 
are representatives of the broader class of topological media. In condensed matter it includes
topological insulators (see reviews \cite{HasanKane2010,Xiao-LiangQi2011}), topological semimetals
(see \cite{Abrikosov1971,Abrikosov1998,Burkov2011,XiangangWan2011,Ryu2002,Manes2007,Vozmediano2010,Cortijo2011}), topological superconductors and superfluids, states which experience quantum Hall effect, and other topologically nontrivial gapless and gapped phases of matter. 
Topological media have many peculiar properties:  topological stability of gap nodes; topologically protected edge states including Majorana fermions; topological quantum phase transitions occurring at $T=0$; topological quantization of physical parameters including Hall and spin-Hall conductivity;  chiral anomaly; topological Chern-Simons and Wess-Zumino actions; etc.

 It appears that quantum vacuum of Standard Model is topologically nontrivial both in its massless and massive states. In the massless state the quantum vacuum is topologically similar to the superfluid $^3$He-A and gapless semimetal. In the massive state the quantum vacuum is topologically similar to the superfluid $^3$He-B  and 3+1 dimensional topological insulator. This is discussed in Sections VI--VIII.

\section{Quantum vacuum as self-sustained medium}
\label{Quantum_vacuum_self-sustained}

 \subsection{Vacuum energy and  cosmological constant} 
\label{sec:CC}

There is a  huge contribution to the vacuum energy density, which comes from the ultraviolet  (Planckian) degrees of freedom and is of order
$E^4_{\rm P} \approx \big(10^{28}\,\text{eV}\big)^4$.
The observed cosmological is  smaller by many orders of magnitude and corresponds to the
energy density of the vacuum  $\rho_{\rm vac}\sim \big(10^{-3}\,\text{eV}\big)^4$. 
  In general relativity, the cosmological constant  is arbitrary constant, and thus its smallness requires fine-tuning.
If gravitation would be a truly fundamental interaction,
it would be hard to understand why the energies stored
in the quantum vacuum would not gravitate
at all \cite{Nobbenhuis2006}.
If, however, gravitation would be only a low-energy effective interaction,
it could be that the corresponding gravitons as quasiparticles
do not feel \emph{all} microscopic degrees of freedom
(gravitons would be analogous to small-amplitude waves
at the surface of the ocean)
and that the gravitating effect of the vacuum energy density would be
effectively \emph{tuned away} and cosmological constant would be naturally small or zero  \cite{Volovik2003,Dreyer2007}. 

\subsection{Variables for Lorentz invariant vacuum} 

A particular mechanism of nullification of the relevant vacuum energy works for such vacua which have the property of a \emph{self-sustained medium} \cite{KlinkhamerVolovik2008a,KlinkhamerVolovik2008b,KlinkhamerVolovik2009b,KlinkhamerVolovik2008jetpl,KlinkhamerVolovik2009a,KlinkhamerVolovik2010}.
A self-sustained vacuum is a medium with a definite macroscopic
volume even in the absence of an environment. A condensed matter
example is a droplet of quantum liquid at zero temperature in empty space.
The observed near-zero value of the cosmological constant
compared to Planck-scale values 
suggests that the quantum vacuum of our universe
belongs to this class of systems.
As any medium of this kind, the equilibrium vacuum
would be homogeneous and extensive. The homogeneity assumption is
indeed supported by the observed flatness and
smoothness of our universe \cite{de Bernardis2000,Hinshaw2007,Riess2007}.
The implication is that the energy of the equilibrium quantum
vacuum would be proportional to the volume considered.

Usually, a self-sustained medium is characterized by an
\emph{extensive conserved quantity}
whose total value determines the actual volume of the
system \cite{LL1980,Perrot1998}. The quantum liquid at $T=0$ is a self sustained system because of the conservation law for the particle number $N$, and its state is characterized by the particle density $n$
which acquires a non-zero value $n=n_0$ in the equilibrium ground state. As distinct from condensed matter systems, the 
quantum vacuum of our Universe is a relativistic invariant system.
The Lorentz invariance of the vacuum imposes strong constraints on the possible form this variable can take. One must find the relativistic analog of the particle density $n$. 
An example of a possible vacuum variable is a
symmetric tensor $q^{\mu\nu}$, which in a homogeneous vacuum 
is proportional to the metric tensor
\begin{equation}
q^{\mu\nu}=q\,g^{\mu\nu} \,.
\label{eq:2tensor}
\end{equation}
This variable satisfies the Lorentz invariance of the vacuum.
Another example is the 4-tensor $q^{\mu\nu\alpha\beta}$, which in a homogeneous vacuum 
is proportional either to the fully antisymmetric Levi--Civita tensor:
 \begin{equation}
q^{\mu\nu\alpha\beta}=q\,e^{\mu\nu\alpha\beta}\,, 
\label{eq:2tensor}
\end{equation}
or to the product of metric tensors such as:
\begin{equation}
q^{\mu\nu\alpha\beta}=q \left(g_{\alpha\mu} g_{\beta\nu} -g_{\alpha\nu} g_{\beta\mu}\right)
 \,.
\label{eq:mixed}
\end{equation}

Scalar field is also the Lorentz invariant variable, but it does not satisfy another necessary condition of the self sustained system: the vacuum variable $q$ must obey some kind of the conservation law. Below we consider some examples satisfying the two conditions: Lorentz invariance of the perfect vacuum state and the conservation law.

\subsection{Yang-Mills chiral condensate as example}

Let us first consider as an example the chiral condensate of gauge fields. It can be the
gluonic condensate in QCD  \cite{Shifman1992,Shifman1991}, or any other condensate of
Yang-Mills fields, if it is Lorentz invariant.
We assume that the Savvidy vacuum \cite{Savvidy}
is absent, i.e. the vacuum expectation value of the
 color magnetic field is zero (we shall omit  color indices):
\begin{equation}
\left<F_{\alpha\beta}\right>=0 \,,
\label{eq:VEVF}
\end{equation}
while the vacuum expectation value of the quadratic form is nonzero:
\begin{equation}
 \left<F_{\alpha\beta}  F_{\mu\nu} \right>=  \frac{q}{24}  \sqrt{-g}e_{\alpha\beta\mu\nu}  \,.
\label{eq:VEVF^2}
\end{equation}
Here $q$ is  the anomaly-driven topological condensate (see e.g. \cite{HalperinZhitnitsky1998}):
\begin{equation}
q= \left< \tilde F^{\mu\nu}F_{\mu\nu}\right>
= \frac{1}{\sqrt{-g}}e^{\alpha\beta\mu\nu} \left<F_{\alpha\beta}  F_{\mu\nu} \right> \,,
 \label{eq:q}
\end{equation}
In the homogeneous static vacuum state, the $q$-condensate
violates the $P$ and $T$ symmetries of the vacuum,
but it conserves the combined symmetry $PT$  symmetry.

\subsubsection{Cosmological term}

Let us choose the vacuum action in the form
\begin{equation}
 S_q=\int d^4x \sqrt{-g}\epsilon(q)\,,
\label{eq:VacuumAction}
\end{equation}
with $q$ given by \eqref{eq:q}.
The energy-momentum tensor of the vacuum field $q$ is obtained by variation
of the action over $g^{\mu\nu}$:
\begin{equation}
T^q_{\mu\nu}=-\frac{2}{\sqrt{-g}}\: \frac{\delta S_q}{\delta g^{\mu\nu}}=
\epsilon(q)\,  g_{\mu\nu} -
2\,  \frac{\partial\epsilon}{\partial q}\, \frac{\partial q}{\partial g^{\mu\nu}}
 \,.
\label{eq:em_tensor_pot}
\end{equation}
Using \eqref{eq:VEVF^2} and \eqref{eq:q} one obtains
\begin{equation}
 \frac{\partial q}{\partial g^{\mu\nu}}
=
\frac{1}{2} q  g_{\mu\nu}
 \,.
\label{eq:dq/dg}
\end{equation}
and thus
\begin{equation}
T_{\mu\nu}^q= g_{\mu\nu} \rho_{\rm vac}(q)~~,~~ \rho_{\rm vac}(q)=\epsilon(q)
- q  \frac{\partial\epsilon}{\partial q}   \,.
\label{eq:tilde_epsilon}
\end{equation}
In Einstein equations this energy momentum tensor plays the role
of the cosmological term: 
\begin{equation}
T_{\mu\nu}^q= \Lambda g_{\mu\nu} ~~,~~ \Lambda=\rho_{\rm vac}(q)= \epsilon(q)
- q  \frac{\partial\epsilon}{\partial q}   \,.
\label{eq:cosmological_term}
\end{equation}
It is important that the cosmological constant
is given not by the vacuum energy as is usually assumed, but by thermodynamic potential
$\rho_{\rm vac}=\epsilon(q)
 - \mu q$, where $\mu$ is thermodynamically conjugate to $q$ variable, $\mu=d\epsilon/dq$. 
 Below, when we consider dynamcs we shall see that this fact reflects the conservation of the variable $q$. 
 
  The crucial difference between the vacuum energy $\epsilon(q)$ and thermodynamic potential
$\rho_{\rm vac}=\epsilon(q)- \mu q$ is revealed when we consider the corresponding quantities in the ground state of quantum liquids, the energy density of the liquid
 $\epsilon(n)$ and the density of the  grand canonical energy, $\epsilon(n)- \mu n$, which enters macroscopic thermodynamics due to conservation of particle number. The first one, $\epsilon(n)$, has the value dictated by atomic physics, which is equivalent to $E_{\rm P}^4$ in the quantum vacuum. On the contrary, the second one equals minus pressure, $\epsilon(n)- \mu n=-P$, according to the Gibbs-Duhem thermodynamic relation at $T=0$. Thus its value is dictated not by the microscopic physics, but by external conditions. In the absence of environment, the external pressure is zero, and the value  of $\epsilon(n)- \mu n$ in a fully equilibrium ground state of the liquid is zero. This is valid for any self-sustained macroscopic system, including the self-sustained quantum vacuum, which suggests the natural solution of the main cosmological constant problem.

\subsubsection{Conservation law for $q$}

Equation for $q$ in flat space
can be obtained from Maxwell equation, which in turn is obtained by variation 
of the action over the gauge field $A_\mu$:
\begin{equation}
\nabla_\mu \left(\frac{\partial\epsilon}{\partial q}\tilde F^{\mu\nu} \right) =0
 \,,
\label{eq:Maxwell}
\end{equation}
where $\nabla_\mu$ is the covariant derivative.
Since  $\nabla_\mu \tilde F^{\mu\nu}=0$, equation \eqref{eq:Maxwell}
is reduced to
\begin{equation}
\nabla_\mu \left(\frac{\partial\epsilon}{\partial q}\right) =0 \,.
\label{eq:tilde_mu1}
\end{equation}
The solution of this equation is
\begin{equation}
 \frac{\partial\epsilon}{\partial q}=  \mu\,,
\label{eq:tilde_mu2}
\end{equation}
 where $\mu$ is integration constant. In thermodynamics, this $\mu$ will play   the role of the chemical potential, which is thermodynamically conjugate to $q$. This demonstrates that $q$ obeys the conservation law and thus can be the proper variable for description the self-sustained vacuum.

\subsection{4-form field as example}

Another example of the vacuum variable appropriate for the self-sustained vacuum is given by the four-form field
strength~\cite{DuffNieuwenhuizen1980,Aurilia-etal1980,Hawking1984,HenneauxTeitelboim1984,
Duff1989,DuncanJensen1989,BoussoPolchinski2000,Aurilia-etal2004,Wu2008}, which is expressed in
terms of $q$ in the following way:
\bsubeqs\label{eq:EinsteinF-all}
\beqa
F_{\alpha\beta\gamma\delta}  &\equiv&
q\,e_{\alpha\beta\gamma\delta}\, \sqrt{-\det g}
= \nabla_{[\alpha}A_{\beta\gamma\delta]}\,,
\label{eq:Fdefinition1}\\[2mm]
q^2 &=&
- \frac{1}{24}\,F_{\alpha\beta\gamma\delta}\,F^{\alpha\beta\gamma\delta}\,,
\label{eq:q2definition}
\eeqa
\esubeqs
where $e_{\alpha\beta\gamma\delta}$ the Levi--Civita tensor density;
and the square bracket around spacetime indices complete anti-symmetrization.

Originally the quadratic action has been used for this field ~\cite{DuffNieuwenhuizen1980,Aurilia-etal1980}, which corresponds to the special case of  \eqref{eq:VacuumAction} with $\epsilon(q)=\half\,q^2$. For the general $\epsilon(q)$
one obtains the Maxwell equation
\begin{equation}
\nabla_\alpha \left(\sqrt{-\det g} \,\;\frac{F^{\alpha\beta\gamma\delta}}{q}
           \frac{\partial\epsilon(q)}{\partial q} \right)=0\,.
\label{eq:4Maxwell}
\end{equation}
Using \eqref{eq:Fdefinition1} the Maxwell equation is reduced to
\begin{equation}
\nabla_\alpha \left(
           \frac{\partial\epsilon(q)}{\partial q} \right)=0\,.
\label{eq:4Maxwell2}
\end{equation}
The first integral of \eqref{eq:4Maxwell2} with integration constant $\mu$
gives again Eq.\eqref{eq:tilde_mu2}, which reflects the conservation law for $q$.

Variation of the action over $g^{\mu\nu}$ gives again the cosmological constant \eqref{eq:cosmological_term} with $\Lambda=\rho_{\rm vac}=\epsilon(q)- \mu q$.
This demonstrates the universality of the macroscopic description of the self-sustained vacuum: description of the quantum vacuum in terms of $q$ does not depend on the microscopic details of the vacuum and on the nature of the vacuum variable. 

\subsection{Aether field as example}
\label{Aether_field}

Another example of the vacuum variable $q$ may be through
a four-vector field $u^{\mu}(x)$. This vector field could be the
four-dimensional analog of the concept of shift in the deformation
theory of crystals. (Deformation theory can be described in terms of a
metric field, with the role of  torsion and curvature fields played by
dislocations and disclinations, respectively;
see, e.g.,  Ref.~\cite{Dzyaloshinskii1980} for a review.)
A realization of $u^{\mu}$ could be also a 4--velocity field
entering the description of the structure of spacetime.
It is  the 4-velocity of  ``aether''  \cite{Jacobson2007,Gasperini1987,Jacobson2001,WillNordvedt}.

The nonzero value of the 4-vector in the vacuum violates
the  Lorentz invariance of the vacuum. To restore this invariance one may assume that
$u^{\mu}(x)$ is not an observable variable, instead the observables are its covariant derivatives $\nabla_\nu u^{\mu}\equiv u_\nu^\mu$. This means that the action does not depend on $u^{\mu}$ explicitly but only
depends on  $u_\nu^\mu$:
\begin{equation}
S= 
\int_{\mathbb{R}^4} \,d^4x\, 
\epsilon(u_\nu^\mu)\,,
\label{eq:action3}
\end{equation}
with an energy density containing even powers of
$u_\nu^\mu$:
\begin{equation}\label{eq:epsilon-u-mu-nu}
\epsilon(u_\nu^\mu)
= K
+ K_{\mu\nu}^{\alpha\beta}\,u_\alpha^\mu u_\beta^\nu
+ K_{\mu\nu\rho\sigma}^{\alpha\beta\gamma\delta}\,
u_\alpha^\mu u_\beta^\nu u_\gamma^\rho u_\delta^\sigma + \cdots \;.
\end{equation}
According to the imposed conditions, the
tensors $K_{\mu\nu}^{\alpha\beta}$ and
$K_{\mu\nu\rho\sigma}^{\alpha\beta\gamma\delta}$
depend only on $g_{\mu\nu}$ or $g^{\mu\nu}$ and the same holds for the other
$K$--like tensors in the ellipsis of \eqref{eq:epsilon-u-mu-nu}.
In particular, the tensor $K_{\mu\nu}^{\alpha\beta}$ of the quadratic term
in \eqref{eq:epsilon-u-mu-nu} has the following form in the notation of
Ref.~\cite{Jacobson2007}:
\begin{equation}
K_{\mu\nu}^{\alpha\beta}=c_1\, g^{\alpha\beta}g_{\mu\nu}  +
c_2\, \delta^{\alpha}_{\mu}  \delta^{\beta}_{\nu}  +
c_3\, \delta^{\alpha}_{\nu}  \delta^{\beta}_{\mu}  ~,
\label{eq:Kparameters}
\end{equation}
for real constants $c_n$.
Distinct from the original aether theory in Ref.~\cite{Jacobson2007},
the tensor \eqref{eq:Kparameters}
does not contain a term $c_4\, u^\alpha u^\beta g_{\mu\nu}$,
as such a term would depend explicitly on $u^{\mu}$ and
contradict the Lorentz invariance
of the quantum vacuum.

The equation of motion for $u^{\mu}$ in flat space,
\begin{equation}
\nabla_\nu  \, \frac{\partial\epsilon}{\partial u_\nu^\mu} =0\,,
\label{eq:Motion}
\end{equation}
has the Lorentz invariant solution expected for a vacuum-variable $q$--type field:
\begin{equation}
u^q_{\mu\nu}=q\,g_{\mu\nu}\,,\quad q=\text{constant}~.
\label{eq:solution3}
\end{equation}
With this solution, the energy density in the action \eqref{eq:action3} is
simply $\epsilon(q)$
in terms of contracted coefficients $K$, $K_{\mu\nu}^{\mu\nu}$, and
$K_{\mu\nu\rho\sigma}^{\mu\nu\rho\sigma}$ from \eqref{eq:epsilon-u-mu-nu}.
However, just as for previous examples,
the energy-momentum tensor of the vacuum field obtained by
variation over $g^{\mu\nu}$ and evaluated for solution \eqref{eq:solution3}
is expressed again in terms of  the thermodynamic potential:
\beqa
T^q_{\mu\nu}
&=&
\frac{2}{\sqrt{-g}}\;\frac{\delta S}{\delta g^{\mu\nu}}\,
 =              g_{\mu\nu} \left(\epsilon(q) - q\,\frac{d\epsilon(q)}{dq}\right)=\rho_{\rm vac}(q) g_{\mu\nu}\,,
\label{eq:emSolution3}
\eeqa
which corresponds to cosmological constant in Einstein's gravitational field
equations.

\section{Thermodynamics of quantum vacuum}

\subsection{Liquid-like quantum vacuum}

\begin{figure}
\centerline{\includegraphics[width=0.8\linewidth]{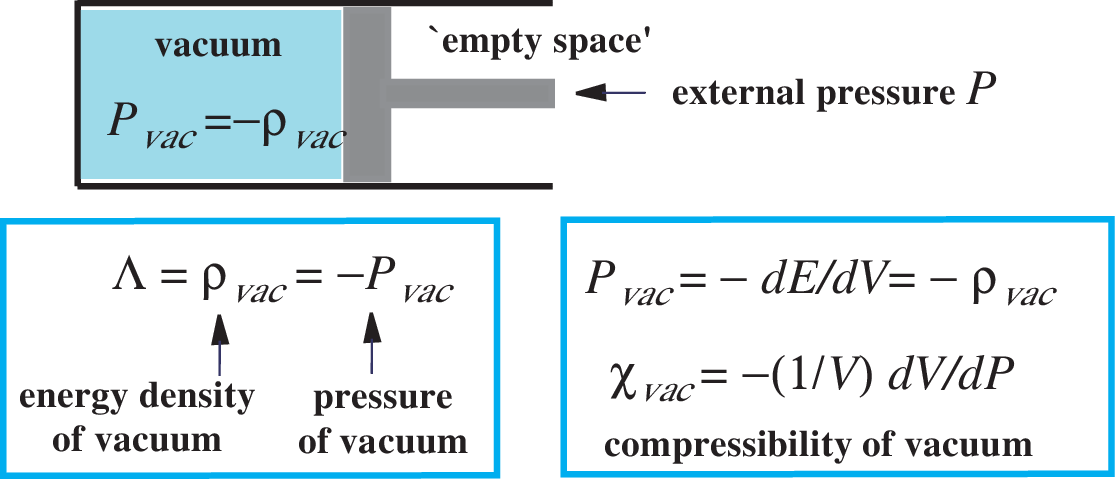}}
\caption{Vacuum as a medium obeying macroscopic thermodynamic laws. Relativistic vacuum possesses energy density, pressure and compressibility but has no momentum.  In equilibrium, the vacuum pressure $P_{\rm vac}$ equals the external pressure $P$ acting from the environment. 
The  ``gas-like'' vacuum may exist only under external pressure. The  ``liquid-like'' vacuum is self-sustained: it can be stable in the absence of external pressure.
The thermodynamic energy density of the vacuum $\rho_{\rm vac}$ which enters the vacuum equation of state $\rho_{\rm vac}= -P_{\rm vac}$ does not coincide with the microscopic vacuum energy $\epsilon$. While the natural value of  $\epsilon$ is determined by the Planck scale, $\epsilon\sim E_{\rm P}^4$,  the natural value of the macroscopic quantity $\rho_{\rm vac}$ is zero for the self-sustained vacuum which may exist in the absence of environment, i.e. at $P=0$. This may explain why the present cosmological constant $\Lambda=\rho_{\rm vac}$ is small.
}  
\label{compressibility} 
\end{figure}

The zeroth order term $K$ in \eqref{eq:epsilon-u-mu-nu} corresponds to a ``bare'' cosmological constant which can be considered as 
cosmological constant in the ``empty'' vacuum -- vacuum with $q=0$:
\begin{equation}
\Lambda_{\rm bare}=\epsilon(q=0)~.
\label{eq:Lambda_bare}
\end{equation}
 The nonzero  value $q=q_0$ in the self-sustained vacuum does \emph{not} violate Lorentz symmetry
but leads to compensation of the bare cosmological constant $\Lambda_{\rm bare}$
in the equilibrium vacuum.
This  illustrates the important difference between the 
two states of vacua. The quantum vacuum with $q=0$ 
can exist only with external pressure $P =-\Lambda_{\rm bare}$.
By analogy with condensed-matter physics,
this kind of quantum vacuum may be called ``gas-like'' (Fig. \ref{compressibility}).
The quantum vacuum with nonzero $q$
is self-sustained: it can be stable at $P=0$, provided that a stable nonzero solution of equation
$\epsilon(q) - q\, d\epsilon/dq=0$ exists.
This kind of quantum  vacuum may then be called ``liquid-like''.

The universal behavior of the self-sustained vacuum in equilibrium suggests that it obeys the same thermodynamic laws as any other 
self-sustained macroscopic system described by the conserved quantity $q$, such as
quantum liquid.
In other words, vacuum can be considered as a special quantum liquid which is Lorentz invariant 
in its ground state.  This liquid is characterized by  the Lorentz invariant ``charge'' density $q$ -- an analog of particle density $n$ in non-relativistic quantum liquids.

Let us consider  a large portion of such  vacuum liquid under external pressure $P$ \cite{KlinkhamerVolovik2008a}. The volume $V$ of quantum vacuum is variable, but its total ``charge''
$Q(t)\equiv \int d^3r~q(\mathbf{r},t)$ must be conserved,
$\mathrm{d}Q/\mathrm{d}t=0$. 
The energy of this portion of quantum vacuum at fixed  total``charge'' $Q=q\, V$
is then given by the thermodynamic potential
\begin{equation}
W=E+P\,V=\int d^3r~\epsilon\left(Q/V\right) + P\,V~,
\label{eq:ThermodynamicPotential}
\end{equation}
where
$\epsilon\left(q\right)$ is the energy density in terms of charge density $q$.
As the volume of the system is a free parameter,
the equilibrium state of the system is obtained by variation over the volume $V$:
\begin{equation}
\frac{d W}{dV}=0~,
\label{eq:Equilibrium}
\end{equation}
This gives an integrated form of the Gibbs--Duhem equation for the vacuum pressure:
\begin{equation}
P_{\rm vac}=-\epsilon(q) +q\,\frac{d\epsilon(q)}{dq}=-\rho_{\rm vac}(q)~,
\label{eq:Gibbs-Duhem}
\end{equation}
whose solution determines the equilibrium value $q=q(P)$
and the corresponding volume  $V=(P,Q)=Q/q(P)$.

\subsection{Macroscopic energy of quantum vacuum} 

Since the vacuum energy density is the vacuum pressure with minus sign, equation (\ref{eq:Gibbs-Duhem}) suggests that the relevant  vacuum energy, which is revealed in thermodynamics and dynamics of the low-energy Universe, is:
\begin{equation}
\rho_{\rm vac}(q)=\epsilon(q) -q\,\frac{d\epsilon(q)}{dq}~.
\label{eq:vev}
\end{equation}
This is confirmed by Eqs.~\eqref{eq:cosmological_term}  and \eqref{eq:emSolution3} for energy-momentum tensor of the 
self-sustained vacuum, which demonstrates
that it is $\rho_{\rm vac}\left(q\right)$ rather than $\epsilon\left(q\right)$, which enters the equation of state for the vacuum and thus corresponds to the cosmological constant:
\begin{equation}
\Lambda=\rho_{\rm vac}=-P_{\rm vac}~.
\label{eq:EoS}
\end{equation}

 \begin{figure}
\centerline{\includegraphics[width=0.8\linewidth]{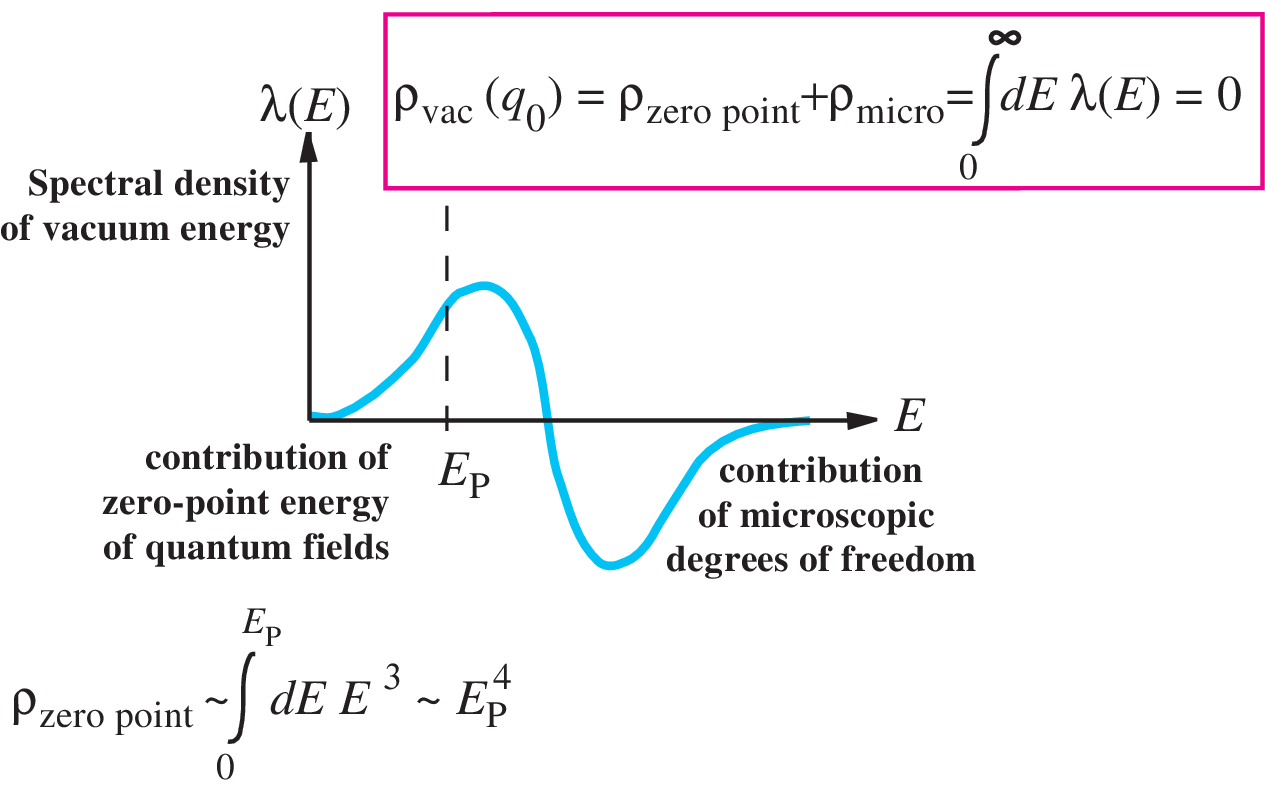}}
\caption{Contribution of different energy scales into the macroscopic energy of the self-sustained system at $T=0$. Zero point energy of the effective bosonic and fermionic quantum fields gives rise to the diverging contribution to the energy  of the system. In quantum vacuum it is of order of $E_{\rm P}^4$. In equilibrium this contribution is compensated without  fine-tuning by microscopic degrees of freedom of the system  (by trans-Planckian degrees of quantum vacuum correspondingly). }  
\label{spectrum} 
\end{figure}

While the energy of microscopic quantity $q$ is determined by the Planck scale,
$\epsilon(q_0) \sim E_{\rm P}^4$, the relevant vacuum energy which sources the effective gravity is determined by a macroscopic quantity -- the external pressure.
In the absence of an environment, i.e. at zero external pressure, $P=0$, one obtains that the pressure of pure and equilibrium vacuum is exactly zero:
\begin{equation}
\Lambda=-P_{\rm vac}=-P=0~.
\label{eq:Null}
\end{equation}
Equation $\rho_{\rm vac}(q)=0$ determines the equilibrium value $q_0$ of the 
equilibrium self-sustained vacuum. 
Thus from the thermodynamic arguments it follows that for any effective theory of gravity the natural value of $\Lambda$ is zero in equilibrium vacuum. 

This result does not depend on the microscopic structure of the vacuum  from which gravity emerges, and is actually the final result of the renormalization dictated by macroscopic physics. In the self-sustained quantum liquid the large contribution of zero-point energy of phonon field is naturally compensated by microscopic (atomic) degrees of freedom of quantum liquid.
In the same manner, the huge contribution of zero-point energy of macroscopic fields to the vacuum energy $\rho_{\rm vac}$ is naturally compensated by microscopic degrees of the self sustained  quantum vacuum:
the vacuum variable $q$ is adjusted automatically to nullify the macroscopic vacuum energy,  $\rho_{\rm vac}(q_0)= \rho_{\rm zero~point} +\rho_{\rm microscopic}=0$. The actual spectrum of the vacuum energy density (meaning the
different contributions to $\epsilon$ from different energy scales)
is not important for the cancellation mechanism, because it is dictated by thermodynamics. The particular example of the spectrum of the vacuum energy density is shown in Fig. \ref{spectrum}, where the positive energy of the quantum vacuum, which comes from the
zero-point energy of bosonic fields, is compensated by negative contribution from trans-Planckian degrees of freedom
\cite{VolovikSpectrum} . 

Using the quantum-liquid counterpart of the self-sustained quantum vacuum as example, one may predict the behavior of the vacuum after  cosmological phase transition, when $\Lambda$ is kicked from its zero value. The vacuum will readjust itself to a new equilibrium state with new $q_0$ so that $\Lambda$ will again approach its equilibrium zero value  \cite{KlinkhamerVolovik2008a}.
The process of relaxation of the system to the equilibrium state depends on details of dynamics of the vacuum variable $q$ and its interaction with matter fields, and later on we shall consider some examples of dynamical relaxation of $\Lambda$.

 \subsection{Compressibility of the vacuum}

Using the standard definition of the inverse of
the isothermal compressibility, $\chi^{-1} \equiv -V\,dP/dV$ (Fig. \ref{compressibility}),  one obtains the compressibility of the vacuum by varying Eq.(\ref{eq:Gibbs-Duhem}) at fixed $Q=qV$ 
\cite{KlinkhamerVolovik2008a}:
\begin{equation}
\chi_\text{vac}^{-1} \equiv -V\frac{dP_{\rm vac}}{dV}=\left[q^2\;\frac{d^2\epsilon(q)}{dq^2}\,\right]_{q=q_0}
> 0~.
\label{eq:Stability}
\end{equation}
A positive value of the vacuum compressibility
is a necessary condition for the stability of the vacuum. It is, in fact, the stability of
the vacuum, which is at the origin of the nullification of the cosmological
constant 
in the absence of an external environment.

From the low-energy point of view, the compressibility of the vacuum $\chi_\text{vac}$ is as fundamental physical constant as the Newton constant $G_N=G(q=q_0)$. It enters equations describing  the response of the quantum vacuum  to different perturbations.  While the natural value of the macroscopic quantity $P_{\rm vac}$ (and $\rho_{\rm vac}$) is zero, the natural values of  the parameters $G(q=q_0)$ and $\chi_\text{vac}(q=q_0)$ are determined by the Planck physics and are expected to be of order $1/E^{2}_{\rm P}$ and $1/E^{4}_{\rm P}$ correspondingly.

 \subsection{Thermal fluctuations of $\Lambda$ and the volume of Universe}

The compressibility of the vacuum $\chi_\text{vac}$, though not measurable at the moment, can be used for estimation of  the lower limit for the volume $V$ of the Universe. This estimation follows from the upper limit for thermal fluctuations of cosmological constant \cite{Volovik2004}. The mean square of thermal  fluctuations of $\Lambda$ equals the mean square of thermal  fluctuations of the vacuum pressure, which in turn is determined by thermodynamic equation \cite{LL1980}:
\begin{equation}
\left <\left(\Delta\Lambda\right)^2\right>=\left <\left(\Delta P\right)^2\right>=\frac{ T}{V\chi_{\rm vac}}~.
\label{Fluctuations}
\end{equation}
 Typical fluctuations of the cosmological constant $\Lambda$ should not exceed the observed value:  $\left <\left(\Delta\Lambda\right)^2\right>< \Lambda_{\rm obs}^2$. Let us assume, for example, that the temperature of the Universe is determined by the temperature $T_{\rm CMB}$ of the cosmic microwave background radiation. Then, using our estimate for vacuum compressibility $\chi_{\rm vac}^{-1}\sim E^4_{\rm P}$, one obtains that the volume $V$ of our Universe highly exceeds the Hubble volume $V_H=R_H^3$ --  the volume of visible Universe inside the present cosmological horizon:
\begin{equation}
V> \frac{T_{\rm CMB}} { \chi_{\rm vac} \Lambda_{\rm obs}^2}\sim 10^{28}V_H~.
\label{Volume}
\end{equation}
This demonstrates that the real volume of the Universe is certainly not limited by the present cosmological horizon. 
  
\section{Dynamics of quantum vacuum}

\subsection{Action}\label{sec:Action}

In   section \ref{Quantum_vacuum_self-sustained} a special quantity,
the  vacuum ``charge''  $q$, was introduced to describe the statics and thermodynamics of the self-sustained quantum vacuum. Now we can extend this approach to the dynamics of the vacuum charge.
We expect to find some universal features of the vacuum dynamics, using several realizations 
of this vacuum variable. We start with the 4-form field
strength~\cite{DuffNieuwenhuizen1980,Aurilia-etal1980,Hawking1984,HenneauxTeitelboim1984,
Duff1989,DuncanJensen1989,BoussoPolchinski2000,Aurilia-etal2004,Wu2008} expressed in
terms of $q$. 
The low-energy effective action takes the following general form: 
\bsubeqs\label{eq:EinsteinF-all} \beqa S=- \int_{\mathbb{R}^4}
\,d^4x\, \sqrt{|g|}\,\left(\frac{R}{16\pi G(q)} +\epsilon(q)
+\mathcal{L}^\text{M}(q,\psi)\right) \,,  
\label{eq:actionF}\\[2mm]
q^2 \equiv- \frac{1}{24}\,
F_{\kappa\lambda\mu\nu}\,F^{\kappa\lambda\mu\nu}\,,\quad
F_{\kappa\lambda\mu\nu}\equiv
\nabla_{[\kappa}\!\!A_{\lambda\mu\nu]}\,,  
\label{eq:Fdefinition}\\[2mm]
F_{\kappa\lambda\mu\nu}=q\sqrt{|g|} \,e_{\kappa\lambda\mu\nu}\,,\quad
F^{\kappa\lambda\mu\nu}=q \,e^{\kappa\lambda\mu\nu}/\sqrt{|g|}\,. \quad
\label{eq:Fdefinition2} 
\eeqa \esubeqs
where $R$ denotes the Ricci curvature scalar; and $\mathcal{L}^\text{M}$ is matter action.
Throughout, we use the same conventions as in Ref.~\cite{Weinberg1988},
in particular, those for the Riemann curvature tensor
and the metric signature $(-+++)$.

The vacuum energy density $\epsilon$ in \eqref{eq:actionF}
depends on the vacuum variable $q$ which in turn is expressed via the 3-form field 
$A_{\lambda\mu\nu}$ and metric field $g_{\mu\nu}$ in \eqref{eq:Fdefinition}.
The field $\psi$ combines all the matter fields of the Standard Model. All possible constant terms in matter action (which includes the zero-point energies
from the Standard Model fields) are absorbed in the vacuum energy $\epsilon(q)$.

Since $q$ describes the state of the vacuum, the parameters of the effective action -- the Newton constant $G$ and parameters which enter the matter action -- must depend on $q$.  This
dependence results in particular in the interaction between the matter fields and the vacuum. There are
different sources of this interaction. For example, in the gauge field sector of Standard Model, the running
coupling contains the ultraviolet cut-off and thus depends on $q$:
\begin{equation}
\mathcal{L}^\text{{\bf G},q}= \gamma(q)F^{\mu\nu}F_{\mu\nu}
\,, \label{RunningCupling}
\end{equation}
where $F_{\mu\nu}$ is the field strength of the particular gauge field (we omitted the color indices).
In the fermionic sector, $q$ should enter parameters
of the Yukawa interaction and fermion masses.

\subsection{Vacuum dynamics}

The variation of the action \eqref{eq:actionF} over the three-form
gauge field $A$ gives the generalized Maxwell equations for $F$-field,
\begin{equation}
\nabla_\nu \left(\sqrt{|g|} \;\frac{F^{\kappa\lambda\mu\nu}}{q} \left(
\frac{d\epsilon(q)}{d q}+\frac{R}{16\pi} \frac{dG^{-1}(q)}{d q}
+ \frac{d \mathcal{L}^\text{M}(q)}{d q}
\right)\right)=0\,.
\label{eq:Maxwell}
\end{equation}
Using \eqref{eq:Fdefinition2} for $F^{\kappa\lambda\mu\nu}$,
we find that the solutions of Maxwell equations \eqref{eq:Maxwell} 
are still determined by the integration constant $\mu$
\begin{equation}
\frac{d\epsilon(q)}{d q}+\frac{R}{16\pi} \frac{dG^{-1}(q)}{d q}
+ \frac{d \mathcal{L}^\text{M}(q)}{d q}
=\mu \,.
\label{eq:Maxwell2}
\end{equation}

\subsection{Generalized Einstein equations}

The variation over the metric $g^{\mu\nu}$ gives the
generalized Einstein equations,
\begin{eqnarray}
&&
\frac{1}{8\pi G(q)}
\left( R_{\mu\nu}-\frac{1}{2}\,R\,g_{\mu\nu}\right)
+\frac{1}{16\pi}\, q\,\frac{d G^{-1}(q)}{d q}\, {R}\,g_{\mu\nu}
\nonumber\\[2mm]
&&+ \frac{1}{8\pi} \Big( \nabla_\mu\nabla_\nu\, G^{-1}(q) - g_{\mu\nu}\,
\Box\, G^{-1}(q)\Big) -\left( \epsilon(q) -q\,\frac{d\epsilon(q)}{d q}\right)g_{\mu\nu}
\nonumber\\[2mm]
&&+  q\frac{\partial \mathcal{L}^\text{M}}{\partial q} g_{\mu\nu}
 +T^\text{M}_{\mu\nu} =0\,, \label{eq:EinsteinEquationF}
\end{eqnarray}
where $\Box$ is the invariant d'Alembertian;
and
$T^\text{M}_{\mu\nu}$ is the  energy-momentum tensor of the matter
fields, obtained by variation over $g^{\mu\nu}$ at constant $q$,
i.e. without variation over
 $g^{\mu\nu}$, which enters $q$.

Eliminating $dG^{-1}/dq$ and $\partial \mathcal{L}^\text{M}/\partial q$ from \eqref{eq:EinsteinEquationF} by use of
\eqref{eq:Maxwell2}, the generalized Einstein equations become
\begin{equation}
\frac{1}{8\pi G(q)}\Big( R_{\mu\nu}-\half\,R\,g_{\mu\nu} \Big) 
+ \frac{1}{8\pi}
\Big( \nabla_\mu\nabla_\nu\, G^{-1}(q) - g_{\mu\nu}\, \Box\, G^{-1}(q)\Big)
- \rho_{\rm vac} g_{\mu\nu}+T^\text{M}_{\mu\nu} =0\,,
\label{eq:EinsteinEquationF2}
\end{equation}
where
\begin{equation}
\rho_{\rm vac}=\epsilon(q)-\mu\, q \,.
\label{eq:Lambda}
\end{equation}
For the special case when the dependence of the Newton constant and matter action on $q$ is ignored, \eqref{eq:EinsteinEquationF2}
reduces to the standard Einstein equation of general relativity with the constant cosmological constant $\Lambda= \rho_{\rm vac}$.

\subsection{Minkowski-type solution and Weinberg problem}

Among different solutions of equations \eqref{eq:Maxwell} and \eqref{eq:EinsteinEquationF}
 there is the solution corresponding to perfect equilibrium Minkowski vacuum without matter.
It is characterized by
the constant in space and time values $q=q_0$ and $\mu=\mu_0$ obeying
the following two conditions:
\bsubeqs\label{eq:equil-eqs}
\beqa
\Bigg[\, \frac{\dd \epsilon(q)}{\dd q} - \mu\, \Bigg]_{\mu=\mu_0\,,\,q=q_0} &=&0\,,
\label{eq:equil-eqs-mu}
\\[2mm]
\Big[\, \epsilon(q)    -  \mu\, q\,\Big]_{\mu=\mu_0\,,\,q=q_0} &=&0\,.
\label{eq:equil-eqs-GDcondition}
\eeqa \esubeqs
The two conditions \eqref{eq:equil-eqs-mu}--\eqref{eq:equil-eqs-GDcondition}
can be combined into a  \emph{single} equilibrium condition for $q_0$:
\beq
\Lambda_0 \equiv
\Bigg[\, \epsilon(q)  -  q\,\frac{\dd \epsilon(q)}{\dd q}\, \Bigg]_{\,q=q_0} = 0\,,
\label{eq:equil-eqs-q0}
\eeq
with the \emph{derived} quantity 
\beq
\mu_0 = \Bigg[\, \frac{\dd \epsilon(q)}{\dd q}\,    \Bigg]_{\,q=q_0}\,.
\label{eq:equil-eqs-mu0}
\eeq
In order for the Minkowski vacuum to be stable, there is the further condition:
$\chi(q_0)>0$ where $\chi$ corresponds to the isothermal vacuum
compressibility \eqref{eq:Stability}~\cite{KlinkhamerVolovik2008a}.
In this equilibrium vacuum the gravitational constant $G(q_0)$ can be identified
with Newton's constant $G_{N}$.

Let us compare the conditions for the equilibrium  self-sustained vacuum, \eqref{eq:equil-eqs-q0} and \eqref{eq:equil-eqs-mu0}, with the two conditions
 suggested by Weinberg, who used the fundamental scalar field $\phi$ for the description of the vacuum.   In this description there are two constant-field equilibrium conditions for Minkowski vacuum, 
$\partial \mathcal{L} / \partial g_{\alpha\beta}=0$
and $\partial \mathcal{L} / \partial \phi =0$, see Eqs.~(6.2) and (6.3) in ~\cite{Weinberg1988}.
These two conditions turn out to be inconsistent,
unless the potential term in $\mathcal{L}(\phi)$ is fine-tuned  (see also Sec.~2 of Ref.~\cite{Weinberg1996}). In other words, the Minkowski vacuum solution may exist only for the  fine-tuned action. This is the Weinberg formulation of the cosmological constant problem.

The self-sustained vacuum naturally bypasses this problem \cite{KlinkhamerVolovik2010}.
Equation~
$\partial \mathcal{L} / \partial g_{\alpha\beta}=0$ corresponds to the   equation~\eqref{eq:equil-eqs-q0}. However, the equation $\partial \mathcal{L} / \partial \phi =0$  is \emph{relaxed} in the $q$-theory of self-sustained vacuum.
Instead of the condition $\partial \mathcal{L} / \partial q =0$,
the conditions are $\nabla_\alpha (\partial \mathcal{L} / \partial q)=0$,
which allow for having $\partial \mathcal{L} / \partial q=\mu$
with an \emph{arbitrary} constant $\mu$.  This is the crucial difference between a fundamental scalar field $\phi$
and the variable $q$ describing the self-sustained vacuum. As a result, the
equilibrium conditions for $g_{\alpha\beta}$ and $q$ can be consistent
without fine-tuning of the original action. For Minkowski vacuum to exist only one condition \eqref{eq:equil-eqs-q0} must be satisfied. In other words, the Minkowski vacuum solution exists for arbitrary action provided that solution of equation \eqref{eq:equil-eqs-q0} exists.

\begin{figure}[t]
\begin{center}
\includegraphics[width=.6\textwidth]{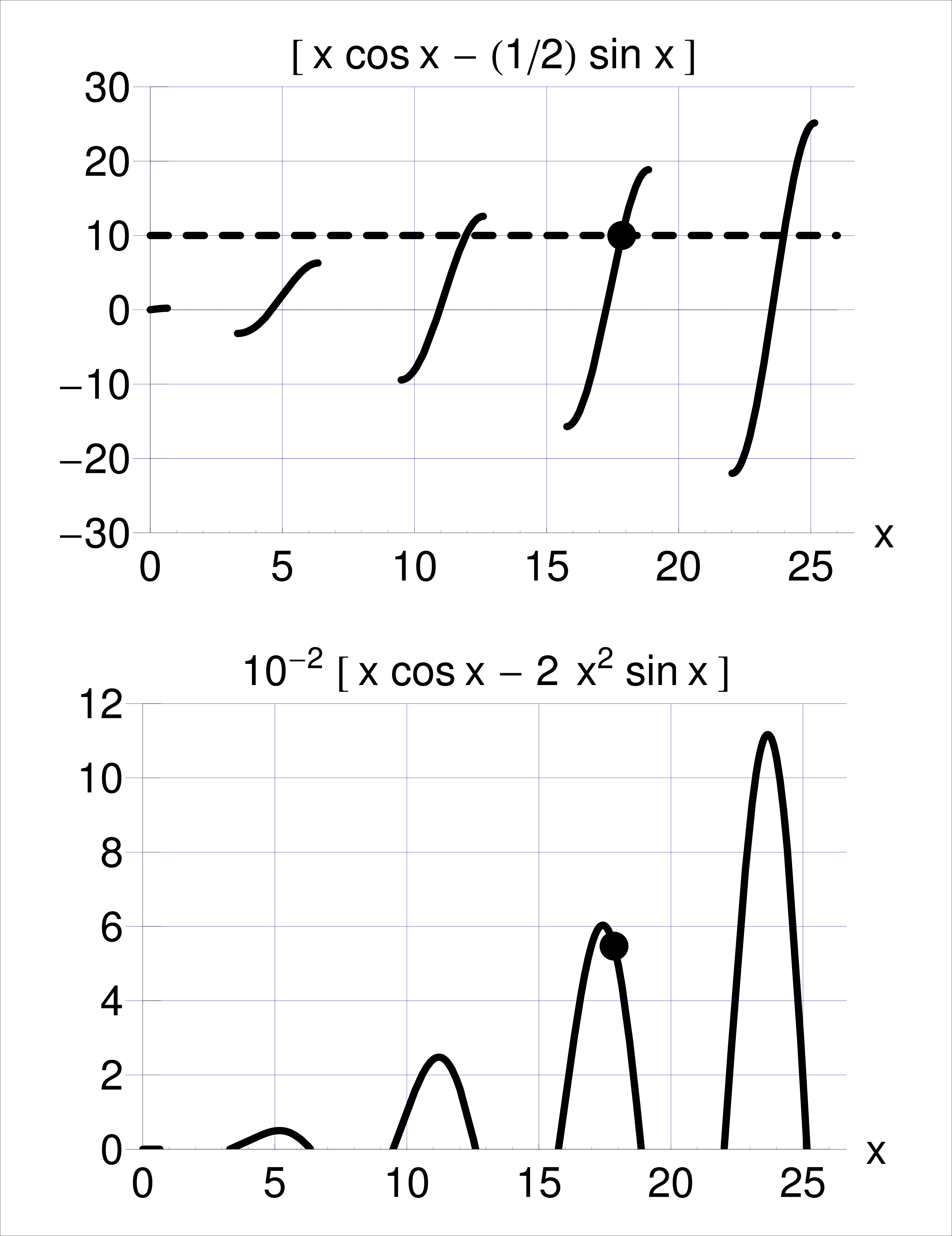}
\end{center}
\caption{ 
A set of Minkowski equilibrium vacua emerging for a particular choice of
the vacuum energy density function in \eqref{eq:epsilon-example}.
In each vacuum the huge bare cosmological constant $\Lambda_\text{bare} \sim E_\text{P}^4$ is compensated 
by the $q$-field.
The curves of the top panel show the left-hand side
of \eqref{eq:q0-eq} for those
values of $x \equiv q^2/E_\text{P}^4$ that obey the stability condition
\eqref{eq:q0-cond}. The curves of the bottom panel show the corresponding
positive segments of the inverse of the dimensionless vacuum
compressibility ${\chi}E_\text{P}^4$. Minkowski-type vacua  are
obtained at the intersection points of the curve of the top panel with a
horizontal line at the value $\lambda_\text{bare}\equiv \Lambda_\text{bare}/E_\text{P}^4$
[for example, the dashed line at
$\lambda_\text{bare}=10$ gives the value $x_0 \approx 17.8$
corresponding to the heavy dot in the top panel]. Each such vacuum is
characterized, in part, by the corresponding value of the inverse vacuum
compressibility from the bottom panel 
shown by the heavy dot for the case chosen in the top panel].
Minkowski vacua with positive compressibility are stable and become
attractors in a dynamical context (cf. next Section).
}
\label{fig:GibbsDuhemLHSandChi}
\end{figure}

\subsection{Multiple Mnkowski vacua}

It is instructive to illustrate this using
a concrete example. The particular choice for the vacuum energy density
function  is considered in \cite{KlinkhamerVolovik2010}:
\beq
\epsilon(q)= \Lambda_\text{bare} + (1/2)\,(E_\text{P})^4\,
\sin\big[\,q^2/(E_\text{P})^4\,\big]\,.
\label{eq:epsilon-example}
\eeq
It contains the higher-order terms in addition to the standard
quadratic term $\half\,q^2$.
With \eqref{eq:epsilon-example}, the expressions for
the equilibrium condition \eqref{eq:equil-eqs-q0} and
the stability condition \eqref{eq:Stability} become
\bsubeqs\label{eq:q0-eq-cond}
\beqa
x\,\cos x-(1/2)\,\sin x                   &=& \lambda_\text{bare} \,,
\label{eq:q0-eq}\\[3mm]
{\chi}^{\;-1} E_\text{P}^{-4}=
x \cos x-2\, x^{\;2}\sin x    &>& 0\,,
\label{eq:q0-cond}
\eeqa
\esubeqs
where  dimensionless quantities
$x \equiv q^2/E_\text{P}^4$
and $\lambda_\text{bare}\equiv \Lambda_\text{bare}/E_\text{P}^4$ are introduced.
A straightforward  graphical analysis (Fig.~\ref{fig:GibbsDuhemLHSandChi})
shows that, for any $\lambda_\text{bare} \in \mathbb{R}$,
there are infinitely many equilibrium states of quantum vacuum, i.e. infinitely many values $q_0 \in \mathbb{R}$
which obey both \eqref{eq:q0-eq} and \eqref{eq:q0-cond}. Each of these vacua has its own values 
of the Newton constant $G(q_0)$ and Standard Model parameters. But all these vacua 
have zero cosmological constant: the Planck-scale bare cosmological constant $\Lambda_\text{bare}$ is compensated by the
$q$ field in any  equilibrium vacuum. The top panel of
Fig.~\ref{fig:GibbsDuhemLHSandChi} shows that the $q$ values
on the one segment singled-out by the heavy dot already allow for a complete
cancellation of \emph{any} $\Lambda_\text{bare}$ value
between $-15\, E_\text{P}^4$ and $+18\, E_\text{P}^4$.

\section{Cosmology as approach to equilibrium} \label{sec:Dynamics}

\subsection{Energy exchange between vacuum and gravity+matter}

In the curved Universe and/or in the presence of matter, $q$ becomes space-time
dependent due to interaction with gravity and matter  (see \eqref{eq:Maxwell2}). As a result the vacuum energy can be  transferred to the energy of gravitational field and/or to the energy of matter fields.
This also means that the energy of matter is not conserved.
The energy-momentum tensor of matter $T^\text{M}_{\mu\nu}$, which enters the generalized Einstein equations \eqref{eq:EinsteinEquationF2},  is
 determined by variation over $g^{\mu\nu}$ at constant $q$. That is why it is not conserved:
\begin{equation}
\nabla_\nu T^{\text{M}\mu\nu} =-\frac{\partial \mathcal{L}^\text{M}}{\partial q}
\nabla_\mu q
 \,.
\label{eq:non-conservation}
\end{equation}
The matter energy can be transferred to the vacuum energy due to interaction
with $q$-field. Using \eqref{eq:Maxwell2} and
the equation \eqref{eq:Lambda} for cosmological constant
one obtains that the vacuum energy is transferred both to gravity
and matter with the rate:
\begin{equation}
\nabla_\mu \Lambda\equiv  \nabla_\mu\rho_{\rm vac}=\left(
\frac{d\epsilon(q)}{d q}-\mu\right)\nabla_\mu q
 =-\frac{R}{16\pi} \frac{dG^{-1}(q)}{d q}   \nabla_\mu q
 +\nabla_\nu T^{\text{M}\mu\nu}
 \,.
\label{eq:non-conservation_vacuum}
\end{equation}

The energy exchange between the vacuum and gravity+matter allows for the relaxation of the vacuum energy and  cosmological ``constant''.

\subsection{Dynamic relaxation of vacuum energy}
\label{Dynamic_relaxation}

Let us assume that 
we can make a sharp kick of the system from its equilibrium state.
For quantum liquids (or any other quantum condensed matter) we know the result of the kick: the liquid or superconductor starts to move back to the equilibrium state, and with or without oscillations it finally approaches
the equilibrium
\cite{VolkovKogan1974,Barankov2004,Yuzbashyan2005,Yuzbashyan2008,Gurarie2009}. 
The same should happen with the quantum vacuum.
Let us consider this behavior using  the realization of the vacuum $q$ field in terms of the 4-form field,
when $\mu$ serves as the overall  integration constant. We start with the fully equilibrium vacuum state, which is characterized by the values $q=q_0$ and $\mu=\mu_0$ in \eqref{eq:equil-eqs}.  The kick moves the variable
$q$ away from its equilibrium value, while $\mu$  still remains the same being the overall integration constant, $\mu=\mu_0$. In the non-equilibrium state which arises  immediately after the kick, the vacuum energy is non-zero and big. If the
kick is very sharp, with the time scale of order $t_{\rm P}=1/E_{\rm P}=\sqrt{G_\text{N}}$, the energy density of the vacuum can reach the Planck-scale value,
$\rho_\text{vac}\sim E_\text{P}^4$.

For simplicity we ignore  the interaction between the vacuum and matter. Then from the solution
of dynamic equations \eqref{eq:Maxwell2} and \eqref{eq:EinsteinEquationF2} with $\mu=\mu_0$ one 
finds  that after the kick $q$ does  return to  its equilibrium value $q_0$ in  the Minkowski vacuum. At
late time the relaxation has the following asymptotic behavior: \cite{KlinkhamerVolovik2008b} 
\beq\label{eq:time-dependence-q}
q(t)-q_0\sim\,
        q_0  \frac{  \sin\omega\, t }{ \omega\,  t} ~~,~~ \omega \,t\gg 1\,,
\eeq
where oscillation frequency $\omega$ is of the order of the Planck-energy scale
$E_\text{P}$. The gravitational constant $G$ approaches
its Newton value $G_\text{N}$ also with the power-law modulation:
\beq\label{eq:time-dependence-G}
G(t)-G_\text{N} \sim\,
        G_\text{N}    \frac{  \sin\omega\, t }{ \omega\,  t} ~~,~~ \omega \,t\gg 1\,,
\eeq
The vacuum energy relaxes to zero in the following way:
\bsubeqs\label{eq:VacuumEnergyDecay}
\beqa
 \rho_\text{vac}(t)
 \propto
 \frac{\omega^2}{t^2}\;\sin^2 \omega\,  t~~,~~\omega\,  t\gg 1\,,
\label{eq:VacuumEnergyOscillating-dimensionfull}
\eeqa
For the Planck scale kick,  the vacuum energy density after the kick, i.e.  at $t\sim 1/E_\text{P}$, has a Planck-scale value,
$\rho_\text{vac}\sim E_\text{P}^4$. According to \eqref{eq:VacuumEnergyOscillating-dimensionfull}, 
at present time it must reach the value
\begin{equation}
\overline \rho_\text{vac}(t_{\rm present})
 \propto
 \frac{E_\text{P}^2}{t^2_{\rm present}}\sim E_\text{P}^2H^2 \,,
\end{equation}
where $H$ is the Hubble parameter. This value approximately  corresponds to the
measured value of the cosmological constant.

This, however, can be considered as an illustration of the dynamical reduction of the
large value of the cosmological constant, rather than the real scenario of the evolution of the Universe.
We did not take into account quantum
dissipative effects and the energy exchange between vacuum and matter. 
Indeed, matter field radiation (matter quanta emission)
by the oscillations of the vacuum can be expected to lead to
faster relaxation of the initial vacuum energy~\cite{Starobinsky1980},
\beqa
 \rho_\text{vac}(t)
 \propto \Gamma^4  \exp(-\Gamma\, t)\,,
\label{eq:VacuumEnergyQuantum-dimensionful}
\eeqa
\esubeqs
with a decay rate $\Gamma\sim \omega \sim E_\text{P}$.

Nevertheless, the cancellation mechanism and example of relaxation provide the following lesson. The Minkowski-type solution appears without fine-tuning of the parameters of the action,
precisely because the vacuum is characterized by a constant derivative
of the vacuum field rather than by a constant vacuum field itself.
As a result, the parameter $\mu_0$ emerges in \eqref{eq:equil-eqs-mu}
as an \emph{integration constant}, i.e., as a parameter of the solution
rather than a parameter of the action. Since after the kick the integration constant remains 
intact, the Universe will  return to its equilibrium Minkowski state with $\rho_\text{vac}=0$, even if in the non-equilibrium state after the kick the  vacuum energy could reach
$\rho_\text{vac}\sim E_\text{P}^4$.
The idea that the constant derivative of a field may be important
for the cosmological constant problem has been suggested earlier by
Dolgov~\cite{Dolgov1985,Dolgov1997} and
Polyakov~\cite{Polyakov1991,PolyakovPrivateComm}, where the latter explored
the analogy with the Larkin--Pikin effect~\cite{LarkinPikin1969}
in solid-state physics.

\subsection{Minkowski vacuum as attractor} \label{sec:Attractor}

The example of relaxation of the vacuum energy in Sec. \ref{Dynamic_relaxation} has the principle drawback. 
Instead of the fine-tuning of the action, which is bypassed in the self-sustained vacuum,
we have the fine-tuning  of the integration constant. We assumed that originally, before the kick, the Universe was in its Minkowski ground state,
and thus the specific value of the integration constant $\mu=\mu_0$ has been chosen, that fixes the value $q=q_0$ of the original Minkowski equilibrium vacuum.
In the  4-form realization of the vacuum field, any other choice of the integration constant ($\mu \ne \mu_0$) leads asymptotically to a de-Sitter-type solution~\cite{KlinkhamerVolovik2008b}.

To avoid this fine-tuning and obtain the natural relaxation
of $\mu$ to $\mu_0$, which as we know occurs in quantum liquids,
we must relax the condition on $\mu$. It should not serve
as an overall integration constant, while remaining the conjugate
variable in thermodynamics. Then using the condensed matter experience one may expect
that the Minkowski equilibrium vacuum becomes an attractor and the  de-Sitter solution with $\mu \neq\mu_0$ will
inevitably relax to Minkowski vacuum with $\mu=\mu_0$.
This expectation is confirmed in the aether type realization of the vacuum variable
in terms of a vector field  as discussed in Sec. \ref{Aether_field}.

The constant vacuum field $q$ there appears 
as the derivative
of a vector field in the specific solution $u^q_\beta$
corresponding to the equilibrium vacuum,
$q\,g_{\alpha\beta} \equiv \nabla_\alpha\, u^q_{\beta}=
u^q_{\alpha\beta}$.
In this realization, the effective chemical potential
$\mu \equiv d\epsilon(q)/d q$ plays a role only for the equilibrium states
(i.e., for their thermodynamical properties),
but $\mu$ does not appear as an integration constant for the dynamics.
Hence, the fine-tuning problem of the integration constant is overcome,
simply because there is no integration constant.

The instability of the de-Sitter solution towards the Minkowski one has been already demonstrated by
Dolgov~\cite{Dolgov1997}, who considered  the simplest
quadratic choices of the Lagrange density of $u_\beta(x)$. 
But his result also holds for the generalized Lagrangian with a generic
function $\epsilon(u_{\alpha\beta})$  in Sec. \ref{Aether_field} \cite{KlinkhamerVolovik2010}.

\begin{figure}[t]
\includegraphics[width=.6\textwidth]{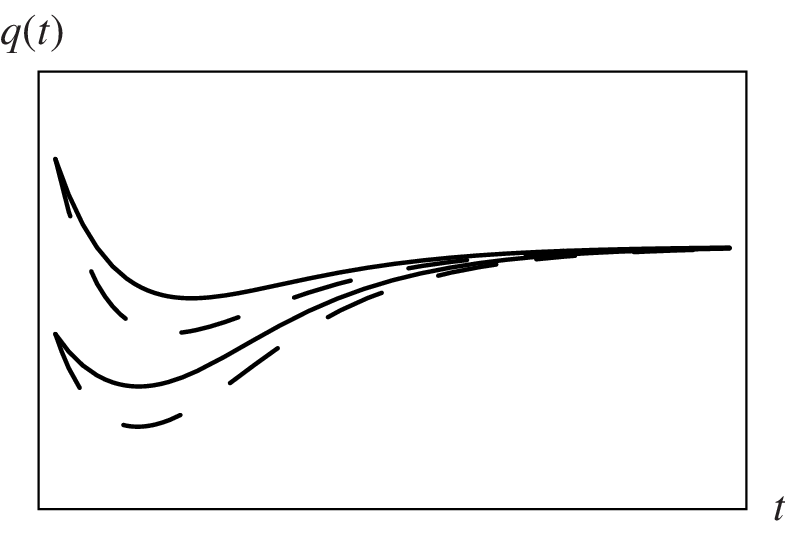}
\caption{
Aaether-field $q$ evolution and Minkowski attractor
 in a spatially flat Friedmann--Robertson--Walker universe in Dolgov model \cite{Dolgov1997} (see \cite{KlinkhamerVolovik2010} for details). 
The  bare cosmological constant is $\Lambda_\text{bare} \sim E_\text{P}^4$.   Four numerical solutions correspond to different boundary conditions, but all approach the Minkowski-spacetime
solution \eqref{eq:asymp_solution}.
The Minkowski vacuum is an attractor
because the vacuum compressibility \eqref{eq:Stability} is positive,
$\chi(q_0)>0$.}
\label{fig:attractor}
\end{figure}

The Dolgov scenario does not require the variable gravitational coupling parameter, so that
we use $G(q)=\text{const}$.
In this scenario, for a spatially flat Robertson--Walker metric with
cosmic time $t$ and scale factor $a(t)$,
the initial de-Sitter-type expansion evolves towards the Minkowski attractor by the
following $t\rightarrow \infty$ asymptotic solution for the aether-type field
$u_\beta=(u_0(t),0)$:
\beq
u_0(t)\rightarrow q_0\,t \,,\quad
H(t) \rightarrow 1/t\,,
\label{eq:asymp_solution}
\eeq
where  the Hubble parameter $H\equiv [da/dt]/a$. 
At large cosmic times $t$, the curvature terms
decay as $R\sim H^2 \sim 1/t^2$ and the Einstein equations
lead to the nullification of the energy-momentum tensor of the $u_{\beta}$ field:
$T_{\alpha\beta}[u]=0$. Since \eqref{eq:asymp_solution}
with $d u_0/dt = H\,u_0$  satisfies the  $q$--theory \emph{Ansatz}
$u_{\alpha\beta} = q\,g_{\alpha\beta}$,
the  energy-momentum tensor is completely expressed by the single constant $q$:
$T_{\alpha\beta}(q) =[\epsilon(q)  -  q\, d\epsilon(q)/d q]\,g_{\alpha\beta}$.
As a result, the equation  $T_{\alpha\beta}(q)=0$
leads to the equilibrium condition \eqref{eq:equil-eqs-q0}
for the Minkowski vacuum and to the equilibrium value $q=q_0$
in \eqref{eq:asymp_solution}. 

Figure~\ref{fig:attractor} shows explicitly the attractor behavior for the simplest case of Dolgov action,
with the numerical value of $q_0$ in \eqref{eq:asymp_solution} appearing
\emph{dynamically}.  This simple version of Dolgov scenario does not appear to give a realistic description of the present Universe~\cite{RubakovTinyakov1999} and requires an appropriate modification \cite{EmelyanovKlinkhamer2011}.
It nevertheless demonstrates that the compensation of a
large initial vacuum energy density can occur dynamically
and that Minkowski spacetime can emerge
spontaneously, without setting a chemical potential.
In other words, an ``existence proof'' has been given for the conjecture
that the appropriate Minkowski value $q_0$ can result from an attractor-type
solution of the field equations.
The only condition for the Minkowski vacuum to be an attractor
is a positive vacuum compressibility \eqref{eq:Stability}.

\begin{figure}[t]
\includegraphics[width=.7\textwidth]{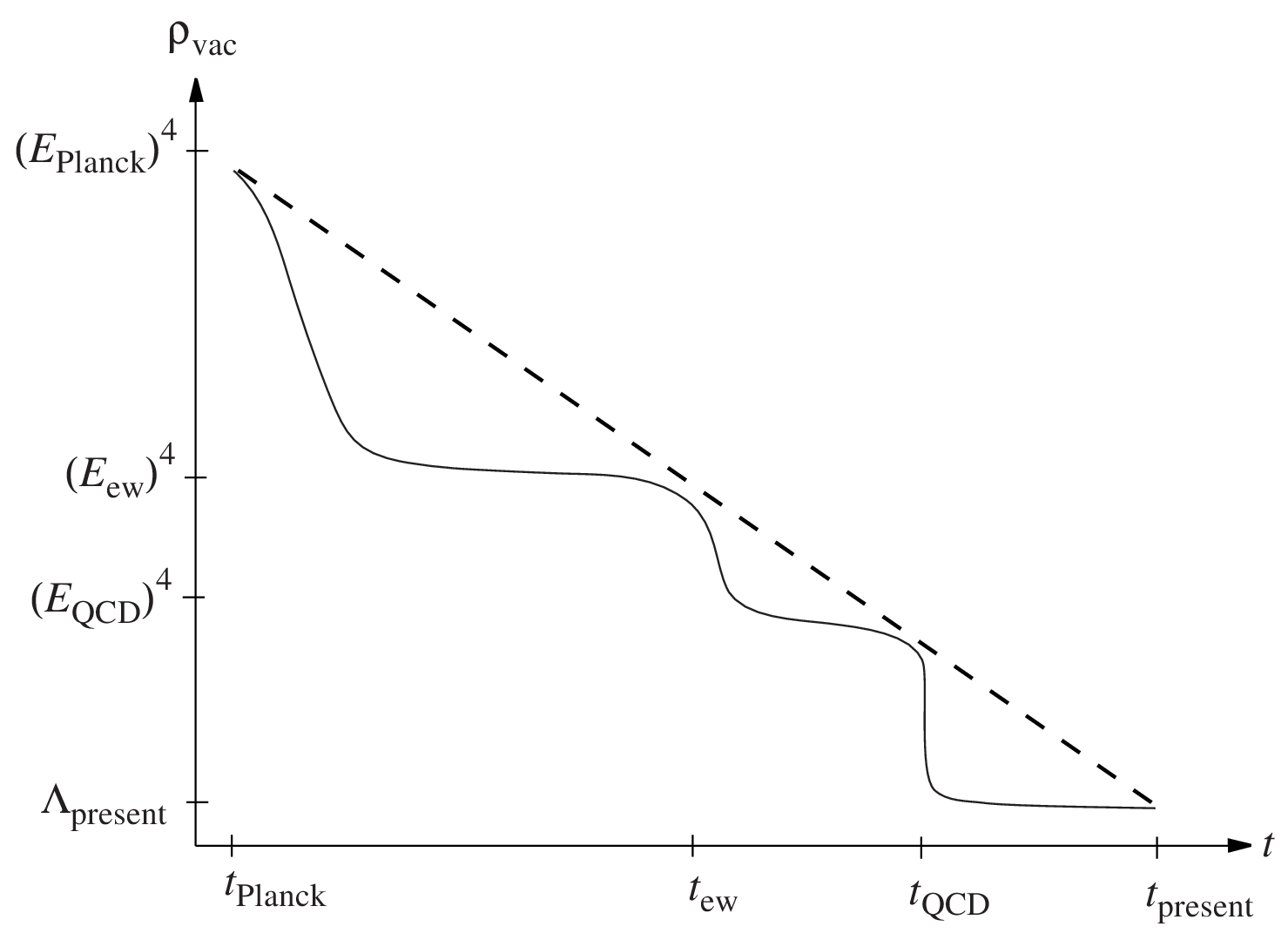}
\caption{
{\it Dashed curve}:  relaxation according to the relation
$<\rho_{\rm vac}(t)>  \;\sim\, (E_{\rm Planck})^2/t^2$ in 
Eq.(\ref{eq:VacuumEnergyOscillating-dimensionfull}).
{\it Full curve}: Sketch of the relaxation of the vacuum energy density
during the evolution of the Universe according to Ref.~\cite{KlinkhamerVolovik2011a}.
 The origin of the current plateau in the vacuum energy $\Lambda_{\rm present}$ 
is discussed in Sec. \ref{sec:Remnant}.
}
\label{fig:StepRelaxation}
\end{figure}
 
\subsection{Remnant cosmological constant} \label{sec:Remnant}

Figure~\ref{fig:StepRelaxation} demonstrates the possible more
realistic scenario with a
step-wise relaxation of the vacuum energy density \cite{KlinkhamerVolovik2011a}.
The vacuum energy density moves from plateau to plateau responding to the 
possible phase transitions or crossovers in the Standard Model vacuum and
follows, on average, the steadily decreasing matter energy density.
The origin of the current plateau with a small positive value of the
vacuum energy density  $\Lambda_{\rm present}=\rho_{\rm vac}\sim \big(10^{-3}\,\text{eV}\big)^4$ is still not clear. It
may result from the phenomena, which occur in the infrared. It may come for example from anomalies in the neutrino sector of the quantum vacuum, such as non-equilibrium contribution
of the light massive neutrinos to the quantum vacuum \cite{KlinkhamerVolovik2011a}; 
reentrant violation of Lorentz invariance \cite{Volovik2001} and Fermi point splitting in the neutrino sector
\cite{KlinkhamerVolovik2005b,KlinkhamerVolovik2011b} (see Sec. \ref{Higgs_vs_splitting}).
The other possible sources include the QCD anomaly 
\cite{Schutzhold2002,KlinkhamerVolovik2009a,UrbanZhitnitsky2009,Ohta2011,Holdom2011};  torsion \cite{Poplawski2011}; relaxation effects during the electroweak crossover \cite{KlinkhamerVolovik2009b}; etc. Most of these scenarios are determined by the momentum space  topology of the quantum vacuum.

\subsection{Summary and outlook}\label{sec:Summary}

To study the problems related to quantum vacuum one must search for the proper extension
of the current theory of elementary particle physics -- the Standard Model.
However, many properties of the quantum vacuum can be understood by extending of our experience
with self-sustained macroscopic systems to the quantum vacuum.
A simple picture of quantum vacuum is based on three assumptions:
(i) The quantum vacuum is a self-sustained medium -- the system which is stable at zero external pressure, like quantum liquids.
(ii)  The quantum vacuum is characterized by a conserved charge $q$, which is analog of the particle density $n$ in quantum liquids and which is non-zero in the ground state of the system, $q=q_0\neq 0$.
(iii) The quantum vacuum with $q=q_0$ is a Lorentz-invariant state. This is the only property which distinguishes the quantum vacuum from the condensed-matter quantum liquids.

These assumptions naturally solve the main cosmological constant problem without fine-tuning.
In any self-sustained system, relativistic or non-relativistic, in thermodynamic equilibrium at $T=0$ the zero-point energy of quantum fields is fully compensated by the microscopic degrees of freedom, so that the relevant energy density is zero in the ground state. This consequence of thermodynamics is automatically fulfilled in any system, which may exist  without external environment. This leads to the trivial result for gravity: the cosmological constant in any equilibrium vacuum state is zero.
The zero-point energy  of the Standard Model fields
is automatically compensated by the $q$--field that describes
the degrees of freedom of the deep quantum vacuum.

These assumptions  allow us to suggest that  cosmology is the process of equilibration. From the condensed matter experience we know that the ground state of the system serves as  an attractor:
 starting far away from equilibrium, the quantum liquid finally reaches its ground state. The same should occur for the particular case of our Universe:  starting far away from equilibrium in a very early phase of universe, the vacuum  is moving towards the Minkowski attractor. We are now close to this attractor, simply because our Universe is old. This is a possible reason of the small remnant cosmological constant measured in present time. 
 
The $q$--theory transforms the standard cosmological constant problem into the search for the proper decay mechanism of the vacuum energy density
and for the proper mechanism of formation of small remnant cosmological constant. For that we need the theory of dynamics of quantum vacuum. The latter is a new topic in physics waiting for input from theory and observational cosmology. Using several possible realozation of the vacuum variable  $q$ we are able to model some features of the vacuum dynamics in a hope that this will allow us to find the generic features and construct the phenomenology of equilibration.

\section{Vacuum as topological medium}

\subsection{Topological media} 

There are two schemes for the classification of
states in condensed matter physics and relativistic quantum fields:
classification by symmetry and classification  by topology. 

For the first classification method, a given state of the
system is characterized by a symmetry group $H$ which is a subgroup
of the symmetry group $G$ of the relevant physical laws. 
The thermodynamic phase transition
between equilibrium states is usually marked by a change of the
symmetry group $H$. This classification reflects the
phenomenon of spontaneously broken symmetry. In relativistic quantum
fields the chain of successive phase transitions, in which the large
symmetry group existing at high energy is reduced at low energy, 
is in the basis of the Grand
Unification models (GUT) \cite{UnificationModel,Unification}. In
condensed matter  the spontaneous symmetry breaking is a typical
phenomenon, and the thermodynamic states are also classified in terms of 
the subgroup $H$ of the relevant group $G$  (see e.g, the classification 
of superfluid and superconducting states  in Refs.
\cite{VolovikGorkov1985,VollhardtWoelfle}). The groups $G$ and $H$  are
also responsible for classification of topological defects,  which are determined by the
nontrivial elements of the homotopy groups $\pi_n(G/H)$ \cite{TopologyReview1}. 

The second classification method  deals with the ground states of the system at zero temperature
($T=0$). In particle physics it is the classification of quantum vacua. Topological media are systems whose properties are protected by topology and thus are robust to deformations of the action.
The universality classes of topological media are determined by momentum-space
topology. The latter is also responsible for the type of the effective
theory which emerges at low energy. In this sense, topological classification reflects the tendency  opposite to GUT , which is called
the anti Grand Unification (anti-GUT).  In 
the GUT scheme,  the fundamental symmetry of the vacuum
state is primary and the phenomenon of spontaneous symmetry breaking gives rise to topological defects. In the anti-GUT scheme the topology  is primary, while  effective symmetry gradually emerges at low energy  \cite{Froggatt1991,Volovik2003}.

Different aspects of physics of topological matter have been discussed, including topological stability of gap nodes; classification of fully gapped vacua;
edge states; Majorana fermions; influence of disorder and interaction; topological quantum phase transitions; intrinsic quantum Hall and spin-Hall effects; quantization of physical parameters; experimental realization; connections with relativistic quantum fields; chiral anomaly; topological Chern-Simons and Wess-Zumino actions; etc.

\subsection{Gapless topological media}

There are two big groups of topological media: with fully gapped fermionic excitations and with gapless fermions.
 
In 3+1 spacetime, there are four basic universality classes
of gapless fermionic vacua protected by  topology in momentum space
\cite{Volovik2003,Horava2005}: 
 
(i) Vacua with fermionic excitations
characterized by Fermi points (Dirac points, Weyl points, Majorana points, etc.) -- points in 3D momentum space at which the
energy  of fermionic quasiparticle vanishes. Examples
are provided by the spin triplet $p$-wave superfluid $^3$He-A, Weyl semimetals, and also by the quantum vacuum of
Standard Model above the electroweak transition, where all elementary
particles are Weyl fermions with Fermi points in the spectrum. This
universality class manifests the phenomenon of emergent relativistic
quantum fields at low energy: close to the Fermi points the
fermionic quasiparticles behave as massless Weyl fermions, while the
collective modes of the vacuum interact with these fermions as gauge and
gravitational fields.

(ii) Vacua with fermionic excitations characterized by lines in  
3D momentum space or points in 2D momentum space. We shall characterize zeroes by
their co-dimension --  the dimension of ${\bf p}$-space minus the
dimension of the manifold of zeros.  Lines in  
3D momentum space and points in 2D momentum space have  co-dimension 2: 
since
$3-1=2-0=2$; compare this with zeroes of class (i) which have 
co-dimension
$3-0=3$. 
Zeroes of co-dimension 2  are
topologically stable only if some special symmetry is obeyed. Examples are 
provided by the vacuum of the high
$T_c$  cuprate superconductors where the Cooper pairing into a $d$-wave state 
occurs \cite{Campuzano2008} and graphene 
\cite{Volovik2007,Manes2007,Vozmediano2010,Cortijo2011}. 
Nodes in spectrum are stabilized there by the combined effect of momentum-space
topology and discrete symmetry.

(iii) Vacua with fermionic excitations characterized by Fermi surfaces. 
The representatives of this universality class are normal metals and
normal liquid $^3$He.  This universality class also manifests the
phenomenon of emergent physics, though non-relativistic: at low
temperature all the metals behave in a similar way, and this behavior is
determined by the Landau theory of Fermi liquid -- the effective theory
based on the existence of Fermi surface.  Fermi surface has co-dimension
1: in 3D system it is the surface (co-dimension $=3-2=1$), in 2D system
it is the line (co-dimension $=2-1=1$), and in 1D system it is the point
(co-dimension $=1-0=1$; in one dimensional system the Landau Fermi-liquid
theory does not work, but the Fermi surface survives). 

(iv) The  Fermi band class, where the energy vanishes in
the finite region of the 3D momentum space,  and thus zeroes have
co-dimension 0. The possible states of this kind  has been discussed
in \cite{Khodel1990,NewClass,Shaginyan2010}.  
In particle physics, the Fermi band or the Fermi ball appears  in  a 2+1 dimensional nonrelativistic quantum field theory which is dual to a gravitational theory in the anti-de Sitter  background with a charged black hole
\cite{Sung-SikLee2009}.
Topologically stable flat band exists on the surface of the materials with lines
of zeroes in bulk 
\cite{SchnyderRyu2010,HeikkilaKopninVolovik2011,SchnyderBrydonTimm2011} and
in the spectrum of fermion zero modes localized in the core of some vortices 
\cite{KopninSalomaa1991,Volovik1994,Volovik2011a}.

\subsection{Fully gapped topological media} 

The gapless and gapped vacuum states are interrelated. For example, the quantum phase transition between the fully gapped states with different topology occurs via the intermediate gapless state. The related phenomenon is that the interface between the  fully gapped states with different values of topological invariant contains gapless fermions.

The most popular examples of the fully gapped  topological matter are topological insulators   \cite{Kane2005,HasanKane2010,Xiao-LiangQi2011}. The first discussion of the possibility of  3+1 topological insulators can be found in Refs. \cite{Volkov1981,VolkovPankratov1985}. The main feature of such materials is that they are insulators in bulk, where electron spectrum  has a gap, but there are  2+1 gapless edge states of electrons on the surface or  at the interface between topologically different bulk states as discussed in Ref.  \cite{VolkovPankratov1985}.  The spin triplet $p$-wave superfluid $^3$He-B is another example the fully gapped 3+1 matter with nontrivial topology in momentum space. It has 2+1 gapless quasiparticles living at interfaces between vacua with different values of the topological invariant describing the bulk states of $^3$He-B \cite{SalomaaVolovik1988,Volovik2009}.  
The only difference from the topological insulators is that the gapless fermions living
on the surface of the topological superfluid and superconductor or at the interface are Majorana fermions. The quantum vacuum of Standard Model below the electroweak transition, i.e. in its massive phase,  is the relativistic counterpart of the topological insulators and gapped topological superfluids \cite{Volovik2010a}.

Examples of the  2+1  topological fully gapped systems are provided by the 
films of superfluid $^3$He-A with broken time reversal symmetry \cite{VolovikYakovenko1989,Volovik1992b} and by the  planar phase which is time reversal invariant
\cite{VolovikYakovenko1989,Volovik1992b}. The topological invariants for 2+1 vacua give rise to  quantization of the Hall and spin-Hall conducticity in these films in the
absence of external magnetic field (the so-called intrinsic qauntum and spin-quantum Hall effects) \cite{VolovikYakovenko1989,SQHE}, see Sec. \ref{Quantum_spin_Hall_effect}.

\subsection{Green's function as an object}

For study the topological properties of condensed matter systems, the ideal noninteracting systems
are frequently used. Sometimes this is justified, if one can find  the effective single-particle Hamiltonian, which emerges at  low energy and which reflects the topological properties of the real interacting many-body system.  However, in general the primary object for the topological classification
of the real systems is the one-electron  propagator -- Green's function $G(\omega,{\bf p})$. 
 In principle one can construct the effective Hamiltonian by proper simplification of the Green's function
 at zero frequency,  $H=G^{-1}(\omega=0,{\bf p})$. Though in the interacting case the propagator $G({\bf p},\omega=0)$  determines correctly only the zero energy states, see e.g. \cite{Haldane2004}, in some cases it can be used for the construction of the topological invariants alongside with the full Green's function $G(\omega,{\bf p})$.
On the other hand there are situations when the Green's function does not have poles (see \cite{Volovik2007,FaridTsvelik2009,Giamarchi2004}). In these cases  no well defined energy spectrum exists, and the effective low energy Hamiltonian cannot be introduced. 
In particle physics, interaction may also  lead to the anomalous infrared behavior of propagators. For example, the pole in the Green's function is  absent for the so-called unparticles \cite{Georgi2007,LuoZhu2008}; the phenomenon of quark confinement  in QCD can lead to the anomalous infrared behavior of the quark and gluon propagators  \cite{Gribov1978,Chernodub2008,Burgio2009}; marginal Green's function of fermions may occur at the black hole horizon \cite{Faulkner2010}; etc.
Thus in the interacting systems, all the information on the topology  is encoded in the topology of the Green's function matrix, and also in its  symmetry. The latter is important, because symmetry supports additional topological invariants, which are absent in the absence of symmetry, see below. 

Green's function topology has been used in particular  for classification of topologically protected nodes in the quasiparticle energy spectrum of systems of different dimensions
including the vacuum of Standard Model in its gapless state \cite{Froggatt1991,Volovik2003,Horava2005,Volovik2007}; for the classification of the topological ground states in the fully gapped $2+1$ systems, which experience intrinsic quantum Hall and spin-Hall effects  \cite{VolovikYakovenko1989,Yakovenko1989,SenguptaYakovenko2000,ReadGreen2000,Volovik2003,Volovik2007};   in relativistic quantum field theory  of $2+1$ massive Dirac fermions 
  \cite{So1985,IshikawaMatsuyama1986,IshikawaMatsuyama1987,Matsuyama1987,Jansen1996}
and $3+1$ massive Dirac fermions \cite{Volovik2010}; etc. (see also recent papers
\cite{EssinGurarie2011,ZubkovVolovik2012}).

 For the topological classification of the gapless vacua, the Green's function is considered at imaginary frequency $\omega=ip_0$.
This allows us to consider only the relevant singularities in the Green's function and to avoid the singularities on the mass shell, which exist in any vacuum, gapless or fully gapped. 

\subsection{Fermi surface as topological object}

\begin{figure}[t]
\centerline{\includegraphics[width=0.5\linewidth]{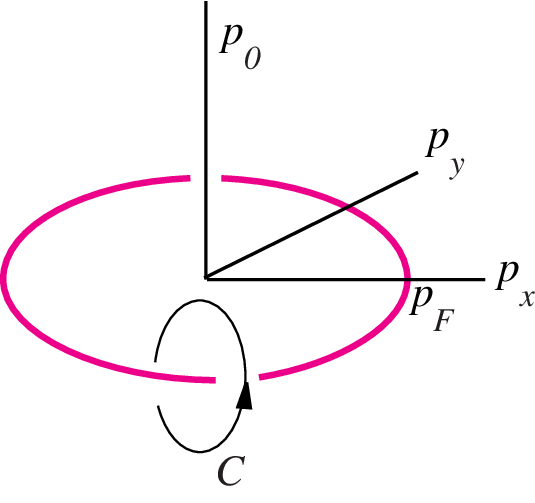}}
\medskip
\caption{Fermi surface in 2+1 systems represents the nodes of co-dimension 1. In this case, the Green's function has singularities on line $p_0=0$, $p_x^2+p_y^2=p_F^2$ in the three-dimensional space $(p_0,p_x,p_y)$.
 Stability of Fermi surface is protected by the invariant \eqref{InvariantForFS} which is represented by  integral  over an arbitrary contour $C$ around the Green's function singularity.
 This is applicable to nodes of co-dimension 1 in any $D$+1 dimension. For  $D=3$ the nodes form conventional Fermi surface in metals and in normal $^3$He.}
\label{FermiSurfaceQPTFig}
\end{figure}

Let us start with gapless vacua. 
The Green's function is generally a matrix with spin
indices. In addition, it may have the band indices (in the case of 
electrons in the periodic potential of crystals).   The general
analysis \cite{Horava2005} demonstrates that topologically stable  
nodes of co-dimension 1 (Fermi
surface in 3+1 metal, Fermi line in 2+1 system or Fermi point in 1+1 system) are described by the group $Z$ of integers. The  winding number
$N_1$, which is responsible for the topological stability of these node, is expressed analytically  in terms of the Green's function
\cite{Volovik2003}:
\begin{equation}
N_1={\bf tr}~\oint_C {dl\over 2\pi i}  G(p_0,{\bf p})\partial_l
G^{-1}(p_0,{\bf p})~.
\label{InvariantForFS}
\end{equation}
Here the integral is taken over an arbitrary contour $C$ around the Green's function singularity
in  the $D+1$
momentum-frequency space. See Fig. \ref{FermiSurfaceQPTFig} for D=2. Example of the Green's function in any dimension $D$ is scalar function
$G^{-1}(\omega,{\bf p})=ip_0 - v_F(|{\bf p}|-p_F)$. For $D=2$, the singularity with winding number $N_1=1$ is on the line $p_0=0$, $p_x^2+p_y^2=p_F^2$, which represents the one-dimensional Fermi surface.

Due to nontrivial topological invariant, Fermi surface survives the perturbative interaction and exists even in marginal and Luttinger liquids without poles in the Green's function, where quasiparticles are not well defined.

 \section{Vacuum in a semi-metal state}

For our Universe, which obeys the Lorentz invariance,  only those vacua  are important that are either Lorentz invariant, or acquire the Lorentz invariance as an effective  symmetry emerging at low energy. This excludes the vacua with Fermi surface and Fermi lines and leaves the class of vacua with Fermi point of chiral type, in which fermionic excitations behave as left-handed or right-handed Weyl fermions   
\cite{Froggatt1991,Volovik2003}, and the class of vacua with the nodal point obeying $Z_2$ topology, where fermionic excitations behave as massless Majorana neutrinos  \cite{Horava2005,Volovik2007}.

\subsection{Fermi points in 3+1 vacua}

For relativistic quantum vacuum of our 3+1 Universe the Green's function singularity of co-dimension 3 is relevant.  They are described by the following topological invariant expressed via integer valued integral over the surface $\sigma$ around the singular point   in the 4-momentum space   $p_\mu=(p_0,{\bf p})$\cite{Volovik2003}:
 \begin{equation}
N_3 = \frac{e_{\alpha\beta\mu\nu}}{24\pi^2}~
{\bf tr}\int_\sigma   dS^\alpha
~ G\partial_{p_\beta} G^{-1}
G\partial_{p_\mu} G^{-1} G\partial_{p_\nu}  G^{-1}\,.
\label{MasslessTopInvariant3D}
\end{equation}
 
If the invaraint is nonzero, the Green's function has point singularity inside the surface $\sigma$ -- the Fermi point. If the topological charge is $N_3 =+1$ or $N_3 =-1$,  the Fermi point represents the so-called  conical Dirac point, but actually describes the chiral Weyl fermions. 
 This is the consequence of the  so-called Atiyah-Bott-Shapiro construction \cite{Horava2005},  which leads to the following general form of expansion of the inverse fermionic propagator near the Fermi point  with $N_3 =+1$ or $N_3 =-1$:
\begin{equation}
G^{-1}(p_\mu)=e_\alpha^\beta\Gamma^\alpha(p_\beta-p_\beta^{(0)})+ \cdots \,.
\label{Atiyah-Bott-Shapiro}
\end{equation}
Here $\Gamma^\mu=(1,\sigma_x,\sigma_y,\sigma_z)$ are Pauli matrices (or Dirac matrices in the more general case); the expansion parameters are the vector 
$p_\beta^{(0)}$ indicating the position of the Fermi point in momentum space where the Green's function has a singularity, and  the matrix $e_\alpha^\beta$; ellipsis denote higher order terms in expansion.  

 \subsection{Emergent fermionic matter}

The equation (\ref{Atiyah-Bott-Shapiro}) can be continuously deformed to the simple one, which describes the relativistic Weyl fermions
\begin{equation}
G^{-1}(p_\mu)=ip_0 +N_3\boldsymbol{\sigma}\cdot{\bf p} + \cdots \,,
\label{Weyl}
\end{equation}
where the position of the Fermi point is shifted to $p_\beta^{(0)}=0$ and ellipsis denote higher order terms in $p_0$ and ${\bf p}$; the matrix $e_\alpha^\beta$ is deformed to unit matrix. This means that close to the Fermi point with $N_3=+1$, the low energy fermions behave as right handed relativistic particles, while the Fermi point with $N_3=-1$ gives rise to the left handed particles. 

The equation \eqref{Weyl} suggests the effective Weyl Hamiltonian
\begin{equation}
H_{\rm eff}= N_3\boldsymbol{\sigma}\cdot{\bf p} \,.
\label{WeylEffective}
\end{equation}
However, the infrared divergences may violate the simple pole structure of the propagator in Eq.(\ref{Weyl}). In this case  in the vicinity of Fermi point one has
\begin{equation}
G(p_\mu)\propto \frac{-ip_0 +N_3\boldsymbol{\sigma}\cdot{\bf p}}{\left(p^2+p_0^2\right)^{\gamma}} ~,
\label{WeylUnparticle}
\end{equation}
with  $\gamma\neq 1$.  This modification does not change the topology of the propagator: the topological charge of singularity is $N_3$ for arbitrary parameter  $\gamma$  \cite{Volovik2007}. For fermionic unparticles one has  
$\gamma=5/2 -d_U$, where $d_U$ is the scale dimension of the quantum field \cite{Georgi2007, LuoZhu2008}.

For $N_3=\pm 2$, the spectrum of (quasi)particles in the vicinity of singularity depends on symmetry.     One may obtain either  two Weyl fermions or exotic massless fermions with nonlinear dispersion at low energy: 
semi-Dirac fermions with linear dispersion in one direction and quadratic dispersion in the two others
     \cite{Volovik2001,Volovik2003,Volovik2007}. 
  \begin{equation}
E({\bf p})\approx \pm \sqrt{c^2p_z^2 + \left(\frac{p_\perp^2}{2m}\right)^2 } \,.
\label{SemiDirac}
\end{equation}
Similar consideration for the 2+1 systems may lead  to semi-Dirac fermions and to fermions with quadratic dispersion at low energy \cite{Volovik2007,Dietl-Piechon-Montambaux2008,Banerjee2009}
\begin{equation}
E({\bf p})\approx \pm \frac{p^2}{2m} \,.
\label{WeylQuadratic}
\end{equation}
 For the higher values of topological charge, the spectrum becomes even more interesting
 (see e.g. Refs. \cite{HeikkilaVolovik2010,HeikkilaVolovik2011} for 2+1 systems). But if the relativistic invariance is obeyed, or under the special discrete symmetry, the nonzero invariant   $N_3$ corresponds to $N_3$ species of Weyl fermions near the Fermi point.

 The main property of the vacua with Dirac points is that according to \eqref{Weyl}, close to the Fermi points the massless relativistic fermions emerge. This is consistent with the fermionic content of our Universe, where all the elementary particles -- left-handed and right-handed quarks and leptons -- are Weyl fermions. Such a coincidence demonstrates that the vacuum of Standard Model in its massless phase is the topological medium of the  Fermi point universality class. This solves the hierarchy problem, since the  value 
of the masses of elementary particles in the vacua of this universality class is strictly zero. 

Let us suppose for a moment, that there is no topological invariant which protects massless fermions.
Then the Universe is fully gapped and the natural masses of  fermions must be on the order of Planck energy scale:  $M \sim E_{\rm P}\sim 10^{19}$ GeV.  In such a natural  Universe, where all masses are of order $E_{\rm P}$, all fermionic degrees of freedom are completely frozen out because of the  Bolzmann factor $e^{-M/T}$, which is about $e^{-10^{16}}$   at the temperature corresponding to the  highest energy reached in accelerators. There is no  fermionic matter in such a Universe at low energy.  That we survive in our Universe is not the result of the anthropic principle (the latter chooses the Universes which are fine-tuned for life but have an extremely low probability). Our Universe is also natural and its vacuum is generic, but it belongs to a different universality class of vacua --  the vacua with Fermi points. In such vacua the masslessness of fermions is protected by topology (combined with symmetry, see below).

 \subsection{Emergent gauge fields}

The vacua with Fermi-point suggest a particular mechanism for emergent symmetry. The Lorentz symmetry is simply the result of the linear expansion: this symmetry becomes better and better when the Fermi point is approached and the non-relativistic higher order terms in Eq.(\ref{Weyl}) may be neglected. This expansion  demonstrates the emergence of the relativistic spin, which is described by the Pauli matrices. It also demonstrates how gauge fields and gravity emerge together with chiral fermions. 
The expansion parameters  
$p_\beta^{(0)}$ and $e_\alpha^\beta$ may depend on the space and time coordinates and they actually represent collective dynamic bosonic fields in the vacuum with Fermi point. 
The vector field  $p_\beta^{(0)}$ in the expansion plays the role of   the effective $U(1)$ gauge field $A_\beta$ acting on  fermions.

For the Fermi points with topological charge $N_3>1$ the situation depends on the symmetry of the system. In the case, when the spectrum corresponds to several species of relativistic Weyl fermions,   the shift $p_\beta^{(0)}$ becomes the matrix field; it gives rise to effective non-Abelian (Yang-Mills)    $SU(N_3)$ gauge fields emerging in the vicinity of Fermi point, i.e. at low energy \cite{Volovik2003}. For example, the Fermi point with $N_3=2$ may give rise to the effective $SU(2)$ gauge field in addition
to the effective $U(1)$ gauge field
\begin{equation}
G^{-1}(p_\mu)=e_\alpha^\beta\Gamma^\alpha \left(p_\beta-g_1A_\beta -g_2 {\bf A}_\beta\cdot\boldsymbol{\tau}\right)+~{\rm higher~order~terms}  \,,
\label{SU2}
\end{equation}
where 
$\boldsymbol{\tau}$ are Pauli matrices corresponding to the emergent isotopic spin.
This is what happens in superfluid $^3$He-A.  In the case, when the symmetry leads to
exotic fermions with the non-linear spectrum $E\sim \pm p^{N_3}$,  the quantum electrodynamics with anisotropic scaling emerges \cite{KatsnelsonVolovik2012,Zubkov2012b}, which is similar to the quantum gravity with
anisotropic scaling suggested by Ho\v{r}ava    
 \cite{HoravaPRL2009,HoravaPRD2009,Horava2010}.

 \subsection{Emergent gravity} 

 The matrix field $e_\alpha^\beta$ in \eqref{SU2} acts on the (quasi)particles as  the field of vierbein, and thus describes the emergent dynamical gravity field. As a result, close to the Fermi point,  matter fields 
(all ingredients of Standard Model: chiral fermions and Abelian and non-Abelian gauge fields) 
emerge  together with geometry, relativistic spin, Dirac  matrices,  and physical laws:  Lorentz and gauge  invariance, equivalence principle, etc.
 In such vacua, gravity emerges together with matter.   If this Fermi point mechanism of emergence of physical laws works for our Universe, then the so-called  ``quantum gravity''  does not exist. The gravitational degrees of freedom can  be separated from all other degrees of freedom of quantum vacuum only at low energy.  
 
In this scenario, classical gravity is a natural macroscopic phenomenon emerging in the low-energy corner of the microscopic quantum vacuum, i.e. it is a typical and actually inevitable consequence of the  coarse graining procedure. It is possible  to quantize gravitational waves to obtain their quanta -- gravitons, since in the low energy corner the results of microscopic and effective theories coincide. It is also possible to obtain some (but not all) quantum corrections to Einstein equation and to extend classical gravity to the semiclassical level.  But one  cannot  obtain ``quantum gravity'' by  quantization of Einstein equations, since all other degrees of freedom of quantum vacuum will be missed in this procedure.

\subsection{Topological invariant for specific Fermi surface}
\label{FS}

\begin{figure}[t]
\centerline{\includegraphics[width=0.5\linewidth]{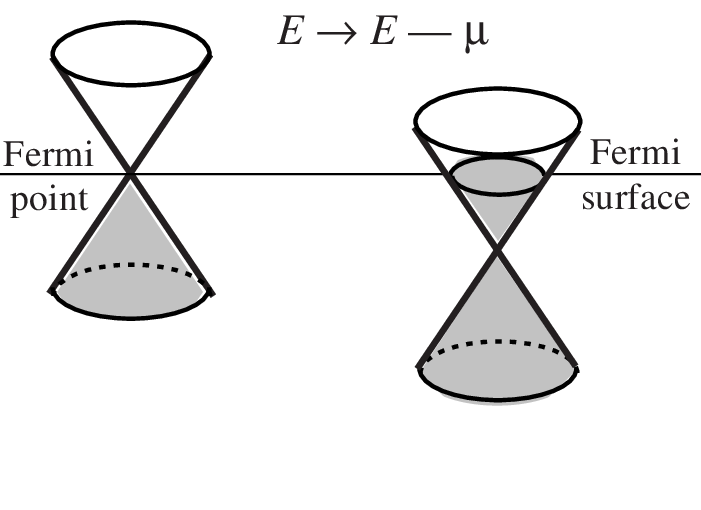}}
\caption{Fermi surface is formed from the Fermi point at finite
chemical potential of chiral fermions, when the Fermi point is moved away  from the zero energy level. }
\label{FermiSurfFromFermiPointFig}
\end{figure}

If the symmetry which fixes the position of conical (Dirac) point at zero energy level is violated, the conical point moves from the zero energy position upward or downward from the chemical potential and the Fermi surface is formed. This is shown in Fig. \ref{FermiSurfFromFermiPointFig}. This Fermi surface has specific property:  in addition to the local charge $N_1$ in \eqref{InvariantForFS}, which characterizes singularities at the Fermi surface,  it is described by the global charge $N_3$ in \eqref{MasslessTopInvariant3D}. The integral in \eqref{MasslessTopInvariant3D} is now over the surface  $\sigma$ which embrace whole Fermi sphere. The Fermi surface with the global topological charge  appears in superfluid $^3$He-A in the presence of mass flow \cite{Volovik2003}; it is also  discussed for the 2+1 systems in relation to the gapless states on the surface of 3+1 insulators \cite{Fu2007a}.

\begin{figure}[t]
\centerline{\includegraphics[width=0.8\linewidth]{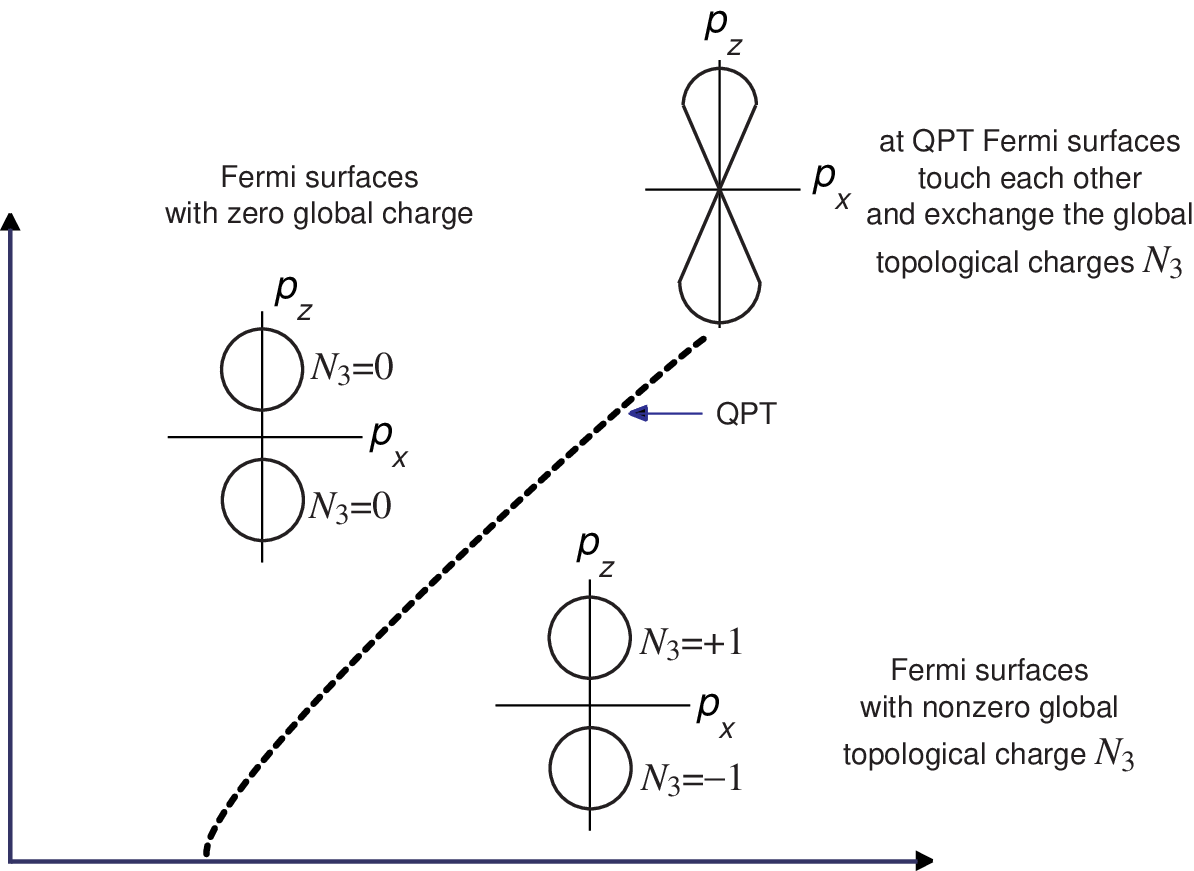}}
\medskip
\caption{Dashed line represents the topological quantum phase transition in the model
\cite{KlinkhamerVolovik2005b}. The vacua on both sides of the transition have Fermi
surfaces. On the right side of transition, Fermi surfaces have nonzero global
topological  charges  $N_3=+1$ and  $N_3=-1$.   At the transition, the
Fermi surfaces ``collide'', and their topological global charges $N_3$ annihilate
each other. On the left side of transition, Fermi surfaces become globally trivial, $N_3=0$, but retain their local topological charge $N_1$ in \eqref{InvariantForFS}.
}
\label{FSTouchFig}
\end{figure}

The ``collision''  of the Fermi surfaces in momentum space leads to the redistribution of the global topological charges $N_3$ between the Fermi surfaces when they touch each other, $(+1)+(-1)\rightarrow  0+0$  \cite{KlinkhamerVolovik2005b,Volovik2007}.  Such collision, at which the  Fermi surface looses its  global charge $N_3$, represents the topological quantum phase transition (see Fig. \ref{FSTouchFig}), one of numerous types of  transitions induced by topology in momentum space \cite{Volovik2007}.

\subsection{Topological invariant protected by symmetry in semi-metal state}
\label{TopSemi-metal}

We assume that Standard Model contains equal number  of right and  left Weyl fermions, $n_R =n_L =8n_g$, where $n_g$  is the number of generations (we do not consider Standard Model with Majorana fermions, and assume that in the insulating state of Standard Model neutrinos are Dirac fermions).
For such Standard Model the topological charge in \eqref {MasslessTopInvariant3D} vanishes, $N_3=8n_g-8n_g=0$. Thus the masslessness of the Weyl fermions is not protected by the invariant \eqref{MasslessTopInvariant3D}, and arbitrary weak interaction may result in massive particles.

However, there is another topological invariant, which takes into account the symmetry of the vacuum. The gapless state of the vacuum with $N_3=0$ can be  protected by the following
integral   \cite{Volovik2003}:
 \begin{equation}
N_3^K = {e_{\alpha\beta\mu\nu}\over{24\pi^2}}~
{\bf tr}\left[K\int_\sigma   dS^\alpha
~ G\partial_{p_\beta} G^{-1}
G\partial_{p_\mu} G^{-1} G\partial_{p_\nu}  G^{-1}\right]\,.
\label{MasslessTopInvariantStandard Model}
\end{equation}
where $K_{ij}$ is the matrix of  some symmetry transformation, which either commutes or anticommutes
with the Green's function matrix. 
In Standard Model there are two relevant symmetries, both are the  $Z_2$ groups, $K^2=1$. One of them is the center subgroup of $SU(2)_L$ gauge group of weak rotations of left fermions, where the element $K$ is the gauge rotation by angle $2\pi$, $K=e^{i\pi {\check \tau}_{3L}}$. The other one is the group of the hypercharge rotation be angle $6\pi$, 
$K=e^{i6\pi Y}$. In the  $G(224)$ Pati-Salam extension of the $G(213)$ group of Standard Model, this symmetry comes as combination of the $Z_2$ center group of the $SU(2)_R$ gauge group for right fermions, $e^{i\pi {\check \tau}_{3R}}$, and  the element $e^{3\pi i(B-L)}$ of the
$Z_4$ center group of the $SU(4)$ color group -- the $P_M$ parity (on the importance of the discrete  groups in particle physics see 
\cite{PepeaWieseb2007,Kadastik2009} and references therein).
Each of these two $Z_2$ symmetry operations   changes sign of left spinor, but does not influence the right particles. Thus these matrices are diagonal, $K_{ij}={\rm diag}(1,1,\ldots, -1,-1,\ldots)$, with eigen values 1 for right fermions and  $-1$ for left fermions.

In the symmetric phase of Standard Model, both matrices commute with the Green's function matrix $G_{ij}$, 
as a result $N^K_3$ is topological invariant: it is robust to deformations of Green's function which preserve the symmetry $K$.  The value of this invariant $N^K_3= 16 n_g$, which means that all
$16 n_g$ fermions are massless if the symmetry $K$ is obeyed.

\begin{figure}
\centerline{\includegraphics[width=0.7\linewidth]{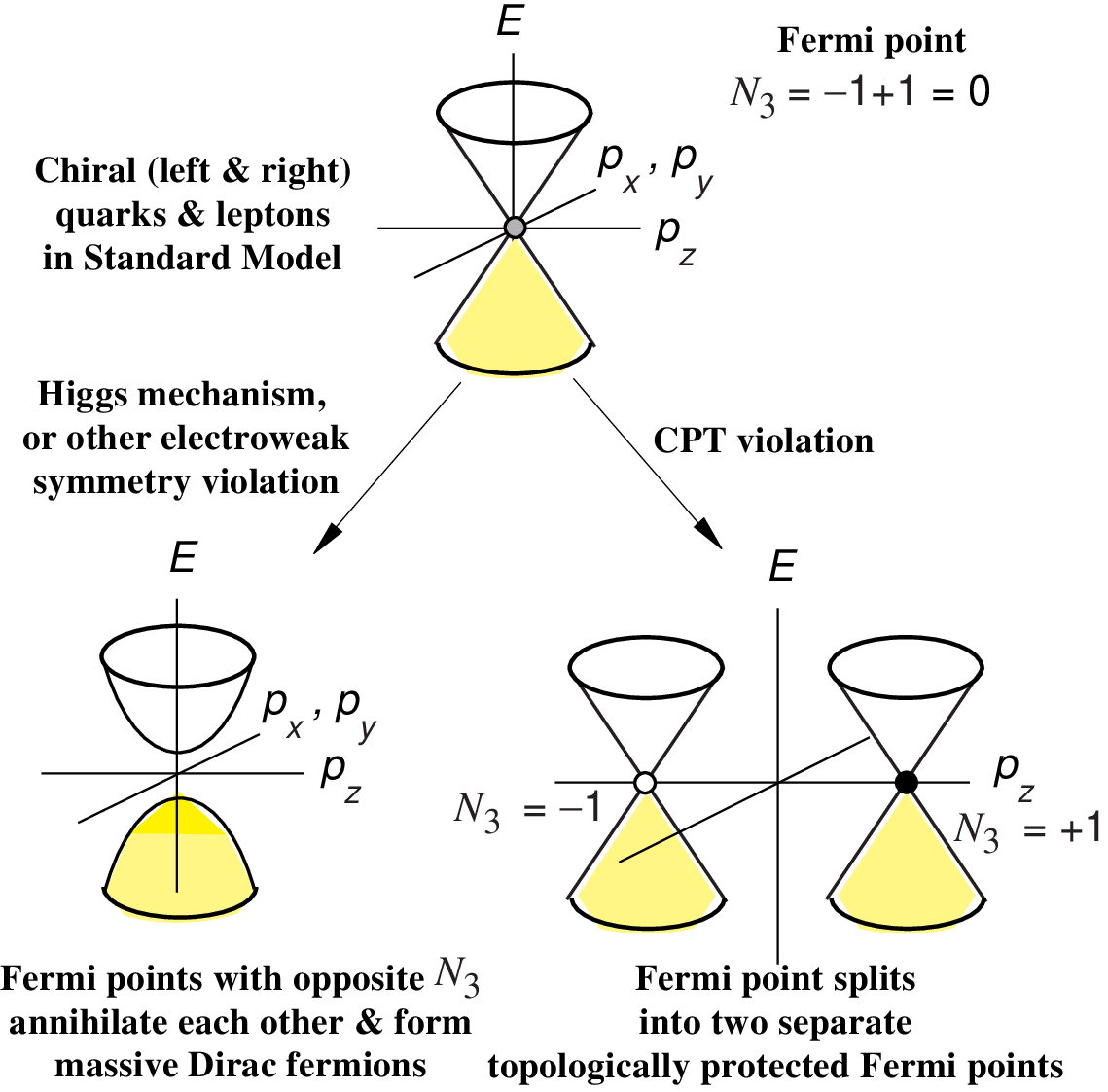}}
\caption{({\it top}): In Standard Model the Fermi points with positive $N_3=+1$ and negative $N_3=-1$ topological charges are at the same point ${\bf p}=0$, forming the marginal Fermi point with $N_3=0$. Symmetry $K$ between the Fermi points prevents their mutual annihilation giving rise to the topological invariant (\ref{MasslessTopInvariantStandard Model}) with $N_3^K=16n_g$. Figure illustrates the simple case with 
two Weyl fermions, one with  $N_3=+1$ and  another with  $N_s=-1$, when the invariant  $N_3^K=2$ protects 
the marginal Fermi point with  $N_3=+1-1=0$. 
 ({\it bottom left}): If symmetry $K$ is
violated or spontaneously broken,  Weyl points annihilate each other and  Dirac mass is formed. ({\it bottom right}): If Lorentz invariance is violated or spontaneously broken, the marginal Fermi point splits \cite{KlinkhamerVolovik2005a}. The topological quantum phase transition between the state with Dirac mass and the state with splitted Dirac points have been observed in cold Fermi gas
\cite{Tarruell2011}.
}  
\label{TwoScenarios} 
\end{figure}

 \subsection{Higgs mechanism vs splitting of Fermi points} 
 \label{Higgs_vs_splitting}
 
The gapless vacuum of Standard Model is supported by combined action of topology and symmetry $K$, and also by the CPT and Lorentz  invariance which keep all the Fermi points at ${\bf p}=0$.

Explicit violation or spontaneous breaking of one of these symmetries 
transforms the vacuum of the Standard Model
into one of the two possible vacua.  If, 
for example, the $K$  symmetry is broken, the invariant \eqref{MasslessTopInvariantStandard Model} supported by this symmetry ceases to exist, and
 the Fermi point
disappears. All $16n_g$  fermions become massive  
(Fig.~\ref{TwoScenarios} {\it bottom left}). This is assumed to happen below the
symmetry breaking electroweak transition caused by Higgs mechanism
where quarks and charged leptons acquire the Dirac masses.  

If, on the other hand, the CPT symmetry is violated, the
marginal Fermi point splits into topologically stable Fermi
points with non-zero invariant $N_3$, which protects massless chiral fermions
(Fig.~\ref{TwoScenarios} {\it bottom right}). Since the invariant $N_3$ does not depend on symmetry, the further symmetry breaking cannot destroy the nodes.
One can speculate that in the  Standard Model the latter may happen
with the electrically neutral leptons, the neutrinos
\cite{KlinkhamerVolovik2005b}.  Fermi point splitting in the neutrino sector may serve as an example of spontaneous breaking of Lorentz symmetry
\cite{KlinkhamerVolovik2011b,Klinkhamer2011b}. It
may also provide a new source of T and CP violation
in the leptonic sector, which may be relevant
for the creation of the observed cosmic
matter-antimatter asymmetry \cite{Klinkhamer2006}.
Examples of splitting of Fermi and Majorana points in condensed matter are discussed in the review paper \cite{Volovik2007}.

 \subsection{Fermi points in condensed matter} 

 \subsubsection{Chiral superfluid $^3$He-A} 

The discovery of superfluid $^3$He in 1972 \cite{Osheroff1972,VollhardtWoelfle} marked the first condensed matter realization of topological medium. Both phases of superfluid $^3$He (gapless $^3$He-A and fully gapped $^3$He-B) are topological superfluids.
The chiral superfluid $^3$He-A with broken time reversal symmetry has the following simplified Green's function for each spin projection
\begin{equation}
 G^{-1}(\omega,{\bf p})=i p_0 +\tau_3\left(\frac{p^2}{2m} -\mu\right) + c( \tau_1 p_x+ \tau_2 p_y)
\,,
\label{3He-A_bulk}
\end{equation}
where $\tau_i$ are Pauli matrices of Bogolyubov-Nambu spin. 
For $\mu>0$ there are two Weyl points at $p_x=p_y=0$ and $p_z = \pm\sqrt{2m\mu}$ with
$N_3=\pm 1$. For $\mu<0$ the vacuum is fully gapped. Thus at $\mu=0$ there is a topological quantum phase transition from the gapless to gapped vacuum \cite{Volovik1992b}. At the transition, i.e. at  $\mu=0$, there is a marginal (topologically trivial) Fermi point with $N_3=0$, situated at ${\bf p}=0$, just as in  Fig.~\ref{TwoScenarios} {\it top}. At $\mu >0$,  this marginal Fermi point splits in two topologically protected Weyl points with $N_3=\pm 1$, Fig.~\ref{TwoScenarios} {\it bottom right}.

\subsubsection{Cube of Fermi points} 

\begin{figure}
\centerline{\includegraphics[width=0.7\linewidth]{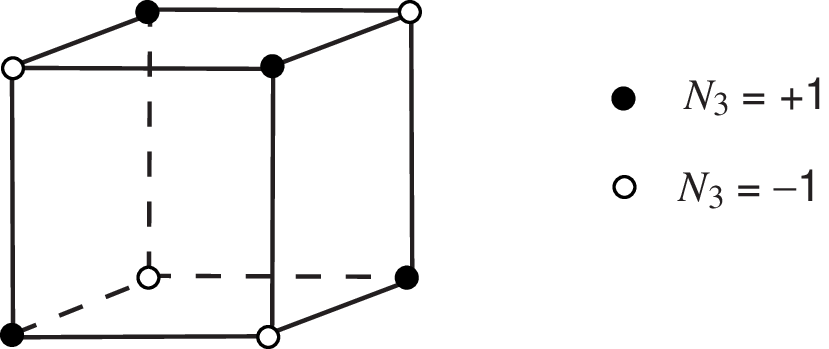}}
\caption{
Cube of Fermi points (Weyl points) in momentum space in the $\alpha$-phase of spin-triplet $p$-wave superfluid \cite{VollhardtWoelfle} and spin-singlet $d$-wave superconductor \cite{VolovikGorkov1985}. Filled circles denote Weyl points  with positive topological charge $N_3=+1$ and thus with right-handed fermions living in the vicinity of these Weyl points. Open circles denote the nodes  with negative topological charge $N_3=-1$ and the left-handed fermions. 
This is analogous of the four-dimensional graphene-like vacuum in a model of relativistic quantum field theory,
which  is characterized by Dirac points on vertices of the 4D hypercube \cite{Creutz2008}.
}  
\label{Cube} 
\end{figure}

The vacuum of the Standard Model contains 16 Weyl points in each generation. Example of the condensed matter system with Weyl Dirac points is provided by the
$\alpha$-phase of spin-triplet $p$-wave superfluid \cite{VollhardtWoelfle} and spin-singlet $d$-wave superconductor \cite{VolovikGorkov1985}.  The latter has the following simplified Green's function:
\begin{equation}
 G^{-1}(\omega,{\bf p})=i p_0 +\tau_3\left(\frac{p^2}{2m} -\mu\right) + \tau_1 (2p_z^2-p_y^2-p_x^2)
+  \tau_2 \sqrt{3}(p_y^2-p_x^2)\,.
\label{ExampleCube}
\end{equation}
The spectrum has 8 point nodes -- Weyl points with $N_3=+1$ and $N_3=-1$ situated on vertices of cube in momentum space in Fig. \ref{Cube} \cite{VolovikGorkov1985}. At $\mu=0$ all eight Weyl points collapse to 
the marginal Fermi point with $N_3=0$ situated at ${\bf p}=0$.

The $\alpha$-phase is the analog of the 3+1 ``graphene'' in relativistic quantum fields \cite{Creutz2008}. 

 \subsubsection{Time reversal invariant planar phase} 

The planar phase of spin-triplet superfluid/superconductor is characterized by the  following simplified $4\times 4$ Green's function matrix
 \begin{equation}
  G^{-1}(\omega,{\bf p})=
 i p_0 +  \tau_3\left(\frac{p^2}{2m} -\mu\right) + \tau_1 (\sigma_x p_x  + \sigma_yp_y) \,.
 \label{planar_state_bulk}
\end{equation}
As distinct from the chiral $^3$He-A and $\alpha$-phase, the planar phase obeys time-reversal invariance. Two  nodes in spectrum, at $p_x=p_y=0$ and $p_z =\pm\sqrt{2m\mu}$, have zero topological charges \eqref{MasslessTopInvariant3D}, $N_3=0$. But these nodes are protected by the topological charge $N_3^K=\pm 2$ in 
\eqref{MasslessTopInvariantStandard Model}, where the correpsonding symmetry of the planar phase is $K=\tau_3\sigma_z$. This matrix $K$ commutes with the Green's function matrix.

\subsubsection{Gapless  2+1 vacua}

In addition to the Fermi surface class of singularities of co-dimension 1, in 2+1 systems there is a class of vacua with singularities of co-dimension 2:  points in 2D momentum space. They corresponds to lines in 3+1 vacua, which also have co-dimension 2.
According to \cite{Horava2005}: if no symmetry is imposed there is no singularity in Green's function, which is topologically stable. For real fermions, the $Z_2$ singularities of co-dimension 2 are possible \cite{Horava2005}. This means that two such singularities may collapse forming the fully gapped state, $1+1=0$. In 2+1 dimension, the fermions near such singularity behave as Majorana fermions. Some symmetries allow to have the Fermi points  of co-dimension 2  with group $Z$. The corresponding  invariant protected by symmetry $K$
\cite{WenZee2002,Volovik2007,Beri2009} is:
\begin{equation}
N^K_2= \frac{1}{4 \pi i}~
{\bf tr}\oint_C   dl 
~ K G(\omega=0,{\bf p})\partial_l G^{-1}(\omega=0,{\bf p}) \,,
\label{MasslessTopInvariant1+1}
\end{equation}
where $C$ is contour around the Fermi point in 2D momentum space $(p_x,p_y)$, or around the Fermi surface if the Fermi point expands to the Fermi surface.
Examples are graphene and $d$-wave cuprate superconductor. For the latter the simplified Green's function has the form
\begin{equation}
 G^{-1}(\omega,p_x.p_y)=ip_0+ \tau_3\left(\frac{p_x^2+p_y^2}{2m} -\mu\right) + \tau_1 (p_y^2-p_x^2)\,,
\label{Example2+1}
\end{equation}
with $K=\tau_2$, which anti-commutes with Green's function at zero frequency.
The $d$-wave superconductor has 4 point nodes at $|p_x|=|p_y|=(m\mu)^{1/2}$ with $N^K_2=\pm 1$.
The nodes do not disappear under deformation which preserves symmetry $K$. For example, the deformation which violates the 4-fold symmetry of \eqref{Example2+1}
\begin{equation}
 G^{-1}(\omega,p_x.p_y)=ip_0+ \tau_3\left(\frac{p_x^2+p_y^2}{2m} -\mu\right) + \tau_1 (p_y^2-ap_x^2)\,,
\label{Example2+1_disturbed}
\end{equation}
does not destroy nodes, but shifts positions of nodes. The nodes disappear only at large deformation, when the deformation parameter $a$ in \eqref{Example2+1_disturbed} crosses zero, and the topological quantum phase transition occurs. At $a=0$ nodes collapse forming two marginal nodes with $N^K_2=0$ at $p_y=0$, $p_x =\pm(2m\mu)^{1/2}$, and at $a<0$ the fully gapped state is formed \cite{Volovik2007}. 

 Note that the inverse propagator at $p_0=0$ has all the properties of a free-fermion Hamiltonian, whose topology was discussed in Ref. \cite{Kitaev2009}. But this is actually the effective Hamiltonian, which emerges in the original interacting system  (see \cite{Haldane2004,Volovik2009b}).  

Another class of $2+1$ Fermi points arise at the boundary
between the $3+1$ gapped systems with different topological charges. Such points are described by the difference
of bulk invariants  \cite{Volovik2009b}. This is analogous to the index theorem
for fermion zero modes on strings \cite{JackiwRossi1981} and vortices \cite{Volovik2003}.

 \section{Vacuum in state of topological insulator}
 
 The special role  in classification of topological  systems is played by dimensional reduction. The dimensional reduction allows us to use for classification of the gapped systems  the scheme, which was suggested by 
 Ho\v{r}ava for the classification of the topologically nontrivial nodes in spectrum \cite{Horava2005}. The fully gapped vacua in $D+1$ space-time are described by the same invariants as nodes of co-dimension $D+1$ \cite{Volovik2003}. For example,  the  winding number
$N_1$ in \eqref{InvariantForFS} which describes zeroes of co-dimension 1 (conventional Fermi surface in $D=3$ momentum space), also describes the $D=0$ gapped systems. The integral  \eqref{InvariantForFS} is now over imaginary frequency:
\begin{equation}
\tilde N_1={\bf tr}~\int {d p_0\over 2\pi i}  G(p_0)\partial_{p_0}G^{-1}(p_0)~.
\label{InvariantForFSreduced}
\end{equation}
This integer-valued index now shows the difference between
the numbers of the positive and negative energy levels of zero-dimensional system. 
 
The classification must be supplemented by the symmetry consideration, which leads to the additional invariants of the type
\begin{equation}
\tilde N_1^K={\bf tr}~\int {d p_0\over 2\pi i}  K G(p_0)\partial_{p_0}G^{-1}(p_0)~,
\label{InvariantForFSreducedK}
\end{equation}
where $K$ is the symmetry operator, which commutes or anti-commutes with the Green's function.
The emergent symmetries might also appear within some topological classes.

 \subsection{2+1 fully gapped vacua}

\subsubsection{$^3$He-A film: 2+1 chiral superfluid}

Let us start with $D=2$. The fully gapped ground states (vacua) in 2+1 or quasi 2+1 thin
films of $^3$He-A are characterized by the invariant obtained by dimensional reduction
from the topological invariant describing the nodes of co-dimension 3. This is the invariant $N_3$
 for the
Fermi point in \eqref{MasslessTopInvariant3D}, which is now over the (2+1)-dimensional momentum-frequency space $(p_x.p_y,p_0)$:
\begin{equation}
 \tilde N_3 = {e_{ijk}\over{24\pi^2}} ~
{\bf tr}\left[  \int    d^2p dp_0
~G\partial_{p_i} G^{-1}
G\partial_{p_j} G^{-1}G\partial_{p_k}  G^{-1}\right].
\label{N2+1}
\end{equation}

\begin{figure}[t]
\centerline{\includegraphics[width=0.7\linewidth]{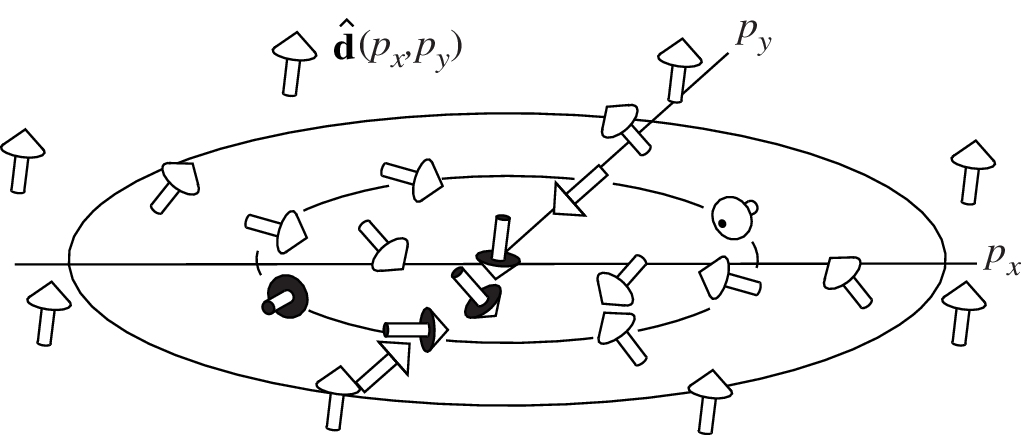}}
\caption{Skyrmion in ${\bf p}$-space with momentum space topological
charge
$\tilde N_3=-1$ in \eqref{N2+1d}. It describes topologically non-trivial fully gapped vacua in 2+1
systems, which  have non-singular Green's function. Vacua with nonzero $\tilde N_3$ 
have topologically protected gapless edge states. The nonzero topological charge leads also to quantization of Hall and spin Hall conductance.}
\label{PSpaceSkyrmionsFig}
\end{figure}

This equation \eqref{N2+1}
was first introduced in relativistic $2+1$ theories \cite{So1985,IshikawaMatsuyama1986,IshikawaMatsuyama1987}
and then independently for the film of $^3$He-A in condensed matter \cite{Volovik1988,VolovikYakovenko1989}, where it was inspired by
the dimensional reduction from the Fermi point, see \cite{Volovik1992b}.
The topological invariant for the general case of the insulating relativistic vacua 
in even space dimension $D=2n$ has been considered in \cite{Kaplan1992,Golterman1993,Kaplan2011}. 
In simple case of the $2\times 2$ matrix, the Green's function can be expressed in terms
of the three-dimensional vector  ${\bf d}(p_x,p_y)$, 
\begin{equation}
 G^{-1}(\omega,p_x.p_y)=ip_0 +{\mbox{\boldmath$\tau$}} \cdot {\bf d}(p_x,p_y)~,
\label{d-vector}
\end{equation}
 Example of the   ${\bf d}$-vector configuration, which corresponds to the  topologically nontrivial vacuum is presented in  Fig. \ref{PSpaceSkyrmionsFig}. This is the momemtum-space analog of the topological object in real space -- skyrmion (skyrmions in real space are described by the relative homotopy groups \cite{MineevVolovik1978}).  The winding number of the momentum-space skyrmion is
 \cite{Volovik1988}
\begin{equation}
\tilde N_3 = \frac{1}{4\pi} ~
   \int    d^2p  
~\hat{\bf d}\cdot \left(\frac{\partial \hat{\bf d}}{\partial p_x}
\times \frac{\partial \hat{\bf d}}{\partial p_y} \right) \,,
\label{N2+1d}
\end{equation}
where $\hat{\bf d}={\bf d}/|{\bf d}|$ is unit vector.
For a single layer of the $^3$He-A film and for one spin projection, the simplified Green's function has the form:
\begin{equation}
  G^{-1}(\omega,{\bf p})=i\ p_0 +{\mbox{\boldmath$\tau$}} \cdot {\bf d}({\bf p}) =
 i p_0 +  \tau_3\left(\frac{p_x^2+p_y^2}{2m} -\mu\right) + \tau_1 p_x  + \tau_2 p_y~,
\label{3He-A_film}
\end{equation}
For $\mu>0$ the topological charge $\tilde N_3=1$ and for $\mu<0$ the topological charge is $\tilde N_3=0$. That is why at $\mu=0$ there is a topological quantum phase transition between the topological superfluid at $\mu>0$ and non-topological superfluid at $\mu<0$ \cite{Volovik1992b}.

In general case of multilayered $^3$He-A,  topological charge $\tilde N_3$ may take any integer value of group $Z$. This charge determines  quantization of Hall and spin-Hall conductance (see Sec. \ref{Quantum_spin_Hall_effect}),
and  the quantum statistics of the topological objects --  real-space skyrmions \cite{Volovik1988,VolovikYakovenko1989,SQHE,Volovik1992b}.
For $\tilde N_3=4k+1$ and   $\tilde N_3=4k+3$, skyrmion is anyon;  
for $\tilde N_3=4k+2$ it is fermion; and for $\tilde N_3=4k$ it is boson \cite{Volovik1992b}. This demonstrates the importance of the $Z_2$ and $Z_4$ subgroups of the group $Z$ in classification of topological matter;
and also  provides an example of the interplay of momentum-space and real-space topologies.  Applications of topology in combined $({\bf p},{\bf r})$-space
see in \cite{GrinevichVolovik1988,SalomaaVolovik1988,Volovik1989c,Volovik2003,TeoKane2010a,TeoKane2010b,SilaevVolovik2010,EssinGurarie2011}, in particular it is responsible for the topologically protected  spectrum of fermions living on topological objects such as walls, strings and monopoles.

\subsubsection{Planar phase: time reversal invariant gapped vacuum}

In case when some symmetry is present, additional invariants  appear, which correspond to dimensional reduction of invariant $N^K_3$ in \eqref{MasslessTopInvariantStandard Model}:
\begin{equation}
 \tilde N^K_3 = {e_{ijk}\over{24\pi^2}} ~
{\bf tr}\left[  \int    d^2p dp_0
~K G\partial_{p_i} G^{-1}
G\partial_{p_j} G^{-1}G\partial_{p_k}  G^{-1}\right],
\label{N2+1prime}
\end{equation}
where as before, the matrix $K$ commutes or anticommutes with the Green's function matrix.
Example of the symmetric $2+1$ gapped state with  $\tilde N^K_3$ is the film of the planar phase of superfluid $^3$He \cite{VolovikYakovenko1989,Volovik1992b}. In the single layer case, the simplest expression for the Green's function is
\begin{equation}
  G^{-1}(\omega,p_x,p_y)=
 i p_0 +  \tau_3\left(\frac{p_x^2+p_y^2}{2m} -\mu\right) + \tau_1 (\sigma_x p_x  + \sigma_yp_y)~,
 \label{planar_state}
\end{equation}
with $K=\tau_3\sigma_z$ commuting with the Green's function. This state is time reversal invariant. 
It has $\tilde N_3=0$ and $\tilde N^K_3=2$. For the general case of the quasi 2D  film with multiple layers of the planar phase, the invariant $ \tilde N^K_3$ belongs to the group $Z$, i.e. $\tilde N^K_3=2k$.

\subsubsection{Quantum spin Hall effect}
\label{Quantum_spin_Hall_effect}

The topological invariants $\tilde N_3$ and $\tilde N^K_3$ give rise to quantization of 
Hall and spin-Hall conductance in 2+1 gapped systems.
There are several types of responses of spin current and electric current to transverse forces
which are quantized in 2+1 systems under appropriate conditions. The most familiar is the conventional quantum Hall effect (QHE) \cite{Klitzing1980}.  It is quantized response of the particle  current  to the transverse force, say to transverse gradient of chemical potential, ${\bf J}= \sigma_{xy} \hat{\bf z}\times \boldsymbol{\nabla} \mu$. In the electrically charged  systems this is the    
quantized response of the electric current ${\bf J}^e$ to transverse electric field 
${\bf J}^e= e^2\sigma_{xy} \hat{\bf z}\times {\bf E}$.

The other effects involve the spin degrees of freedom. 
An example is the mixed spin quantum Hall effect: quantized response of 
the particle current ${\bf J}$ (or electric current ${\bf J}^e$) to transverse gradient  of magnetic field ${\bf H}$ interacting with Pauli spins (Pauli field in short) \cite{VolovikYakovenko1989,Volovik1992b}:
\begin{equation}
{\bf J} = \sigma^{\rm mixed}_{xy} \hat{\bf z}\times \boldsymbol{\nabla} (\gamma H^z) ~~, ~~{\bf J}^e =e{\bf J} \,.
\label{Current_vs_field_gradient}
\end{equation}
Here $\gamma$ is gyromagnetic ratio. 
The related effect, which is determined by the same quantized parameter  $\sigma^{\rm mixed}_{xy}$, is the quantized response of the spin current, say the current ${\bf J}^z$  of the $z$ component of spin, to  the gradient of chemical potential   \cite{SQHE}.
 In the electrically charged  systems this corresponds  to the   quantized response of the spin current
to transverse electric field:
\begin{equation}
{\bf J}^z = \sigma^{\rm mixed}_{xy} \hat{\bf z}\times \boldsymbol{\nabla} \mu  =e \sigma^{\rm mixed}_{xy} \hat{\bf z}\times {\bf E}  \,.
\label{Spin_current_vs_field}
\end{equation}
  This kind of mixed Hall effect is now used in spintronics \cite{Awschalom2009}.   
  
  Finally there is a pure spin Hall effect --  the quantized response of the spin
current to transverse gradient of magnetic field \cite{VolovikYakovenko1989,Volovik1992b,HaldaneArovas1995,Senthil1999}:
\begin{equation}
{\bf J}^z = \sigma^{\rm spin/spin}_{xy} \hat{\bf z}\times \boldsymbol{\nabla} (\gamma H^z)  \,.
\label{Spin_current_vs_field_gradient}
\end{equation}  
All these parameters $\sigma_{xy}$ are quantized being expressed via topological charges  $\tilde N_3$ and $\tilde N^K_3$.

 \subsection{3+1 fully gapped states: $^3$He-B and quantum vacuum}

 In the asymmetric phase of Standard Model, there is no mass protection by topology and all the fermions become massive, i.e. Standard Model vacuum becomes the fully gapped insulator.  In quantum liquids, the fully gapped three-dimensional system with time reversal symmetry and nontrivial topology is represented by another phase of superfluid $^3$He -- the $^3$He-B. Its topology is also supported by symmetry and gives rise  to the 2D gapless quasiparticles living at interfaces between vacua with different values of the topological invariant or on the surface of $^3$He-B \cite{SalomaaVolovik1988,Volovik2009,Volovik2009b,Volovik2010}. It is important that $^3$He-B belongs to the same topological class as the vacuum of Standard Model in its present insulating phase \cite{Volovik2010a}.  The topological classes of the $^3$He-B states can be represented by the following simplified Green's function:
 \begin{equation}
G^{-1}(\omega,{\bf p})=ip_0 +  \tau_3\left(\frac{p^2}{2m} - \mu\right)+  \tau_1\left( \sigma_x c_xp_x + \sigma_y c_yp_y + \sigma_z c_zp_z   \right)
\,.
\label{eq:B-phase}
\end{equation}
In the isotropic $^3$He-B all `speeds of light' are equal, $|c_x|= |c_y|=|c_z|=c$. The vacuum of free Dirac particles is obtained in the limit $1/m=0$.

In the fully gapped  systems, the Green's function has no singularities in the  whole 4-dimensional space $(p_0,{\bf p})$. That is why we are able to use the Green's function at $p_0=0$. The topological invariant relevant  for $^3$He-B and for quantum vacuum with massive Dirac fermions is:
\begin{equation}
N^K = {e_{ijk}\over{24\pi^2}} ~
{\bf tr}\left[  \int_{\omega=0}   d^3p ~K
~G\partial_{p_i} G^{-1}
G\partial_{p_j} G^{-1} G\partial_{p_k} G^{-1}\right]\,.
\label{3DTopInvariant_tau}
\end{equation} 
 with matrix $K=\tau_2$ which anti-commutes with the Green's function at $p_0=0$. In $^3$He-B, the $\tau_2$ symmetry   is combination of time reversal  and particle-hole symmetries; for Standard Model  the matrix  $\tau_2=\gamma_5\gamma^0$.  
Note that at $p_0=0$ the symmetry of the Green's function is enhanced, and thus there are more matrices $K$, which commute or anti-commute with the Green's function, than at $p_0\neq 0$.

 \begin{figure}[top]
\centerline{\includegraphics[width=0.5\linewidth]{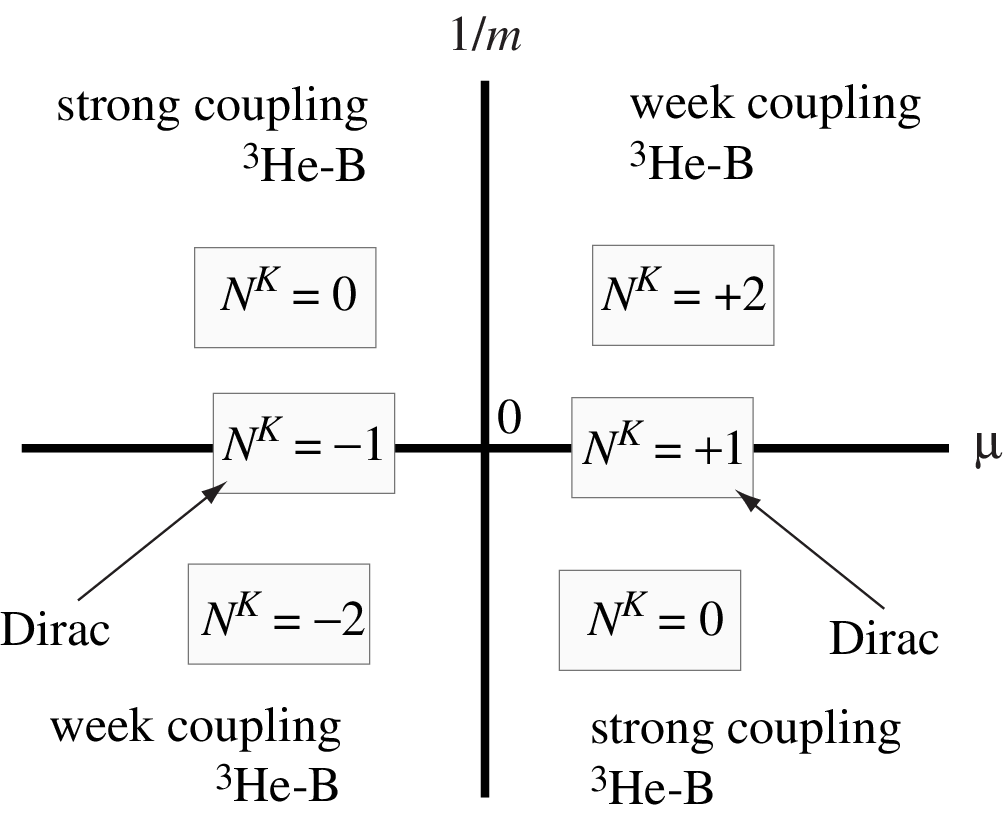}}
  \caption{\label{QPT}  Phase diagram of topological states of $^3$He-B in equation \eqref{eq:B-phase} in the plane $(\mu,1/m)$ for the speeds of light $c_x>0$, $c_y>0$ and $c_z>0$. States on the line 
  $1/m=0$ correspond to the  Dirac vacua, which Hamiltonian is non-compact. Topological charge of the Dirac fermions  is intermediate between charges of compact $^3$He-B states.
The line $1/m=0$ separates the states with different asymptotic behavior of the Green's function at infinity:
$G^{-1}(\omega=0,{\bf p}) \rightarrow \pm \tau_3 p^2/2m$. 
 The line $\mu=0$ marks topological quantum phase transition, which occurs between the weak coupling $^3$He-B  (with $\mu>0$, $m>0$ and topological charge $N^K=2$) and the strong coupling $^3$He-B   (with $\mu<0$, $m>0$ and $N^K=0$).    This transition is topologically equivalent to quantum phase transition between Dirac vacua with opposite mass parameter 
 $M=\pm |\mu|$, which occurs when $\mu$ crosses zero along the line $1/m=0$. 
 The interface which separates two states contains single Majorana fermion in case of $^3$He-B, and single chiral fermion in case of  relativistic quantum fields.  Difference in the nature of the fermions is that in Bogoliubov-de Gennes system the components of spinor are related by complex conjugation. This reduces the number of degrees of freedom compared to Dirac case.
 }
\end{figure}

Fig. \ref{QPT} shows the phase diagram of topological states of $^3$He-B in the plane $(\mu,1/m)$.
On the line  $1/m=0$ one obtains the free Dirac fermions  ( $n_L=n_R=1$)  with the mass parameter $M=-\mu$. The conventional Dirac vacuum of free fermions has topological charge
\begin{equation}
N^K= {\rm sign}(M)
\,.
\label{eq:DiracInvariants}
\end{equation}

The real superfluid $^3$He-B lives in the weak-coupling corner of the phase diagram:  $\mu>0$, $m>0$, $\mu\gg mc^2$. However, in the ultracold Fermi gases with triplet pairing
 the strong coupling limit is possible near the Feshbach resonance \cite{GurarieRadzihovsky2007}.
 When $\mu$ crosses zero the topological quantum phase transition occurs, at which the topological charge $N^K$ changes from  $N^K=2$ to  $N^K=0$. 
 
There is an important difference between $^3$He-B and Dirac vacuum. The space of the Green's function of free  Dirac fermions is non-compact: $G$ has different asymptotes at $|{\bf p}|\rightarrow \infty $ for different directions of momentum ${\bf p}$.   As a result, the topological charge of the interacting Dirac fermions depends on the regularization at large momentum. $^3$He-B can serve as regularization of the Dirac vacuum, which can be made in the Lorentz invariant way \cite{Volovik2010a}. One can see from Fig. \ref{QPT}, that  the topological charge of free Dirac vacuum has intermediate value between the charges of the  $^3$He-B vacua with compact  Green's function. On the marginal behavior of free Dirac fermions see Refs. \cite{Haldane1988,Schnyder2008,Volovik2003,Volovik2009b}.

   The vertical axis in Fig. \ref{QPT} separates the states with the same asymptote of the Green's function at infinity. The abrupt change of the topological charge across the line, $\Delta N^K=2$, with fixed asymptote shows that one cannot cross the transition  line adiabatically. This means that all the intermediate states on the line of this  QPT  are necessarily gapless. For the intermediate state between the free Dirac vacua with opposite mass parameter 
 $M$ this is well known. But this is applicable to the general case with or without relativistic invariance:
 the gaplessness is protected 
by the difference of topological invariants on two sides of transition.
 
\begin{figure}[top]
\centerline{\includegraphics[width=0.4\linewidth]{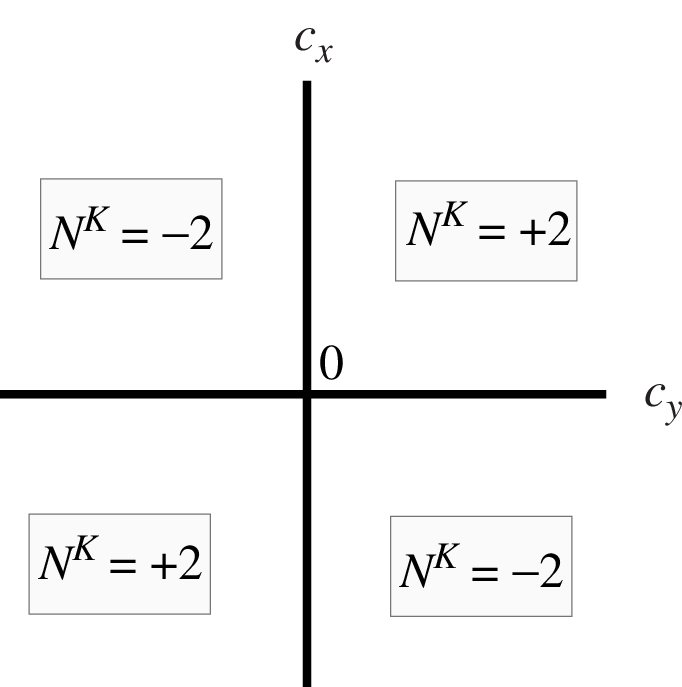}}
  \caption{\label{PD}  Phase diagram of $^3$He-B states at fixed $c_z>0$, $\mu>0$ and $m>0$.
 At the phase boundaries the vacuum is gapless and corresponds to the 3+1 planar phase. The interface between the gapped states with different winding number $N^K$ contains Majorana fermions.   
 In cosmology such interface would correspond to the vierbein wall, where the metric is degenerate
 \cite{Volovik2003}.
}
\end{figure}

Fig. \ref{PD} shows phase diagram of topological states of $^3$He-B in the plane $(c_x,c_y)$
 at fixed $c_z>0$, $\mu>0$ and $m>0$.
When one of the components of speed of light is nullified the state becomes gapless. If, say,  $c_x$ approaches zero, two pairs of point nodes appear in this intermediate state at points ${\bf p}=\pm(0,0,p_F)$, where $p_F =(2\mu m)^{1/2}$. The number of point nodes is related to the difference in topological invariant $N^K$ of the states on two sides of the transition:
\begin{equation}
N_{\rm point~nodes} = |N^K(c_x >0)-N^K(c_x<0)|=4 
\,.
\label{connection2}
\end{equation}
  The intermediate state is the 3+1 planar phase in \eqref{planar_state_bulk}, and  the corresponding gapless fermions in this state are characterized by the topological invariant $N_3^K$ protected by discrete symmetry.

The gaplessness of the intermediate state on phase diagram leads  to the other related phenomenon. The two-dimensional interface (brane), which separates two domains with different $N^K$, contains fermion zero modes,  2+1 massless fermions. The number of zero modes is determined by the difference $\Delta   N^K$ between the topological charges of the vacua on two sides of the interface   \cite{Volovik2009,Volovik2009b}. This is similar to the index theorem for the number of fermion zero modes on cosmic strings, which is determined by the  topological winding number of the string \cite{JackiwRossi1981}.

Superfluid $^3$He-B demonstrates that the interface between the bulk states with the same  topological charge may also contain fermion zero modes \cite{SalomaaVolovik1988,Volovik2009}.
For example, the state with positive $c_x$, $c_y$  and   the state with negative $c_x$, $c_y$ in Fig. \ref{PD}  have the same topological charge $N^K=+2$. But the interface between these bulk states
is gapless. Moreover, the density of states of these  irregular fermion zero modes is larger than the density of states of regular fermion zero modes living on the interface between the bulk states with different $N^K$. This is not very surprising, since the irregular gapless states of co-dimension 1 may appear in many systems.  Nodes of co-dimension 1 are the most stable topological objects, and they  may emerge even in bulk superconductors \cite{Volovik1989d,Gubankova2005}. They also
may exist on the surface of  non-topological insulators and superconductors, and at the interface with $\Delta N^K=0$ as it happens in  $^3$He-B.

The $^3$He-B topology can be also extended to the lattice models explored in QCD, where the four dimensional Brillouin zone is used.  The lattice model with fully gapped  Wilson fermions is described both by $\tilde N_3$ and $\tilde N_5$
topological invariants, which give rise to two index theorems for gapless fermions in the intermediate states  \cite{ZubkovVolovik2012,Zubkov2012a}. 
The topological invariant $\tilde N_5$ has been also used for description of fully gapped vacua in 4+1 systems
\cite{Kaplan1992,Golterman1993,Volovik2003,Kaplan2011}. They give rise to the 3+1 gapless fermion zero modes living at the 3+1 interfaces between the massive vacua. This provides another topological scenario for emergent  chiral relativistic fermion, accompanied by relativistic quantum field theory and gravity.

\section{Discussion}

 There is a fundamental interplay of symmetry and topology in physics, both in condensed matter and relativistic quantum fields. Traditionally the first role was played by symmetry (symmetry classification of crystals, liquid crystals, magnets, superconductors, superfluids, etc.).  The phenomenon of spontaneously broken symmetry remains one of the major tools in physics. In particle physics the chain of successive phase transitions is suggested in which the large symmetry group existing at high energy is reduced at low energy. This symmetry  principle is in the basis of the Grand Unification Theory (GUT).  It also gives rise to the classification of the topological defects which arise due to spontaneous symmetry breaking. In this approach, symmetry is primary, while topology is secondary being fully determined by the broken symmetry.

The last decades demonstrated the opposite tendency in which topology is primary (the reviews can be found in Refs.  \cite{Volovik1992b,Volovik2003,Volovik2007}). The unconventional properties of superfluid $^3$He-A, which were found in early seventies, demonstrated that the topology in momentum space is the main characteristics of  the ground states of the system at zero temperature ($T=0$), in other words it is the characteristics of quantum vacua. It demonstrated that the gaplessness of   fermions is protected by topology, and thus is not sensitive to the details of the microscopic (trans-Planckian) physics. Irrespective of the deformation of the parameters of the microscopic theory, the value of the gap (mass) in the energy spectrum of these fermions remains strictly zero. This solves the main hierarchy problem: for these classes of  fermionic vacua the masses of elementary particles are naturally small.

 The vacua, which have nontrivial topology in momentum space, are now called the topological matter (topological superfluids, superconductors, insulators, semi-metals, etc.).
The momentum-space topological invariants determine universality classes of the topological matter and the type of the effective theory which  emerges at low energy and low temperature. In many cases they also give rise to emergent symmetry. Examples are provided by the nodes of the energy spectrum in momentum space:  if they are protected by topology they give rise  to emergent symmetries such as Lorentz invariance, and to emergent phenomena such as gauge and gravitational fields. Contrary to the GUT scheme,  in the anti-GUT scheme the symmetry is secondary, being emergent in the low-energy corner due to topology.  If this is true, then it is topology of the quantum vacuum, which gives rise to 
the fermionic matter in our Universe.

However, this is not the whole story. It appears that in many systems (including condensed matter and relativistic quantum vacua), the topological invariants are trivial, i.e. they have zero values. Nevertheless 
these systems remain to be topological: the underlying discrete symmetry of the vacuum may transform the trivial topological invariant into the nontrivial one.  This is the case when both topology and symmetry are equally important. 
 This implies that if the the Standard Model and gravity are effective theories, the underlying physics must contain discrete symmetries. Their role is extremely important. The main role is to prohibit the cancellation of the Fermi points with opposite topological charges in the Standard Model vacuum. As a side effect, in the low-energy corner discrete symmetries are transformed into gauge symmetries and give rise to effective non-Abelian gauge fields. In particular, the underlying $Z_2$ symmetry produces the effective $SU(2)$ gauge field \cite{Volovik2003}. Discrete symmetries also reduce the number of the massless gauge bosons and the number of effective metric fields. To justify the Fermi point scenario of emergent physical laws, one should find such discrete symmetry of the microscopic vacuum which leads in the low energy corner to one of the GUT or Pati-Salam models.

The discovery of the quantum Hall effect (QHE) in 1980  \cite{Klitzing1980} triggered the further consideration of condensed matter systems using topological methods
\cite{NielsenNinomiya1983,AvronSeilerSimon1983,Semenoff1984,NiuThoulessWu1985,Haldane1988}. 
The topological invariant for QHE effect in terms of the Green's function was introduced in relativistic $2+1$ quantum electrodynamics
\cite{So1985,IshikawaMatsuyama1986,IshikawaMatsuyama1987}. 
This invariant is responsible not only for ordinary QHE but also for the intrinsic QHE, when the Hall conductance is quantized even in the absence  of external magnetic field \cite{Matsuyama1987}. 

Such intrinsic QHE can be realized for example in thin films of   superfluid $^3$He. These quasi-two-dimensional systems serve as an example of the fully gapped 2+1 systems, whose integer valued topological invariant is in the origin of quantization of physical parameters \cite{Volovik1988,VolovikYakovenko1989,Yakovenko1989}.  There are two phases  which may exist in a thin film of liquid $^3$He,  both phases are fully gapped in film. These are chiral superfluid $^3$He-A and the planar phase with time reversal symmetry.  This symmetry supports the topological invariant which is responsible for intrinsic quantum Hall and spin-Hall effects   \cite{VolovikYakovenko1989,Volovik1992b}, thus providing us with the important example of the interplay of topology and symmetry in topological medium.  

Recent observation of  topological insulators \cite{Hsieh2009a} gave new impulse for study of the topological phases in 3+1 systems, such as superfluid $^3$He-B  which is the condensed matter counterpart of Standard Model vacuum in its massive phase. 
Superfluid $^3$He-B  has Majorana bound states on the surface of the liquid. Andreev bound states on the surface or interface of $^3$He-B were  discussed theoretically or probed experimentally  
\cite{SalomaaVolovik1988,Kopnin1991,CCGMP,Davis2008,Nagai2008,Murakawa2009c,ChungZhang2009,Nagato2009}.  
However, the Majorana signature of these states has not yet been reported experimentally.
One of the possible tools  is NMR, which requires an external magnetic field. The effect of magnetic field on  Majorana fermions has been recently discussed in \cite{Nagato2009,Volovik2010}. Majorana fermions become massive, with mass being proportional to magnetic field, which violates the time reversal symmetry.

Vacuum of Standard Model is a topological 3+1 medium. Both known states of the quantum vacuum of Standard Model have non-trivial topology.  The insulating state is described by nonzero value of topological invariant $\tilde N_3^K$ and the topology of this state is similar to that of superfluid $^3$He-B. The semi-metal state is described by nonzero value of  topological invariant $N_3^K$ and the topology of this state is similar to that of superfluid planar phase in 3+1 dimensions.  Both invariants are supported by symmetry. Superfluid phases of $^3$He ($^3$He-B, $^3$He-A and planar phase) thus provide a close connection with relativistic quantum fields.

These phases also give generic examples of different classes of topological media  in dimensions $2+1$ and $3+1$ which experience topological stability of gap nodes; topological edge states on the surface of fully gapped insulators; Majorana fermions;  topological quantum phase transitions; intrinsic quantum Hall and spin-Hall effects; quantization of physical parameters; chiral anomaly; topological Chern-Simons and Wess-Zumino actions; etc.   

These examples of topological medium demonstrate the important role of both topology and symmetry. That is why we need the general classification of topological matter in terms of its symmetry and  topology. The attempt of such classification was made in \cite{Schnyder2009a,Kitaev2009}. However, only non-interacting systems were considered, and also not all the symmetries were exploited, including the approximate symmetries. Examples of latter are again provided by the superfluid phases of  $^3$He, where the symmetry is enhanced due to relative smallness of the spin-orbit interactions. One task should be to consider  symmetry classes, including crystal symmetry  classes, magnetic classes, superconductivity classes, etc., and to find out   what are the topological classes of Green's function which are allowed within a given symmetry class. Then one should find out what happens when the symmetry is smoothly violated, etc.
The Green's function matrices with spin and band indices must be used for this classification, 
since they take into account interaction. In this way  we may finally obtain the full classification of  
topological matter, including insulators, superconductors, magnets, liquids, and vacua of relativistic quantum fields. 

Finally this experience must be used for the investigation of the topologically nontrivial
quantum vacua in relativistic quantum field theories, where the topology in momentum space 
is becoming the important tool (see e.g.
\cite{NielsenNinomiya1981,So1985,IshikawaMatsuyama1986,Kaplan1992,Golterman1993,Volovik2003,Horava2005,Creutz2008,Kaplan2011,ZubkovVolovik2012,Zubkov2012a}).

On the other hand, the gravitational properties of quantum vacuum do not depend much on the
topology of the vacuum. In any  self-sustained  vacuum, gapless or gapped, topological or trivial, the gravitating energy of  the vacuum is strictly zero if the vacuum is perfect and is isolated from environment.  This is the consequence of thermodynamics, which is not sensitive to the structure of the quantum vacuum. Nullification of $\rho_{\rm vac}$ occurs in any equilibrium system at $T=0$, relativistic or non-relativistic. This solves the main cosmological constant problem: if the quantum vacuum belongs to self-sustained media, $\Lambda$ is naturally small. According to this view, cosmology is the process of relaxation of the Universe to its equilibrium vacuum state, and this process does depend on the momentum space topology of the quantum vacuum.

\vspace{5mm}

\textbf{Acknowledgements}. This work is supported in part by the Academy of Finland, Centers of
excellence program 2006-2011 and the Khalatnikov--Starobinsky leading scientific school (Grant No. 4899.2008.2)


\begin{thebibliography}{30}


\bibitem{Einstein1917}
A. Einstein
Kosmologische Betrachtungen zur allgemeinen Relativit\"{a}tstheorie,
Sitzungsber. Preuss. Akad. Wiss.   \textbf{1}, 142--152 (1917).

\bibitem{Bronstein1933}
M.P. Bronstein, 
On the expanding universe,
 Phys. Z. Sowjetunion \textbf{3}, 73--82 (1933).

\bibitem{Zeldovich1967}
Y.B. Zel'dovich,  
Cosmological constant and elementary particles,
 JETP Lett.  \textbf{6}, 316--317 (1967).

\bibitem{Riess-etal1998}
A.G. Riess {\it et al.}  [Supernova Search Team Collaboration],
 Observational evidence from supernovae for an accelerating universe and a 
cosmological constant,
Astron. J. {\bf 116}, 1009 (1998),
arXiv:astro-ph/9805201.

\bibitem{Perlmutter-etal1998}
S. Perlmutter {\it et al.}  [Supernova Cosmology Project Collaboration],
Measurements of $\Omega$ and $\Lambda$ from 42 high-redshift supernovae,
Astrophys. J.  {\bf 517}, 565 (1999), arXiv:astro-ph/9812133.

\bibitem{Komatsu2008}
E. Komatsu {\it et al.},
Five-year Wilkinson Microwave Anisotropy Probe (WMAP) observations: Cosmological 
interpretation,
Astrophys. J. Suppl.  {\bf 180}, 330 (2009), arXiv:0803.0547.

\bibitem{Froggatt1991}
C.D. Froggatt   and  H.B. Nielsen,
{\it Origin of Symmetry}, 
World Scientific, Singapore, 1991.


\bibitem{Volovik2003} 
G.E. Volovik, 
{\it The Universe in a Helium Droplet}, 
Clarendon Press,  Oxford (2003).

\bibitem{HasanKane2010}   
M.Z. Hasan and C.L. Kane, 
Topological Insulators,
Rev. Mod. Phys. {\bf 82}, 3045--3067 (2010).

\bibitem{Xiao-LiangQi2011}   
Xiao-Liang Qi and Shou-Cheng Zhang, 
Topological insulators and superconductors,
Rev. Mod. Phys. {\bf 83}, 1057--1110 (2011).

 \bibitem{Abrikosov1971}
A.A. Abrikosov    and S.D. Beneslavskii,
Possible existence of substances intermediate between metals and dielectrics, 
Sov. Phys. JETP {\bf 32}, 699 (1971). 

 \bibitem{Abrikosov1998}
A.A. Abrikosov,
 Quantum magnetoresistance,  
 Phys. Rev. {\bf B 58}, 2788 (1998).

\bibitem{Burkov2011}
A.A. Burkov and L. Balents, 
Weyl semimetal in a topological insulator multilayer,
Phys. Rev. Lett. {\bf 107}, 127205 (2011);
A.A. Burkov, M.D. Hook, L. Balents,
Topological nodal semimetals,
arXiv:1110.1089.

\bibitem{XiangangWan2011}
Xiangang Wan, A.M. Turner,  A. Vishwanath  and S.Y. Savrasov,
 Topological semimetal and Fermi-arc surface states in the electronic structure of pyrochlore iridates,
Phys. Rev. B {\bf 83}, 205101 (2011).

\bibitem{Ryu2002}
S. Ryu and  Y. Hatsugai, 
Topological origin of zero-energy edge states in particle-hole symmetric systems,
 Phys. Rev. Lett. {\bf 89}, 077002 (2002).


 \bibitem{Manes2007}
J.L. Manes, F. Guinea and M.A.H. Vozmediano,
Existence and topological stability of Fermi points in multilayered graphene,
Phys. Rev. B {\bf 75}, 155424 (2007).

\bibitem{Vozmediano2010} 
M. A. H. Vozmediano, M. I. Katsnelson, F. Guinea,
Gauge fields in graphene,
Physics Reports {\bf 496}, 109 (2010).

 \bibitem{Cortijo2011} 
 A. Cortijo, F. Guinea, M.A.H. Vozmediano
Geometrical and topological aspects of graphene and related materials,
 arXiv:1112.2054.

\bibitem{Nobbenhuis2006}
S. Nobbenhuis,   
Categorizing different approaches to the cosmological constant problem,
 Found.  Phys.  \textbf{36}, 613--680 (2006).


\bibitem{Dreyer2007}
O. Dreyer,  
Emergent general relativity.
in: \emph{Approaches to Quantum Gravity -- toward a new understanding of
space, time and  matter},
edited by D. Oriti, Cambridge University Press (2007);
arXiv:gr-qc/0604075.

  
  \bibitem{KlinkhamerVolovik2008a}
F.R. Klinkhamer and G.E. Volovik,
Self-tuning vacuum variable and cosmological constant,'
Phys. Rev. D {\bf 77}, 085015 (2008),
arXiv:0711.3170.

\bibitem{KlinkhamerVolovik2008b}
F.R. Klinkhamer and G.E. Volovik,
Dynamic vacuum variable and equilibrium approach in cosmology,
 Phys. Rev. D {\bf 78}, 063528 (2008), arXiv:0806.2805.

\bibitem{KlinkhamerVolovik2009b}
F.R. Klinkhamer and G.E. Volovik,
Vacuum energy density kicked by the electroweak crossover,'
Phys. Rev. D {\bf 80}, 083001 (2009), arXiv:0905.1919.

\bibitem{KlinkhamerVolovik2008jetpl}
F.R. Klinkhamer and G.E. Volovik,
$f(R)$ cosmology from $q$-theory,''
JETP Lett.\  {\bf 88}, 289 (2008), arXiv:0807.3896.

\bibitem{KlinkhamerVolovik2009a}
F.R. Klinkhamer and G.E. Volovik,
Gluonic vacuum, $q$-theory, and the cosmological constant,
Phys. Rev.  D {\bf 79}, 063527 (2009), arXiv:0811.4347.

\bibitem{KlinkhamerVolovik2010}
F.R. Klinkhamer and G.E. Volovik, 
Towards a solution of the cosmological constant problem,
 Pis'ma ZhETF {\bf 91},  279--285 (2010);
arXiv:0907.4887.
  
\bibitem{de Bernardis2000}
de Bernardis, P.  {\it et al.}  2000 [Boomerang Collaboration],
A flat universe from high-resolution maps of the cosmic microwave
     background radiation.
\textit{Nature} \textbf{404}, 955--959.

\bibitem{Hinshaw2007}
Hinshaw, G.  {\it et al.}  [WMAP Collaboration] 2007
Three-year Wilkinson Microwave Anisotropy Probe (WMAP) observations:
 temperature analysis.
\textit{Astrophys. J. Suppl.} \textbf{170}, 288--334.

\bibitem{Riess2007}
Riess, A. G.  {\it et al.} 2007
New Hubble Space Telescope discoveries of type Ia supernovae at $z >1$:
Narrowing constraints on the early behavior of dark energy,
\textit{Astrophys. J.} \textbf{659}, 98--121.

\bibitem{LL1980}
L.D. Landau and E.M. Lifshitz,   
\textit{Statistical Physics, Part 1},
Oxford: Pergamon Press (1980).

\bibitem{Perrot1998}
P. Perrot,  
\textit{A to Z of Thermodynamics},
Oxford: University Press (1998).

\bibitem{Shifman1992} M.A. Shifman,
Vacuum structure and QCD sum rules,
Elsevier, Amsterdam, 1992;
M.A. Shifman, A.I. Vainstein, and V.I. Zakharov,
Nucl. Phys. {\bf B~147}, 385 (1979); 448 (1979).

\bibitem{Shifman1991} M.A. Shifman,
Anomalies in gauge theories,
Phys. Rep. {\bf 209}, 341--378 (1991).

\bibitem{Savvidy}  
S.G. Matinyan and G.K. Savvidy, 
Vacuum polarization induced by the intense gauge field,
Nucl. Phys. {\bf B~134},  539 (1978);
G.K. Savvidy,
Infrared instability of the vacuum state of gauge theories and asymptotic freedom,
 Phys. Lett. {\bf B~71},  133 (1977).

\bibitem{HalperinZhitnitsky1998} I. Halperin and A. Zhitnitsky,
Can $\theta//N$ dependence for gluodynamics be compatible
with 2$\pi$ periodicity in $\theta$ ? Phys. Rev. {\bf D~58},  054016 (1998).

\bibitem{DuffNieuwenhuizen1980}
M.J. Duff and P. van Nieuwenhuizen,
Quantum inequivalence of different field representations,
Phys. Lett.  B {\bf 94}, 179 (1980).

\bibitem{Aurilia-etal1980}
A. Aurilia, H. Nicolai, and P.K. Townsend,
Hidden constants: The theta parameter of QCD and the cosmological constant of $N=8$ supergravity,
Nucl.\ Phys.\  B {\bf 176}, 509 (1980).

\bibitem{Hawking1984}
S.W. Hawking,
The cosmological constant is probably zero,
Phys. Lett.  B {\bf 134}, 403 (1984).

\bibitem{HenneauxTeitelboim1984}
M. Henneaux and C. Teitelboim,
The cosmological constant as a canonical variable,
Phys.\ Lett.\  B {\bf 143}, 415 (1984).

\bibitem{Duff1989}
M.J. Duff,
The cosmological constant is possibly zero, but the proof is probably wrong,
Phys.\ Lett.\  B {\bf 226}, 36 (1989).

\bibitem{DuncanJensen1989}
M.J. Duncan and L.G. Jensen,
Four-forms and the vanishing of the cosmological constant,
Nucl.\ Phys.\  B {\bf 336}, 100 (1990).

\bibitem{BoussoPolchinski2000}
R. Bousso and J. Polchinski,
Quantization of four-form fluxes and dynamical neutralization of the   cosmological constant,
JHEP {\bf 0006}, 006 (2000), arXiv:hep-th/0004134.

\bibitem{Aurilia-etal2004}
A. Aurilia and E. Spallucci,
Quantum fluctuations of a `constant' gauge field,
Phys.\ Rev.\  D {\bf 69}, 105004 (2004), arXiv:hep-th/0402096.

\bibitem{Wu2008}
Z.C. Wu,
The cosmological constant is probably zero, and a proof is possibly right,
Phys. Lett.  B {\bf 659}, 891 (2008), arXiv:0709.3314.

\bibitem{Dzyaloshinskii1980}
I.E. Dzyaloshinskii and G.E. Volovik,
Poisson brackets in  condensed matter,
Ann. Phys. {\bf 125}, 67 (1980).

\bibitem{Jacobson2007}
T. Jacobson,
Einstein--aether gravity: Theory and observational constraints,
arXiv:0711.3822 [gr-qc].

\bibitem{WillNordvedt}
(a) C.M. Will and K. Nordvedt,
 Conservation laws and preferred frames in relativistic gravity:
 I. Preferred-frame,
Astrophys. J. {\bf 177}, 757 (1972);
(b) K. Nordvedt, and C.M. Will,
Conservation laws and preferred frames in relativistic gravity:
  II. Experimental evidence,
Astrophys. J. {\bf 177}, 775 (1972);
(c) R.W. Hellings and K. Nordvedt,
Vector-metric theory of gravity,
Phys. Rev. D {\bf 7}, 3593 (1973).

\bibitem{Gasperini1987}
(a) M. Gasperini,
Singularity prevention and broken Lorentz symmetry,
Class. Quantum Grav. {\bf 4}, 485 (1987);
(b) M. Gasperini,
Repulsive gravity in the very early Universe,
Gen. Rel. Grav. {\bf 30}, 1703 (1998),
arXiv:gr-qc/9805060.

\bibitem{Jacobson2001}
T. Jacobson and D. Mattingly,
Gravity with a dynamical preferred frame,
Phys. Rev. D {\bf 64}, 024028 (2001),
arXiv:gr-qc/0007031.

\bibitem{VolovikSpectrum}
G.E. Volovik, 
On spectrum of  vacuum energy, 
Journal of Physics: Conference Series {\bf 174}, 012007 (2009); arXiv:0801.2714


\bibitem{Volovik2004}
G.E. Volovik,  
On the thermodynamic and quantum fluctuations of the cosmological constant,
JETP Lett. {\bf 80}, 531--534 (2004); 
gr-qc/0406005.

\bibitem{Weinberg1988}
S. Weinberg,
The cosmological constant problem,
Rev.\ Mod.\ Phys.\  {\bf 61}, 1 (1989).

\bibitem{Weinberg1996}
S. Weinberg,
Theories of the cosmological constant,
in: N. Turok, \emph{Critical Dialogues in Cosmology}
(World Scientific, Singapore, 1997), p. 195,
arXiv:astro-ph/9610044.

\bibitem{VolkovKogan1974}  
A.F. Volkov and  S.M. Kogan,
Collisionless relaxation of the energy gap in superconductors, 
JETP {\bf 38}, 1018--1021 (1974).

\bibitem{Barankov2004}
R. A. Barankov, L. S. Levitov and B. Z. Spivak, 
Collective Rabi oscillations and solitons in a time-dependent BCS pairing problem,
Phys. Rev. Lett. {\bf 93}, 160401 (2004).

\bibitem{Yuzbashyan2005} 
A. Yuzbashyan, B. L. Altshuler, V. B. Kuznetsov and V. Z. Enolskii,
Nonequilibrium cooper pairing in the nonadiabatic regime,
Phys. Rev. B {\bf 72}, 220503 (2005).

\bibitem{Yuzbashyan2008} 
E. A. Yuzbashyan,
Normal and anomalous solitons in the theory of dynamical Cooper pairing,
Phys. Rev. B {\bf 78}, 184507 (2008).

\bibitem{Gurarie2009} 
V. Gurarie, 
Nonequilibrium dynamics of weakly and strongly paired superconductors,
Phys. Rev. Lett. {\bf 103}, 075301 (2009).

\bibitem{Starobinsky1980}
A.A. Starobinsky,
 A new type of isotropic cosmological models without singularity,
Phys.\ Lett.\  B {\bf 91}, 99 (1980).

\bibitem{Dolgov1985}
A.D. Dolgov,
Field model with a dynamic cancellation of the cosmological constant,
JETP Lett.  {\bf 41}, 345 (1985),


\bibitem{Dolgov1997}
A.D. Dolgov,
Higher spin fields and the problem of cosmological constant,
Phys.\ Rev.\  D {\bf 55}, 5881 (1997);
 arXiv:astro-ph/9608175.


\bibitem{Polyakov1991}
(a)  A.M. Polyakov,
Selftuning fields and resonant correlations in 2--D gravity,
Mod. Phys. Lett. A {\bf 6}, 635 (1991);
(b) I. Klebanov and A.M. Polyakov,
Interaction of discrete states in two-dimensional string theory,
Mod. Phys. Lett. A {\bf 6}, 3273 (1991);
 arXiv:hep-th/9109032.

\bibitem{PolyakovPrivateComm}
A.M. Polyakov, private communication.

\bibitem{LarkinPikin1969}
(a)  A.I. Larkin and S.A. Pikin,
Phase transitions of first order close to the second order,
Sov. Phys. JETP {\bf 29}, 891 (1969);
(b) J. Sak,
Critical behavior and compressible magnets,
Phys.\ Rev.\  B {\bf 10}, 3957 (1974).

\bibitem{RubakovTinyakov1999}
V.A. Rubakov and P.G. Tinyakov,
Ruling out a higher spin field solution to the cosmological constant problem,
Phys.\ Rev.\  D {\bf 61}, 087503 (2000); 
arXiv:hep-ph/9906239.



\bibitem{EmelyanovKlinkhamer2011}
V. Emelyanov, F.R. Klinkhamer,
Vector-field model with compensated cosmological constant and radiation-dominated FRW phase,
Int. J. Mod. Phys. D;
 arXiv:1108.1995.

\bibitem{KlinkhamerVolovik2011a}
F.R. Klinkhamer and G.E.Volovik,
Dynamics of the quantum vacuum: Cosmology as relaxation to the equilibrium state,
Journal of Physics: Conference Series {\bf 314}, 012004 (2011);
arXiv:1102.3152.

\bibitem{Volovik2001} 
G.E. Volovik,
Reentrant violation of special relativity in the low-energy corner, 
JETP Lett. {\bf 73}, 162--165  (2001); 
hep-ph/0101286.

\bibitem{KlinkhamerVolovik2005b}
 F.R. Klinkhamer and G.E. Volovik, 
 Emergent CPT violation from the splitting of Fermi points, 
 Int. J.  Mod. Phys. A  {\bf 20}, 2795--2812 (2005); 
hep-th/0403037.

\bibitem{KlinkhamerVolovik2011b}
 F.R. Klinkhamer and G.E.Volovik,
Superluminal neutrino and spontaneous breaking of Lorentz invariance,
Pis'ma ZhETF {\bf 94}, 731--733 (2011);  JETP Lett. {\bf 94},  673 (2011); 
arXiv:1109.6624.

\bibitem{Schutzhold2002}
R. Sch\"{u}tzhold,
Small cosmological constant from the QCD trace anomaly?
Phys. Rev. Lett. {\bf 89} 081302 (2002).

\bibitem{UrbanZhitnitsky2009}
F.R. Urban  and A.R. Zhitnitsky,
The QCD nature of dark energy
Nucl. Phys. B {\bf 835} 135 (2010).

\bibitem{Ohta2011}
N. Ohta, 
Dark energy and QCD ghost, 
Phys. Lett. B {\bf 695}, 41--44 (2011);
arXiv:1010.1339.

\bibitem{Holdom2011}
B. Holdom,
From confinement to dark energy,
Phys. Lett. B {\bf 697}, 351--356 (2011);
arXiv:1012.0551

\bibitem{Poplawski2011}
N.J. Poplawski,
Cosmological constant from quarks and torsion,
Annalen Phys. {\bf 523}, 291 (2011);
 arXiv:1005.0893.

\bibitem{UnificationModel}
H. Georgi, S.L. Glashow:
Unity of all elementary particle forces,
Phys.  Rev. Lett.\ \textbf{32}, 438 (1974).


\bibitem{Unification}
H. Georgi, H.R. Quinn, S. Weinberg:
Hierarchy of interactions in unified gauge theories,
Phys. Rev. Lett. \textbf{33}, 451 (1974).

\bibitem{VolovikGorkov1985}
 G.E. Volovik and L.P. Gorkov, 
 Superconductivity  classes in the heavy fermion systems,
JETP  {\bf 61}, 843--854 (1985).


\bibitem{VollhardtWoelfle}
D. Vollhardt, P. W\"olfle:
{\itshape The Superfluid Phases of Helium 3}
(Taylor and Francis, London, 1990)

\bibitem{TopologyReview1} 
N.D. Mermin:  
The topological theory of defects in ordered media,
Rev. Mod. Phys. \textbf {51}, 591 (1979)


\bibitem{Horava2005}  
P. Ho\v{r}ava,
Stability of Fermi surfaces and $K$-theory,
Phys. Rev. Lett. \textbf{95}, 016405 (2005).

\bibitem{Campuzano2008} 
J. C. Campuzano, M. Norman, and M. Randeria in:
{\it  Superconductivity: Physics of Conventional and Unconventional Superconductors},Vol. 1, edited by K. H. Bennemann and J. B. Ketterson (Springer, Berlin, 2008).

 \bibitem{Volovik2007} 
 G.E. Volovik, 
 Quantum phase transitions from topology in momentum space, 
 in:  "Quantum Analogues: From Phase Transitions to Black Holes and Cosmology",  
 eds.  W.G. Unruh and R. Sch\"utzhold, 
 Springer Lecture Notes in Physics {\bf 718} (2007), pp. 31--73;
cond-mat/0601372.

\bibitem{Khodel1990}  V.A. Khodel, V.R.  Shaginyan:
Superfluidity in system with fermion condensate, 
JETP Lett. \textbf{51}, 553 (1990)

\bibitem{NewClass}  G.E. Volovik:
A new class of normal Fermi liquids,   
JETP Lett. \textbf{53}, 222 (1991)



\bibitem{Shaginyan2010}  
V.R. Shaginyan, M.Ya. Amusia, A.Z. Msezane, K.G. Popov,
Scaling behavior of heavy fermion metals,
 Physics Reports,
DOI: 10.1016/j.physrep.2010.03.001
 

\bibitem{Sung-SikLee2009}     
Sung-Sik Lee, 
Non-Fermi liquid from a charged black hole: A critical Fermi ball,
Phys. Rev. D {\bf 79}, 086006 (2009).

\bibitem{SchnyderRyu2010}
 A.P.  Schnyder and S. Ryu, 
Topological phases and flat surface bands in superconductors without inversion symmetry,
 arXiv:1011.1438;
Phys. Rev. B {\bf 84}, 060504(R) (2011).

\bibitem{HeikkilaKopninVolovik2011} 
T.T. Heikkil\"a, N.B. Kopnin and G.E. Volovik,
Flat bands in topological media, 
Pis'ma ZhETF {\bf 94}, 252-- 258 (2011); JETP Lett. {\bf 94}, 233--239(2011);
 arXiv:1012.0905.

\bibitem{SchnyderBrydonTimm2011}
Andreas P. Schnyder, P. M. R. Brydon, Carsten Timm,
Types of topological surface states in nodal noncentrosymmetric superconductors,
Phys. Rev. B {\bf 85}, 024522 (2012);
 arXiv:1111.1207.


\bibitem{KopninSalomaa1991} 
N. B. Kopnin and M. M. Salomaa,
Mutual friction in superfluid $^3$He: Effects of bound states in the vortex core, 
Phys. Rev. B {\bf 44}, 9667--9677 (1991).

\bibitem{Volovik1994} 
G.E. Volovik,
On Fermi condensate: near the saddle point and within  the vortex core, 
JETP Lett.  \textbf{59}, 830 (1994)

\bibitem{Volovik2011a} 
G.E. Volovik,
Flat band in the core of topological defects: bulk-vortex correspondence in topological superfluids with Fermi points,
Pis'ma ZhETF {\bf 93}, 69--72 (2011);  JETP Lett. {\bf 93}, 66--69 (2011);
arXiv:1011.4665.

\bibitem{Kane2005}  
C.L. Kane  and E. Mele,   
$Z_2$ topological order and the quantum spin Hall effect,
Phys. Rev. Lett. {\bf 95}, 146802 (2005).

 \bibitem{Volkov1981} 
B.A. Volkov, A.A. Gorbatsevich, Yu.V. Kopaev and V.V. Tugushev,
Macroscopic current states in crystals,
JETP {\bf 54}, 391--397 (1981). 

\bibitem{VolkovPankratov1985}  
B.A. Volkov and O.A. Pankratov,
Two-dimensional massless electrons in an inverted contact,
JETP Lett. {\bf 42}, 178--181 (1985).

 \bibitem{SalomaaVolovik1988} 
 M.M. Salomaa and  G.E. Volovik, 
 Cosmiclike domain walls in superfluid $^3$He-B: Instantons and diabolical points in (${\bf k}$,${\bf r}$) space, Phys. Rev.  {\bf B~37}, 9298--9311 (1988).


\bibitem{Volovik2009} 
 G.E. Volovik, 
Fermion zero modes at the boundary of superfluid $^3$He-B,
 Pis'ma ZhETF {\bf 90}, 440--442 (2009); JETP Lett. {\bf 90}, 398--401 (2009);
arXiv:0907.5389.

\bibitem{Volovik2010a}
 G.E. Volovik, 
Topological invariants  for Standard Model: from semi-metal to topological insulator,
 JETP Lett. {\bf 91}, 55--61 (2010);
arXiv:0912.0502.

\bibitem{VolovikYakovenko1989} 
G.E. Volovik   and V.M. Yakovenko,  
Fractional charge, spin and statistics of solitons in superfluid $^3$He film, 
J. Phys.: Condens. Matter {\bf 1},  5263--5274 (1989).


\bibitem{Volovik1992b} 
G.E. Volovik, 
{\it Exotic properties of superfluid $^3$He}, 
World Scientific, Singapore, 1992.

\bibitem{SQHE} G.E. Volovik,
Fractional statistics and analogs
  of quantum Hall effect in superfluid $^3$He films. 
  In:  {\itshape Quantum Fluids and Solids - 1989} 
ed. by   G.G.Ihas, Y.Takano  
(AIP Conference  Proceedings , 1989)  \textbf{194},  pp. 136--146.



\bibitem{Haldane2004}  
F.D.M. Haldane,
Berry curvature on the Fermi surface: anomalous Hall effect as a topological Fermi-liquid property,
Phys. Rev. Lett. {\bf 93}, 206602 (2004).



\bibitem{FaridTsvelik2009}
B. Farid and A.M. Tsvelik,
Comment on "Breakdown of the Luttinger sum rule within the Mott-Hubbard insulator", 
arXiv:0909.2886.

\bibitem{Giamarchi2004}
T. Giamarchi, 
 \textit{Quantum physics in one dimension},
Oxford:  University Press (2004).

\bibitem{Georgi2007}
H. Georgi, 
Another odd thing about unparticle physics,
Phys. Lett. {\bf B650}, 275--278 (2007);
arXiv:0704.2457.

\bibitem{LuoZhu2008}
M. Luo and  G. Zhu,  
Some phenomenologies of unparticle physics.
\textit{Phys. Lett.}  B \textbf{659}, 341 (2008).
 
 \bibitem{Gribov1978}
 V.N. Gribov, 
 Quantization of non-Abelian gauge theories,
 Nucl. Phys. {\bf B~139}, 1--19 (1978).

\bibitem{Chernodub2008}
M.N. Chernodub and V.I. Zakharov,
Combining infrared and low-temperature asymptotes in Yang-Mills theories;
Phys. Rev. Lett. {\bf 100}, 222001 (2008).

\bibitem{Burgio2009} 
G. Burgio, M. Quandt and H. Reinhardt,
Coulomb-gauge gluon propagator and the Gribov formula,
Phys. Rev. Lett. {\bf 102}, 032002 (2009) 

\bibitem{Faulkner2010} T. Faulkner, N. Iqbal, Hong Liu, J. McGreevy, D. Vegh,
From black holes to strange metals,
arXiv:1003.1728.

\bibitem{Yakovenko1989} 
V.M. Yakovenko,
Spin, statistics and charge of solitons in (2+1)-dimensional theories,   
Fizika (Zagreb) {\bf 21}, suppl. 3, 231 (1989); 
arXiv:cond-mat/9703195.

\bibitem{SenguptaYakovenko2000} 
K. Sengupta and V.M. Yakovenko,
Hopf invariant for long-wavelength skyrmions in quantum Hall systems for integer and fractional fillings,
Phys. Rev. {\bf B~62}, 4586--4604 (2000).

\bibitem{ReadGreen2000} 
N. Read and D. Green,
Paired states of fermions in two dimensions with breaking of parity and time-reversal symmetries and the fractional quantum Hall effect,
Phys. Rev. {\bf B~61}, 10267--10297 (2000).

\bibitem{So1985} 
H. So,
Induced topological invariants by lattice fermions in odd dimensions,
Prog. Theor. Phys. {\bf 74}, 585--593 (1985).

\bibitem{IshikawaMatsuyama1986} 
K. Ishikawa  and T. Matsuyama,
Magnetic field induced multi component QED in three-dimensions and quantum Hall effect,
Z. Phys. C {\bf 33}, 41--45 (1986). 

\bibitem{IshikawaMatsuyama1987} 
K. Ishikawa and T. Matsuyama,
A microscopic theory of the quantum Hall effect, 
Nucl. Phys. {\bf B~280}, 523--548  (1987).

\bibitem{Matsuyama1987} 
T. Matsuyama,
Quantization of conductivity induced by topological structure of energy-momentum space in generalized QED$_3$,
Progr. Theor. Phys. {\bf 77}, 711--730 (1987).

\bibitem{Jansen1996} 
K. Jansen,
Domain wall fermions and chiral gauge theories,
Phys. Rept. {\bf 273}, 1--54 (1996).

 \bibitem{Volovik2010} 
  G.E. Volovik, 
  Topological superfluid $^3$He-B in magnetic field and Ising variable,
  Pis'ma ZhETF {\bf 91},  215--219 (2010);  JETP Lett. {\bf 91}, 201--205 (2010);
arXiv:1001.1514.

 \bibitem{EssinGurarie2011}
A.M. Essin and V. Gurarie,
Bulk-boundary correspondence of topological insulators from their Green's functions,
Phys. Rev. B {\bf 84}, 125132 (2011).

 \bibitem{ZubkovVolovik2012} 
M.A. Zubkov and  G.E. Volovik,
Topological invariants for the $4D$ systems with mass gap,
Nuclear Physics B {\bf 860}, Issue 2,  295--309 (2012);
arXiv:1201.4185.


\bibitem{Dietl-Piechon-Montambaux2008} 
P. Dietl, F. Piechon  and G. Montambaux,
New magnetic field dependence of Landau levels in a graphenelike structure,
Phys. Rev. Lett. 100, 236405 (2008). 

 \bibitem{Banerjee2009}  
S. Banerjee, R. R. Singh, V. Pardo and W. E. Pickett,
Tight-binding modeling and low-energy behavior of the semi-Dirac point,
Phys. Rev. Lett. {\bf 103}, 016402 (2009).

 \bibitem{HeikkilaVolovik2010} T.T. Heikkil\"a and G.E. Volovik,
Fermions with cubic and quartic spectrum,
Pis'ma ZhETF {\bf 92}, 751--756 (2010); JETP Lett. {\bf 92}, 681--686 (2010);
 arXiv:1010.0393.

 \bibitem{HeikkilaVolovik2011} 
T.T. Heikkil\"a and G.E. Volovik,
Dimensional crossover in topological matter: Evolution of the multiple Dirac point in the layered system to the flat band on the surface,
Pis'ma ZhETF {\bf 93}, 63--68 (2011); JETP Lett. {\bf 93}, 59--65 (2011);
arXiv:1011.4185.



\bibitem{KatsnelsonVolovik2012}  
M.I. Katsnelson and  G.E. Volovik,
Quantum electrodynamics with anisotropic scaling:
Heisenberg-Euler action and Schwinger pair production in the
bilayer graphene,
Pis'ma ZhETF {\bf 95}, (2012);
arXiv:1203.1578.

\bibitem{Zubkov2012b}
M.A. Zubkov,
Schwinger pair creation in multilayer graphene,
arXiv:1204.0138.

\bibitem{HoravaPRL2009}  
P. Ho\v{r}ava, 
Spectral dimension of the Universe in quantum gravity at a Lifshitz point,
Phys. Rev. Lett. {\bf 102}, 161301 (2009).

\bibitem{HoravaPRD2009}  
P. Ho\v{r}ava,
Quantum gravity at a Lifshitz point,
Phys. Rev. D {\bf 79}, 084008 (2009).

\bibitem{Horava2010}
Cenke Xu and P. Ho\v{r}ava,
Emergent gravity at a Lifshitz point from a Bose liquid on the lattice,
Phys. Rev. D {\bf 81}, 104033 (2010).

\bibitem{Fu2007a}
L. Fu, C.L. Kane and E.J. Mele,
 Topological insulators in three dimensions,
Phys. Rev. Lett. {\bf 98}, 106803 (2007).


 \bibitem{PepeaWieseb2007}
 M. Pepea and U.J. Wieseb,
 Exceptional deconfinement in $G(2)$ gauge theory,
Nucl. Phys.  {\bf B~768},  21--37 ( 2007).
 
   \bibitem{Kadastik2009}
M. Kadastik, K. Kannike and M. Raidal,
Dark Matter as the signal of Grand Unification,
Phys. Rev. {\bf D~80}, 085020 (2009).

 \bibitem{KlinkhamerVolovik2005a}
 F.R. Klinkhamer and G.E. Volovik, 
 Emergent CPT violation from the splitting of Fermi points, 
 Int. J.  Mod. Phys. A {\bf 20}, 2795--2812 (2005); 
hep-th/0403037.

  \bibitem{Tarruell2011}
L. Tarruell, D. Greif, T. Uehlinger, G. Jotzu and T. Esslinger,
Creating, moving and merging Dirac points with a Fermi gas in a tunable honeycomb lattice,
arXiv:1111.5020.

 \bibitem{Klinkhamer2011b} 
 F.R. Klinkhamer,
Superluminal neutrino, flavor, and relativity,
arXiv:1110.2146.

\bibitem{Klinkhamer2006}
Klinkhamer, F. R. 2006
Possible new source of T and CP violation in neutrino oscillations.
\textit{Phys. Rev.} D \textbf{73}, 057301.

\bibitem{Osheroff1972} 
D.D. Osheroff, R.C. Richardson, and D.M. Lee,  
Evidence for a new phase of solid He3,  
Phys. Rev. Lett. {\bf 28}, 885--888  (1972).

\bibitem{Creutz2008}
M. Creutz,
Four-dimensional graphene and chiral fermions,
JHEP 04 (2008) 017;
arXiv:0712.1201.

\bibitem{WenZee2002} 
X.G. Wen  and A. Zee,
Gapless fermions and quantum order,
Phys. Rev. {\bf B~66}, 235110 (2002).

\bibitem{Beri2009}
B. Beri,
Topologically stable gapless phases of time-reversal invariant superconductors,
arXiv:0909.5680.

\bibitem{Kitaev2009} A. Kitaev,
Periodic table for topological insulators and superconductors,
AIP Conference Proceedings, Volume {\bf 1134}, pp. 22--30 (2009);
  arXiv:0901.2686.

 \bibitem{Volovik2009b}  G.E. Volovik, 
Topological invariant  for superfluid  $^3$He-B and quantum phase transitions,
JETP Lett. {\bf 90}, 587--591 (2009);
arXiv:0909.3084.

\bibitem{JackiwRossi1981} 
R. Jackiw  and  P. Rossi, 
Zero modes of the vortex-fermion system,
Nucl. Phys. {\bf B~190}, 681--691 (1981).

\bibitem{Kaplan1992}
 D.B. Kaplan,
Method for simulating chiral fermions on the lattice,
Phys. Lett.  B {\bf 288}, 342--347 (1992);
arXiv:hep-lat/9206013.

\bibitem{Golterman1993}
 M.F.L. Golterman, K.  Jansen and D.B. Kaplan,
Chern-Simons  currents and chiral  fermions on the lattice,
 Phys.Lett. B {\bf 301}, 219--223 (1993):
arXiv: hep-lat/9209003.
  
\bibitem{Kaplan2011}
D.B. Kaplan and Sichun Sun,
Spacetime as a topological insulator,
arXiv:1112.0302.

\bibitem{MineevVolovik1978} 
V.P. Mineev and G.E. Volovik, 
Planar and linear solitons in superfluid $^3$He,
Phys. Rev. B {\bf 18}, 3197--3203 (1978).

 
\bibitem{Volovik1988} 
G.E. Volovik,  
Analog of quantum Hall effect in superfluid $^3$He film,
JETP  {\bf 67}, 1804--1811 (1988).


\bibitem{Volovik1989c} 
G.E. Volovik,
Gapless fermionic excitations on the  quantized vortices in superfluids and superconductors, 
JETP Lett. {\bf 49}, 391--395 (1989).

\bibitem{GrinevichVolovik1988} 
P.G. Grinevich and G.E. Volovik, 
Topology of gap nodes in superfluid $^3$He: 
$\pi_4$ homotopy group for $^3He-B$ disclination,
J. Low Temp. Phys. {\bf 72}, 371--380  (1988).

\bibitem{TeoKane2010a}
J.C.Y. Teo and C.L. Kane, 
Majorana fermions and non-Abelian statistics in three dimensions,
Phys. Rev. Lett. {\bf 104}, 046401 (2010).

\bibitem{TeoKane2010b}
J.C.Y. Teo  and  C.L. Kane, 
Topological defects and gapless modes in insulators and superconductors,
Phys. Rev. B {\bf 82}, 115120 (2010).

\bibitem{SilaevVolovik2010}
M.A. Silaev and G.E. Volovik,
Topological superfluid $^3$He-B: fermion zero modes on interfaces and in the vortex core,
J. Low Temp. Phys. {\bf 161},  460--473 (2010);
arXiv:1005.4672.

 \bibitem{Klitzing1980}   
 K. v. Klitzing, G. Dorda and  M. Pepper 
 New method for high-accuracy determination of the fine-structure constant based on quantized Hall
 resistance,
Phys. Rev. Lett. {\bf 45}, 494--497 (1980),



 \bibitem{Awschalom2009}   
D. Awschalom and N. Samarth,
Spintronics without magnetism,
Physics {\bf 2}, 50 (2009).

\bibitem{HaldaneArovas1995}   
F.D.M. Haldane and D.P. Arovas,
Quantized spin currents in 2-dimensional chiral magnets,
Phys. Rev. {\bf B~52}, 4223 (1995).

\bibitem{Senthil1999}    
T. Senthil, J.B. Marston and M.P.A. Fisher,
The spin quantum Hall effect in unconventional superconductors,
Phys. Rev. B {\bf 60}, 4245--4254 (1999).

 \bibitem{GurarieRadzihovsky2007}  
V. Gurarie and  L. Radzihovsky,
Resonantly-paired fermionic superfluids,
Ann. Phys. {\bf 322}, 2--119 (2007).


\bibitem{Haldane1988} 
 F.D.M. Haldane,
 Model for a quantum Hall effect without Landau levels: Condensed-matter realization of the "Parity Anomaly",
Phys. Rev. Lett. {\bf 61}, 2015--2018 (1988).

\bibitem{Schnyder2008} 
A.P. Schnyder, S. Ryu, A. Furusaki and A.W.W. Ludwig, 
Classification of topological insulators and superconductors in three spatial dimensions,
Phys. Rev. {\bf B~ 78}, 195125 (2008).


\bibitem{Volovik1989d} 
G.E. Volovik, 
Zeroes in energy gap in superconductors with high transition temperature,
 Phys. Lett. A {\bf 142} 282--284 (1989).
  
\bibitem{Gubankova2005} 
E. Gubankova, A, Schmitt and F. Wilczek,
Stability conditions and Fermi surface topologies in a superconductor,
Phys.Rev. B {\bf 74}, 064505  (2006).

\bibitem{Zubkov2012a} 
 M. A. Zubkov,
Generalized unparticles, zeros of the Green function, and momentum space topology of the lattice model with overlap fermions,
 arXiv:1202.2524.


\bibitem{NielsenNinomiya1983}
H.B. Nielsen and M. Ninomiya, 
The Adler-Bell-Jackiw anomaly and Weyl fermions in a crystal,
Phys. Lett. {\bf 130~B}, 389--396 (1983).


\bibitem{AvronSeilerSimon1983}
J.E. Avron, R. Seiler and B. Simon,
Homotopy and quantization in condensed matter physics,
Phys. Rev. Lett. {\bf 51}, 51--53 (1983).

\bibitem{Semenoff1984}  
G.W. Semenoff,
Condensed-matter simulation of a three-dimensional anomaly,
Phys. Rev. Lett. {\bf 53}, 2449--2452 (1984).

\bibitem{NiuThoulessWu1985} 
Qian Niu, D. J. Thouless, and Yong-Shi Wu,
Quantized Hall conductance as a topological invariant,
Phys. Rev. {\bf B~31}, 3372--3377 (1985).

\bibitem{Hsieh2009a}   
D. Hsieh,
{\it  The experimental discovery of topological insulators},
 Barnes \& Noble, 2009.  

\bibitem{Kopnin1991}   
N.B. Kopnin, P.I. Soininen and M.M. Salomaa, 
Parameter-free quasiclassical boundary conditions
for superfluid $^3$He at rough walls. B-phase order
parameter and density of states,
J. Low Temp. Phys. {\bf 85}, 267--282 (1991).

\bibitem{CCGMP} C.A.M. Castelijns, K.F. Coates, A.M. Gu\'enault, S.G. Mussett and G.R. Pickett,
Landau critical velocity for a macroscopic object moving in superfluid $^3$He-B: evidence for
gap suppression at a moving surface, 
Phys. Rev. Lett.  {\bf 56}, 69--72 (1986).

\bibitem{Davis2008} 
J.P. Davis, J. Pollanen, H. Choi, J.A. Sauls, W.P. Halperin and A.B. Vorontsov,
Anomalous attenuation of transverse sound in $^3$He,
Phys. Rev. Lett. {\bf 101}, 085301 (2008).

\bibitem{Nagai2008} 
K. Nagai,  Y. Nagato, M. Yamamoto and S. Higashitani,
Surface bound states in superfluid $^3$He,
J.  Phys.  Soc.  Jap. {\bf 77},   111003 (2008).

 

  \bibitem{Murakawa2009c} 
S. Murakawa, Y. Tamura, Y. Wada, M. Wasai, M. Saitoh, Y. Aoki, R.
Nomura, Y. Okuda, Y. Nagato, M. Yamamoto, S. Higashitani and K. Nagai,
New anomaly in transverse acoustic impedance of superfluid
$^3$He-B with a wall coated by several layers of $^4$He,
 Phys. Rev. Lett. {\bf 103}, 155301 (2009). 

\bibitem{ChungZhang2009}
Suk Bum Chung, Shou-Cheng Zhang,
Detecting the Majorana fermion surface state of $^3$He-B through spin relaxation,
Phys. Rev. Lett. {\bf 103}, 235301 (2009);
arXiv:0907.4394.


\bibitem{Nagato2009}   Y. Nagato, S. Higashitani and K. Nagai,
   Strong anisotropy in spin suceptibility of superfluid He-3-B film caused by surface bound states,
   J. Phys. Soc. Japan    {\bf 78}, 123603  (2009).

 \bibitem{Schnyder2009a} 
A.P. Schnyder, S. Ryu, A. Furusaki and A.W.W. Ludwig, 
Classification of topological insulators and superconductors,
 AIP Conf. Proc. {\bf 1134}, 10--21 (2009);    arXiv:0905.2029.


 \bibitem{NielsenNinomiya1981}
H.B. Nielsen, M. Ninomiya: Absence of neutrinos on a lattice.  I - Proof by
homotopy theory, Nucl. Phys. B \textbf{185}, 20  (1981); Absence of neutrinos
on a lattice. II - Intuitive homotopy proof, Nucl. Phys. B \textbf{193}, 173
(1981).


 
 








 \end{thebibliography}
\end{document}